
     \magnification=\magstep1
     \hsize=5 in
     \vsize=6.75 in
     \hoffset=.25 in
     \voffset=.5 in
     \def\half{{\textstyle{1\over 2}}}
     \def\tilde{\widetilde}
     \def\bar{\overline}
     \def\hat{\widehat}
     \def\dsl{\raise.15ex\hbox{/}\kern-.57em\partial}
     \def\Dsl{\,\raise.15ex\hbox{/}\mkern-13.5mu D}

\nopagenumbers
\centerline{{\bf Tau functions for the Dirac operator}}
\centerline{{\bf on the Poincar\'e disk}}
\vskip.2in \par\noindent
\centerline{\sl John Palmer }
\centerline{\sl Department of Mathematics }
\centerline{\sl University of Arizona }
\centerline{\sl Tucson, AZ 85721 }
\vskip.1in \par\noindent
\centerline{\sl Morris Beatty }
\centerline{\sl Department of Mathematics }
\centerline{\sl and }
\centerline{\sl Institute of Theoretical Dynamics }

\centerline{\sl University of California }
\centerline{\sl Davis, CA 95616 }
\vskip.1in \par\noindent
\centerline{\sl Craig A.~Tracy }
\centerline{\sl Department of Mathematics }
\centerline{\sl and }
\centerline{\sl Institute of Theoretical Dynamics }
\centerline{\sl University of California }
\centerline{\sl Davis, CA 95616 }
\vfill\eject

\centerline{{\bf Table of Contents}}
\par\noindent
Introduction---1
\vskip.05in \par\noindent\
\S 1 The Dirac operator on the Hyperbolic disk
\vskip.02in \par\noindent
\hskip0.20in A covering of the frame bundle---4
\vskip.02in \par\noindent
\hskip0.20in The Dirac operator---7
\vskip.02in \par\noindent
\hskip0.20in Covariance of the Dirac operator---10
\vskip.05in \par\noindent
\S 2 Local Expansions
\vskip.02in \par\noindent
\hskip0.20in Eigenfunctions for infinitesimal rotations---14
\vskip.02in \par\noindent
\hskip0.20in Differentiating local expansions with respect to the branch
points---21
\vskip.02in \par\noindent
\hskip0.20in Estimates at infinity---24
\vskip.05in \par\noindent
\S 3 $L^2$ existence results
\vskip.02in \par\noindent
\hskip0.20in A model for the simply connected covering $\tilde {D}_
R(a)$---28
\vskip.02in \par\noindent
\hskip0.20in Multivalued solutions with specified branching---30
\vskip.02in \par\noindent
\hskip0.20in Existence for a canonical $L^2$ basis---35
\vskip.02in \par\noindent
\hskip0.20in The response functions---38
\vskip.05in \par\noindent
\S 4 The Green function $G^{a,\lambda}$ for the Dirac operator
\vskip.02in \par\noindent
\hskip0.20in Green functions in the absence of branch points---40
\vskip.02in \par\noindent
\hskip0.20inThe Green function for the Helmholtz operator with branch
points---44
\vskip.02in \par\noindent
\hskip0.20in The Green function for the Dirac operator with branch points---52
\vskip.02in \par\noindent
\hskip0.20in The derivative of the Green function $G^{a,\lambda}$---54
\vskip.05in \par\noindent
\S 5 Deformation Equations
\vskip.02in \par\noindent
\hskip0.20in A holonomic system---58
\vskip.02in \par\noindent
\hskip0.20in The holonomic extension---63
\vskip.02in \par\noindent
\hskip0.20in The deformation equations---64
\vskip.02in \par\noindent
\hskip0.20in Further identities---66
\vskip.02in \par\noindent
\hskip0.20in The two point deformation theory in a special case---71
\vskip.05in \par\noindent
\S 6 The tau function
\vskip.02in \par\noindent
\hskip0.20in The Grassmannian formalism---80
\vskip.02in \par\noindent
\hskip0.20in The log derivative of $\tau$ in the two point case---88
\vskip.05in \par\noindent
Bibliography---91
\vfill\eject

\footline={\hss\tenrm\folio\hss}
\pageno=1\hfill\break
\centerline{{\bf Introduction}}
\vskip.2in \par\noindent
In this paper we introduce $\tau$-functions for the Dirac operator on
the Poincar\'e disk based on the formalism in [12].  The formalism
of the first author
in [11,12] is in turn a geometric reworking of the analysis of Sato,
Miwa, and Jimbo [15] (SMJ henceforth).  The prototypical
examples of the holonomic fields that are the central objects of the
SMJ theory are the scaling limits of the two dimensional Ising model
from above and below the critical point.  The SMJ theory of
holonomic fields provides a beautiful setting for the earlier
result of Wu, McCoy, Tracy, and Barouch [1,8,17,20] that the
scaled two point functions of the two dimensional Ising model
can be expressed in terms of Painlev\'e functions of the third kind.
Our work also owes something to the papers of
[18,6] on monodromy preserving
deformation theory in the Poincar\'e disk.  In the course of explaining
our work we will point out differences with [18].

Our principal goal in this paper is to define and analyze the
correlation functions
for a family of quantum field theories on the Poincar\'e disk that
are analogues of the holonomic quantum field theories defined
in the Euclidean plane.  Below we will sketch how the theory goes
in the Euclidean version in [12] and we will point the reader to the
parallel developments in the hyperbolic case.

Our starting point is a formula for the Schwinger functions of
holonomic fields that can be written
$$\tau (a_1,a_2,\ldots ,a_n)=\hbox{det}(D^{a,\lambda}),\eqno (1)$$
where $\tau (a_1,a_2,\ldots ,a_n)$ is the vacuum expectation for a product
of quantum fields $\Phi_{\lambda_j}(a_j)$ [11].  In this formula the operator $
D^{a,\lambda}$ is
a singular Dirac operator (on Euclidean space) with a domain
that incorporates functions
with prescribed branching at the points $a_j$.  The functions
in the domain of $D^{a,\lambda}$ are to have monodromy multiplier $
e^{i2\pi\lambda_j}$ in
a counterclockwise circuit of $a_j$.  In [12] formula (1) is given rigorous
mathematical sense by first localizing $D^{a,\lambda}$ away from branch cuts
emerging from the points $a_j$.  The localization for $D^{a,\lambda}$ takes
place
in the exterior of a set $S$ which is the union of strips containing the
branch cuts.  The localization of $D^{a,\lambda}$ is completely characterized
by
a family of subspaces $W^{a,\lambda}\subset H^{{1\over 2}}(\partial
S)$ of the Sobolev space on the
boundary, $\partial S$, of $S$.  These subspaces are the restriction to
$\partial S$ of functions which are ``locally'' in the null space of $
D^{a,\lambda}$.
The subspaces $W^{a,\lambda}$ are contained in a
restricted Grassmannian of subspaces of $H^{{1\over 2}}(\partial
S)$ over which there
sits a determinant line bundle [16].  As is explained in more detail in
[11,12], defining a determinant for the family $D^{a,\lambda}$ is morally
equivalent to finding a trivialization $\delta$ for the $\hbox{det}^{
*}$ bundle over the
family of subspaces $W^{a,\lambda}.$  Comparing the canonical section
$\sigma$ of $\hbox{det}^{*}$ with $\delta$ gives us a formula for the
determinant,
$$\hbox{det}(D^{a,\lambda})={{\sigma (W^{a,\lambda})}\over {\delta
(W^{a,\lambda})}}\,.\eqno (2)$$
The analogue of this formula on the Poincar\'e disk is (6.7) below.
Both the canonical section, $\sigma ,$ and the
trivialization, $\delta ,$ are determined by (different) projections on the
subspaces $W^{a,\lambda}$.  An explicit representation of these projections is
obtained in terms of the Green function $G^{a,\lambda}$ for $D^{a
,\lambda}$ .  The
analogue of this projection formula on the Poincar\'e disk is (6.2)
below.  In order to analyze (2) it is useful to compute the
logarithmic derivative
$$d_a\log\tau .$$
The resulting formula simplifies dramatically because the derivative
of the appropriate projection in the $a$ variables is {\it finite rank}.  This
in turn is a consequence of the analogous result for the Green
function, $G^{a,\lambda}$, namely,
$$\eqalign{\partial_{a_{\nu}}G^{a,\lambda}(x,y)&=c_{\nu}\overline {
W_{\nu}^{*}(x)}\otimes W_{\nu}(y),\cr
\bar{\partial}_{a_{\nu}}G^{a,\lambda}(x,y)&=c_{\nu}\overline {W_{
\nu}(x)}\otimes W_{\nu}^{*}(y).\cr}
\eqno (3)$$
For simplicity we have taken some liberties in displaying only a
schematic version of (3).  The precise version of (3) in the Poincar\'e
disk is (4.50) below.  The functions $W_{\nu}(x)$ and $W_{\nu}^{*}
(x)$ are multivalued
solutions to the Dirac equation which are branched at the points
$a_j$, $j=1,2,\ldots ,n$ with monodromy multipliers $e^{2\pi i\lambda_
j}$.  Because of their
role in (3) we refer to the wave functions $W_{\nu}(x)$ and $W_{\nu}^{
*}(x)$ as
{\it response functions}.  The formula for $d\log\tau$ which one obtains is
$$d\log\tau =\sum_{\nu =1}^n\left\{A_{\nu}da_{\nu}+B_{\nu}d\bar a_{
\nu}\right\},\eqno (4)$$
where $A_{\nu}$ and $B_{\nu}$ are low order Fourier coefficients of the
response
functions $W_{\nu}$ and $W_{\nu}^{*}$ in the local expansions of these
functions about
the points $a_{\nu}$.  The version of this formula in the Poincar\'e disk is
(6.8) below.

We assemble the $2n$ response functions into
a single column vector
$${\bf W}(x):=\left[\matrix{W(x)\cr
W^{*}(x)\cr}
\right],$$
with
$$W(x)=\left[\matrix{W_1(x),W_2(x),\cdots ,W_n(x)\cr}
\right]^{\tau},$$
and
$$W^{*}(x)=\left[\matrix{W_1^{*}(x),W_2^{*}(x),\cdots ,W_n^{*}(x)\cr}
\right]^{\tau},$$
where $[\cdot ]^{\tau}$ is the transpose.  Then ${\bf W}(x)$ satisfies a
holonomic
system
$$d_{x,a}{\bf W}(x)=\Omega (x,a){\bf W}(x).\eqno (5)$$
in the variables $(x,a)$, where $\Omega$ is a matrix valued one-form whose
entries are determined by the low order local expansion
coefficients for $W_{\nu}$ and $W_{\nu}^{*}$ near the branch points $
a_j$.  The
analogue of (5) in the Poincar\'e disk is contained in (5.3) and (5.20)
below.  The consistency condition for the holonomic system (5) is the
zero curvature condition
$$d\Omega =\Omega\wedge\Omega .\eqno (6)$$
These are called the {\it deformation equations}.  The analogue in the
Poincar\'e disk is (5.26) below.  These deformation equations
characterize certain of the low order expansion coefficients of
$W_{\nu}$ and $W_{\nu}^{*}$.  In particular it is possible to express the
coefficients
$A_{\nu}$ and $B_{\nu}$ that occur in the formula for the logarithmic
derivative of
the $\tau -$function in terms of the solution to (6).  The version of this
relationship in hyperbolic space can be found in (6.16) and (6.17) below.
The existence of the holonomic system and the resultant
deformation equations is, incidently, the one part of this
theory that depends strongly on the fact that the Poincar\'e
disk is a symmetric space.  The global symmetries of Dirac
operator on the Poincar\'e disk give rise to the holonomic
system for {\bf W. }

In conclusion one sees that it is possible to characterize the
logarithmic derivative of the $\tau -$function in terms of certain solutions
to the deformation equations.  In the special case of the two point
function in the Euclidean problem SMJ showed that it is possible to
integrate the deformation equations explicitly in terms of Painlev\'e
transcendents of the fifth kind.  In the special Ising case a
reduction to Painlev\'e III is possible (see [8]).
In \S 5 and \S 6 of this paper we carry out the
analysis of the two point case in the hyperbolic setting.  We find that
the two point deformation equations can be integrated in terms of
Painlev\'e transcendents of type VI (Theorem 5.0) and that the
logarithmic derivative of the $\tau -$function is a rational function of
the solution to $P_{VI}$ and its derivative (see (6.19)).

We can now explain the principal differences between this paper
and [18].  First, because the response functions are fundamental in
this approach to holonomic fields, the basic finite dimensional space
of wave functions for us is the $2n$ dimensional space spanned by
$W_{\nu}$ and $W_{\nu}^{*}.$  In [18], as in SMJ [15], the basic finite
dimensional space is
the $n$ dimensional space spanned by the wave functions
that are globally in $L^2.$  Formulating the holonomic system in this
setting leads to extra complications, and the deformation equations
are not explicitly worked out in [18].  Because of the difficulties with
a direct formulation of the deformation equations, one of us introduced
a hyperbolic Laplace transform of the associated holonomic system
and found
that the result was a Fuchsian system of ordinary
differential equations in the complex plane.  Tau functions for
this transformed system were introduced by adopting the $\tau -$functions
for the associated Schlesinger theory (see also [15]).  In [6] the behavior of
these
$\tau -$functions was examined in the limit $R\rightarrow\infty$ and shown to
coincide
with the $\tau$-functions for the Euclidean theory.  We have yet to
explore the relation between the $\tau -$functions introduced in [6,18] and the
ones
we consider in this paper. The algebraic complexity of both
sets of deformation equations makes comparisons difficult.

This last point suggests an important question.  Why should one
believe that the $\tau -$functions we have introduced in this paper are the
Schwinger functions for some quantum field theory on the Poincar\'e
disk?  The reason we believe that there is such a relation is to be
found in [12].  In that paper a transfer matrix formalism was
established that allowed one to connect formula (2) above with
the scaling limit of some lattice field theories defined in [10].  It is
also this connection that suggested the ``minimal singularity''
definition for the domain of $D^{a,\lambda}$ that we adopt here.  We do not
think that it would be economical to try to produce a lattice theory
which would scale  to the theory we present here.  Because of the
connection with the Ising model, however, this is an interesting
problem.  We should remark at this point that the analogue of
the Ising model on the Poincar\'e disk is not considered in this
paper.  There is such an analogue and the deformation theory
in \S 5 is relevant to its analysis, but the singular Dirac operator
whose determinant gives the correlation functions (for the
$T<T_c$ scaling limit) is different from any of the minimal
singularity theories for $\lambda_j=\pm 1/2$.  We expect to discuss this
analogue in another place.  To return to the quantum field
theory connection; there is a preprint of
Segal on conformal field theory which sketches a ``vertex
operator'' formalism that appears well suited to address such
questions as Osterwalder-Schrader positivity.  One of us is
currently investigating this formalism.

Finally we would like to mention that this paper is organized in
almost precisely the reverse of the logical progression (1)---(6) above.
The reason for this is that the existence theory for the response
functions and the Green function proceeds most simply in such
reverse order.  We have included a table of contents to help the
reader obtain some perspective.

The authors would like
to express their gratitude to R.~Narayanan whose notes on
local expansions on the disk were very helpful in the early
stages of this work.  The second two authors wish to thank
T.~Miwa for his hospitality and support at RIMS, Kyoto
University and to the National Science Foundation whose
support through grants DMS--9001794, INT--9106953, and
DMS--9303413 is gratefully acknowledged.
\vfill\eject
\centerline{\S {\bf 1 The Dirac Operator on the Hyperbolic Disk }}
\vskip.2in \par\noindent
{\bf A covering of the frame bundle}. We will begin by identifying a
Dirac operator on the hyperbolic
disk, ${\bf D}_R$, of radius $R$.  The Dirac operator will be defined in the
usual fashion except that we will use the simply connected covering
of SO(2) rather than the two fold covering in our construction.  This
is done principally because the extra
parameter $k$ associated with the representation theory of
the reals does not significantly complicate any of our analysis
and we believe its inclusion might prove illuminating.  Incidently,
the radius $R$ is carried along in all the subsequent calculations
precisely because the zero curvature limit $R\rightarrow\infty$ is interesting
[6].

The additive group of real numbers,
${\bf R}$, is the simply connected covering group of the special orthogonal
group, SO(2), in two dimensions with
the familiar covering map
$${\bf R}\ni\lambda\rightarrow R(\lambda ):=\left[\matrix{\cos\lambda&
-\sin\lambda\cr
\sin\lambda&\cos\lambda\cr}
\right]\in\hbox{SO}(2).$$
To emphasize the role {\bf R} plays as a covering group for SO(2) we will
sometimes write
$${\bf R}=\tilde{\hbox{SO}}(2).$$
Next we define a principal $\tilde{\hbox{SO}}(2)$ bundle on ${\bf D}_
R$, which we call
$S({\bf D}_R)$ together with a bundle map,
$$\pi :S({\bf D}_R)\rightarrow F({\bf D}_R),$$
onto the frame bundle, $F({\bf D}_R)$, that is compatible with the map $
R$.
If $g\in\tilde{\hbox{SO}}(2)$ and $s\rightarrow sg$ is the right action of
$\hbox{$
\tilde{\hbox{SO}}(2)$}$ on the
bundle $S({\bf D}_R)$ then the compatibility we refer to is,
$$\pi (sg)=\pi (s)R(g),\eqno (1.1)$$
where $R(g)\in\hbox{SO}(2)$ acts on the frame bundle in the usual fashion.
By fixing a trivialization of the frame bundle for ${\bf D}_R$ we will
be able to use (1.1) to define the bundle S(${\bf D}_R).$  The metric
we wish to consider on ${\bf D}_R$ is,
$$ds^2:={{R^4\left(dx_1^2+dx_2^2\right)}\over {\left(R^2-x_1^2-x_
2^2\right)^2}},\eqno (1.2)$$
with the associated invariant measure
$$d\mu :={{R^4dx_1dx_2}\over {\left(R^2-x_1^2-x_2^2\right)^2}}\,.\eqno
(1.3)$$
We fix an orthonormal frame,
$$e_1:=\left(1-{{x_1^2+x_2^2}\over {R^2}}\right){{\partial}\over {
\partial x_1}},$$
$$e_2:=\left(1-{{x_1^2+x_2^2}\over {R^2}}\right){{\partial}\over {
\partial x_2}},$$
$$ $$
which identifies the frame bundle, $F({\bf D}_R)$ with
$${\bf D}_R\times\hbox{SO}(2).$$
Relative to this trivialization of the frame bundle we identify
$S({\bf D}_R)={\bf D}_R\times\tilde{\hbox{SO}}(2)$ and the map $\pi$ with the
covering given by
$${\bf D}_R\times\tilde{\hbox{SO}}(2)\ni (p,\lambda )\rightarrow\pi
(p,\lambda )=(p,R(\lambda ))\in {\bf D}_R\times\hbox{SO}(2)$$
for $p\in {\bf D}_R.$  There is a covariance of this construction with
respect to global rigid motions of ${\bf D}_R$ that it will be useful for
us to understand.  Since the (orientation preserving) rigid motions
of ${\bf D}_R$ are simplest
to understand as fractional linear transformations we will introduce
the complex notation $z=x_1+ix_2$.  Consider the fractional linear
transformation
$${w\over R}={{A{z\over R}+B}\over {\bar {B}{z\over R}+\bar {A}}}\,
.\eqno (1.4)$$
This will be an orientation preserving rigid motion of ${\bf D}_R$ provided
that
$$\left[\matrix{A&B\cr
\bar {B}&\bar {A}\cr}
\right]\in\hbox{SU}(1,1),$$
where $g\in\hbox{SU}(1,1)$ if and only if
$$g^{*}\left[\matrix{1&0\cr
0&-1\cr}
\right]g=\left[\matrix{1&0\cr
0&-1\cr}
\right]$$
and $\hbox{det}(g)=1$.  The action of a transformation (1.4) on the frame
$[e_1,e_2]$ is simplest in the complexified tangent space
and so we also introduce the complex vector fields
$$e:=\half\left(e_1-ie_2\right)$$
and
$$\bar {e}:=\half\left(e_1+ie_2\right).$$
For $g=\left[\matrix{A&B\cr
\bar {B}&\bar {A}\cr}
\right]$ write
$$g(z)=R{{A{z\over R}+B}\over {\bar {B}{z\over R}+A}}\,.$$
For $g\in\hbox{SU}(1,1)$ the pair $(g,dg)$ induces a transformation of the
frame bundle which we calculate in the trivialization $\{e_1,e_2\}$ by first
determining the action of $dg$ on the complex vector fields $e$ and
$\bar {e}$.  One finds
$$dg_z(e(z))=u_g(z)e(g(z)),\eqno (1.5)$$
and
$$dg_z(\bar {e}(z))=\overline {u_g(z)}\bar {e}(g(z)),\eqno (1.6)$$
where
$$u_g(z):={{B{{\bar {z}}\over R}+A}\over {\bar {B}{z\over R}+\bar {
A}}}\eqno (1.7)$$
is a number of absolute value 1.  Now write
$$g'(z)={1\over 2}\left[\matrix{u_g(z)+\overline {u_g(z)}&i(u_g(z
)-\overline {u_g(z)})\cr
-i(u_g(z)-\overline {u_g(z)})&u_g(z)+\overline {u_g(z)}\cr}
\right].$$
Then without difficulty one sees that (1.5) and (1.6) imply
$$(dg_ze_1(z),dg_ze_2(z))=(e_1(g(z),e_2(g(z)))g'(z).\eqno (1.8)$$
So that in the trivialization of the frame bundle determined
by $\{e_1,e_2\}$ the map induced by $(g,dg)$ is
$$(z,r)\rightarrow (g(z),g'(z)r),\eqno (1.9)$$
where $r\in\hbox{SO}(2)$.  Next we will show that the map (1.9) is covered
by a map on $S({\bf D}_R)$.  For fixed $g\in\hbox{SU}(1,1)$ this clearly comes
down
to defining a continuous logarithm for  ${\bf D}_R\ni z\rightarrow
u_g(z)$ .  Since
the map $u_g(z)$ is continuous $(|A|^2-|B|^2=1$ for $g\in\hbox{SU}
(1,1)$ implies
that $\bar {B}{z\over R}+\bar {A}$ is never 0 for $z\in {\bf D}_R
)$ it follows that the image of
the map $u_g(z)$ is simply connected and one can find a continuous
map $z\rightarrow\lambda_g(z)$ so that
$$u_g(z)=e^{i\lambda_g(z)}.\eqno (1.10)$$
The bundle map,
$$(z,\lambda )\rightarrow (g(z),\lambda +\lambda_g(z)),\eqno (1.1
1)$$
on ${\bf D}_R\times\tilde{\hbox{SO}}(2)$ then clearly covers the the
transformation
(1.9) on the frame bundle.  Of course, the map $\lambda_g(z)$ is not unique;
for
any integer $n$ the map $\lambda_g(z)+2n\pi$ would serve as well.  The
fundamental group of SU(1,1) is {\bf Z} and so we do not expect to make
a canonical choice of such a logarithm for all elements of SU(1,1).
We do expect that the group of bundle transformations of $S({\bf D}_
R)$
which covers the action of SU(1,1) on $F({\bf D}_R)$ is the simply connected
covering group $\tilde{\hbox{SU}}(1,1)$.  However, since there is no simple
(i.e.
matrix) model for $\tilde{\hbox{SU}}(1,1)$ and since we are more interested in
the
lifts of individual elements and of one parameter subgroups in SU(1,1)
we will not pursue this matter.  However, it will be convenient for
descriptive purposes to refer to a pair $(g,\lambda_g(z))$ where $
\lambda_g$ satisfies
(1.10) as an element of $\tilde{\hbox{SU}}(1,1).$
\vskip.1in \par\noindent
{\bf The Dirac operator}.  We will now introduce
a Dirac operator and observe that the lift of the action of an element
in SU(1,1) to the bundle $S({\bf D}_R)$ also acts on the
solution space of the Dirac equation.

The Dirac operator we define will act in the space of sections of an
associated bundle for $S({\bf D}_R)$.  The bundle we
have in mind is one associated with a complex unitary
representation of $\tilde{\hbox{SO}}(2)$ given by
$$\tilde{\hbox{SO}}(2)\ni\lambda\rightarrow r_k(\lambda ):=\left[\matrix{
e^{-i\left(k+{1\over 2}\right)\lambda}&0\cr
0&e^{-i\left(k-{1\over 2}\right)\lambda}\cr}
\right],\eqno (1.12)$$
where $k$ is an arbitrary real number.  We write
$$S({\bf D}_R)\times_k{\bf C}^2$$
for this associated bundle.  One can view the elements of this bundle
in the usual fashion as equivalence classes of pairs $(s,v)$ where
$s\in S({\bf D}_R)$ and $v\in {\bf C}^2$ and $(s,v)$ is equivalent to $
(s\lambda ,r_k(\lambda )v)$ for any
$\lambda\in\tilde{\hbox{SO}}(2)$.  Because we are working with a fixed
trivialization of
the bundle $S({\bf D}_R)$, the associated bundle $S({\bf D}_R)\times_
k{\bf C}^2$ can be trivialized
by the map
$$(z,\lambda ,v)\rightarrow (z,r_k(\lambda )^{-1}v)\in {\bf D}_R\times
{\bf C}^2.$$
To avoid confusion we will always work in this trivialization of the
associated bundle.

The reason for choosing the representation (1.12) to define
a bundle of spinors for the Dirac operator has to do with the
existence of a complex representation of the Clifford algebra of
${\bf R}^2$ on ${\bf C}^2$ which behaves well with respect to the
representation (1.12).
The representation is generated by
$$\gamma_1:=\left[\matrix{0&1\cr
1&0\cr}
\right]$$
and
$$\gamma_2:=\left[\matrix{0&-i\cr
i&0\cr}
\right].$$
If we define
$$\gamma (v):=\gamma_1v_1+\gamma_2v_2,$$
then one can easily confirm that
$$r_k(\lambda )\gamma (v)r_k(\lambda )^{-1}=\gamma (R(\lambda )v)
.\eqno (1.13)$$
This is the relation that allows one to define a Dirac operator that
will have the appropriate covariance with respect to $\tilde{\hbox{SU}}
(1,1)$.

In order to write down the Dirac operator we first recall some
facts about connections.  On a principal bundle, once a local frame
has been picked, a connection is represented by a Lie algebra valued
one form over the domain of the trivialization in the base.  In our
case we start with the global frame $\{e_1,e_2\}.$ The Levi-Civita
connection form $\omega$ for this frame (which is determined by the
torsion 0 condition $\nabla_{e_1}e_2-\nabla_{e_2}e_1-[e_1,e_2]=0)$ is
$$\omega (e_1)=-{{2x_2}\over {R^2}}\left[\matrix{0&-1\cr
1&0\cr}
\right]$$
and
$$\omega (e_2)={{2x_1}\over {R^2}}\left[\matrix{0&-1\cr
1&0\cr}
\right].$$
Since the map $\pi :S({\bf D}_R)\rightarrow F({\bf D}_R)$ is a covering map on
the fiber
one can pull back the connection on $F({\bf D}_R)$ to a connection, $
\hat{\omega },$ on
$S({\bf D}_R)$.  One finds the connection one form relative to our
standard trivialization of $S({\bf D}_R)$ is
$$\hat{\omega }(e_1)=-{{2x_2}\over {R^2}},$$
$$\hat{\omega }(e_2)={{2x_1}\over {R^2}},$$
where the Lie algebra of $\tilde{\hbox{SO}}(2)$ is identified with {\bf R} in
the usual fashion.
The connection $\hat{\omega}$ on the principal bundle $S({\bf D}_
R)$ induces a connection
on the associated bundle $S({\bf D}_R)\times_k{\bf C}^2$.  In the
trivialization of the
associated bundle defined above one finds the connection one form
$$dr_k\hat{\omega }(e_1)=\left[\matrix{i\left(k+{1\over 2}\right)&
0\cr
0&i\left(k-{1\over 2}\right)\cr}
\right]{{2x_2}\over {R^2}},$$

$$dr_k\hat{\omega }(e_2)=-\left[\matrix{i\left(k+{1\over 2}\right
)&0\cr
0&i\left(k-{1\over 2}\right)\cr}
\right]{{2x_1}\over {R^2}}\,.$$
The Dirac operator (in the trivialization ${\bf D}_R\times {\bf C}^
2$ defined above for
the associated bundle) is then

$$D_k=\sum_{j=1}^2\gamma_j\left(e_j+dr_k\hat\omega (e_j)\right)\eqno
(1.14)$$
or in complex notation
$$D_k=\left[\matrix{0&K_{k-{1\over 2}}\cr
-K_{k-{1\over 2}}^{*}&0\cr}
\right]$$
where
$$K_{k-{1\over 2}}:=2\left\{\left(1-{{|z|^2}\over {R^2}}\right)\partial_
z-{{(k-{1\over 2})}\over {R^2}}\bar z\right\}\eqno (1.15)$$
and
$$K^{*}_{k-{1\over 2}}:=-2\left\{\left(1-{{|z|^2}\over {R^2}}\right
)\bar\partial_z+{{(k+{1\over 2})}\over {R^2}}z\right\}.\eqno (1.1
6)$$
We have written $K_k^{*}$ for the formal adjoint of $K_k$ with respect to
the inner product,
$$\int_{{\bf D}_R}\overline {f(x)}g(x)\,d\mu (x),$$
on ${\bf D}_R\times {\bf C}$.  Thus the Dirac operator (1.14) is
formally skew symmetric with respect to the inner product,
$$\int_{{\bf D}_R}\overline {f(x)}\cdot g(x)\,d\mu (x),\eqno (1.1
7)$$
on ${\bf D}_R\times {\bf C}^2$ where $u\cdot v=u_1v_1+u_2v_2.$  The indexing by
$
k$ in (1.14) and
(1.15) has been done so that $k=0$ corresponds to the Dirac operator
that one would obtain from the usual two fold covering of SO(2)
by itself given by
$$g\rightarrow g^2.$$
We remark that $K_{\kappa}$ is the familar Maass operator (see, e.g., [2]).
\vskip.1in \par\noindent
{\bf Covariance of the Dirac operator}.
One of the principal objects of study for us will be multivalued
solutions to the Dirac equation $D_k\psi =0$ (or more precisely, solutions
to the massive version of the Dirac equation $\left(mI-D_k\right)
\psi =0$)
that are branched at
points $a_j\in {\bf D}_R$ for $j=1,2,\ldots n$.  In order to understand these
solutions to the Dirac equation in a neighborhood of the point
$a_j$ it will be convenient to employ coordinates for ${\bf D}_R$ which
are centered at $a_j$.  Of course, we can use a fractional linear
map,
$$w=R{{z-a_j}\over {\bar {a}_jz-R}},$$
to make such a change of coordinates.  However, the function
$w\rightarrow\psi (z(w))$ no longer satisfies the Dirac equation in the $
w$ variables.
This inconvenience can be overcome by using the covariance of
solutions to the Dirac equation under $\tilde{\hbox{SU}}(1,1)$ that is a
consequence
of the fact that the Dirac operator is a natural operator on the
associated bundle of spinors which depends on a choice of spin
structure and on the representation of the spin group $r_k$ but not
on the choice of local frame or base coordinates used to realize
(1.14) as a differential operator.  We will now explain this covariance
from the passive point of view in which fractional linear maps
on ${\bf D}_R$ are regarded as a change of coordinates on ${\bf D}_
R$.  Let
$$w=g(z)=R{{Az+BR}\over {\bar {B}z+\bar {A}R}}$$
denote a fractional linear map with
$$\left[\matrix{A&B\cr
\bar {B}&\bar {A}\cr}
\right]\in\hbox{SU}(1,1).$$
The inverse coordinate change is
$$z(w)=R{{-\bar {A}w+BR}\over {\bar {B}w-AR}}.$$
Also
$$e(z):=\left(1-{{|z|^2}\over {R^2}}\right)\partial_z=v(w)\left(1
-{{|w|^2}\over {R^2}}\right)\partial_w=v(w)e(w)$$
where
$$v(w):={{\bar {B}w-AR}\over {B\bar {w}-\bar {A}R}}\,.$$
Taking complex conjugates one has
$$\bar {e}(z)=\overline {v(w)}\bar {e}(w)$$
as well.  Thus the coordinate representation of the Dirac
operator in the $w$ coordinates is
$$D_k^{(w)}=\left[\matrix{0&v(w)e(w)-{{(k-{1\over 2})}\over {R^2}}
\bar {z}(w)\cr
\overline {v(w)}\bar {e}(w)+{{(k+{1\over 2})}\over {R^2}}z(w)&0\cr}
\right].$$
Now choose a continuous logarithm, $i\lambda (w),$ for the function $
w\rightarrow v(w)$
so that
$$v(w)=e^{i\lambda (w)},$$
and define
$$v^l:=e^{il\lambda},$$
for $l\in {\bf R}.$  Let
$$M_k:=\left[\matrix{v^{k+{1\over 2}}&0\cr
0&v^{k-{1\over 2}}\cr}
\right].$$
Then the covariance property of the Dirac operator under the
coordinate change $z\leftarrow w$ is simply expressed by
$$M_k^{-1}D_k^{(w)}M_k=\left[\matrix{0&e(w)-{{(k-{1\over 2})}\over {
R^2}}\bar {w}\cr
\bar {e}(w)+{{(k+{1\over 2})}\over {R^2}}w&0\cr}
\right].\eqno (1.18)$$
Observe that the right hand side of (1.18) has precisely the same
form in the $w$ coordinates that the Dirac operator has in the $z$
coordinates.  Thus if $\psi (z)$ is a solution to the Dirac equation
$D_k\psi =0$ in the $z$ coordinates, then $M_k^{-1}(w)\psi (z(w))$ will be a
solution
to exactly the same equation in the $w$ coordinates.  An observation
which is useful in checking (1.18) with a minimum of computation is
$$e(w)\left(v^l\right)=lv^le(w)\left(i\lambda\right)=lv^lv^{-1}e(
w)\left(v\right),$$
with
$$e(w)\left(v\right)=\left(1-{{|w|^2}\over {R^2}}\right)\partial_
w\left({{\bar {B}w-AR}\over {B\bar {w}-\bar {A}R}}\right)=\left(1
-{{|w|^2}\over {R^2}}\right)\left({{\bar {B}}\over {B\bar {w}-\bar {
A}R}}\right),$$
and
$$\bar {e}(w)\left(v^l\right)=lv^l\bar {e}(w)\left(i\lambda\right
)=-lv^lv\bar {e}(w)\left(v^{-1}\right),$$
with
$$\bar {e}(w)\left(v^{-1}\right)=\left(1-{{|w|^2}\over {R^2}}\right
)\bar{\partial}_w\left({{B\bar {w}-\bar {A}R}\over {\bar {B}w-AR}}\right
)=\left(1-{{|w|^2}\over {R^2}}\right)\left({B\over {\bar {B}w-AR}}\right
).$$
$$ $$
We can rephrase this observation in a slightly different form that
will be useful for us:
\vskip.1in \par\noindent
{\bf Proposition} 1.0. {\sl\ Suppose that $\Psi (z)$ is a solution to the Dirac
equation
$$(m-D_k)\Psi =0.$$
If $g=\left[\matrix{A&B\cr
\bar {B}&\bar {A}\cr}
\right]\in\hbox{SU(1,1)}$ and $z\rightarrow\lambda_g(z)$ is a continuous
function of $z$ defined so that }
$$v(g,z):={{B\bar {z}+AR}\over {\bar {B}z+\bar {A}R}}=e^{i\lambda_
g(z)},$$
{\sl\ then the function, }
$$z\rightarrow\tilde{\Psi }(z)=\left[\matrix{v(g,z)^{k+{1\over 2}}&
0\cr
0&v(g,z)^{k-{1\over 2}}\cr}
\right]\Psi (g^{-1}z),\eqno (1.20)$$
{\sl\ remains a solution to the same equation $(m-D_k)\tilde{\Psi }
=0.$ }
\vskip.2in \par\noindent\
It will also be useful to identify the infinitesimal symmetries
of the Dirac operator associated with lifts of the one parameter
subgroups of SU(1,1) into $\tilde{\hbox{SU}}(1,1).$  It will be convenient to
introduce the notation,
$$\eqalign{\hbox{ch}\,t&=\cosh t,\cr
\hbox{sh}\,t&=\sinh t,\cr
\hbox{th}\,t&=\tanh t\,.\cr}
$$
One choice of three one parameter
subgroups of SU(1,1) that generate the Lie algebra su(1,1) is given
by
$$t\rightarrow\left[\matrix{\hbox{ch}\,t&\hbox{sh}\,t\cr
\hbox{sh}\,t&\hbox{ch}\,t\cr}
\right]$$
with generator
$$X_1=\left[\matrix{0&1\cr
1&0\cr}
\right],$$
and
$$t\rightarrow\left[\matrix{\hbox{ch}\,t&i\hbox{sh}\,t\cr
-i\hbox{sh}\,t&\hbox{ch}\,t\cr}
\right]$$
with generator,
$$X_2=\left[\matrix{0&i\cr
-i&0\cr}
\right],$$
and
$$t\rightarrow\left[\matrix{e^{it}&0\cr
0&e^{-it}\cr}
\right]$$
with generator,
$$X_3=\left[\matrix{i&0\cr
0&-i\cr}
\right].$$
For
$$g_j(t):=e^{-tX_j}$$
one finds
$$v(g_1(t),z)={{R-\hbox{th}\,(t)\bar {z}}\over {R-\hbox{th}\,(t)z}}\,
,$$
$$v(g_2(t),z)={{R-i\hbox{th}\,(t)\bar {z}}\over {R+i\hbox{th}\,(t
)z}}\,,$$
and
$$v(g_3(t),z)=e^{2it}.$$
For $j=1,2$ we choose the function $\lambda_{g_j(t)}(z)$ by normalizing it
so that $\lambda_{g_j(t)}(0)=0$.  For $j=3$ we choose $\lambda_{g_
3(t)}(z$ $)=2t$.
Define
$$F_j\Psi :={d\over {dt}}\left\{\left[\matrix{v(g_j(t),z)^{k+{1\over
2}}&0\cr
0&v(g_j(t),z)^{k-{1\over 2}}\cr}
\right]\Psi (g_j(-t)z)\right\}_{t=0}.\eqno (1.21)$$

Then the $F_j$ are infinitesimal symmetries for the Dirac operator.  It
is very convenient to introduce the complexified infinitesimal
symmetries,
$$M_1:=\half\left(F_1-iF_2\right),\eqno (1.22)$$
$$M_2:=\half\left(F_1+iF_2\right),\eqno (1.23)$$
and
$$M_3:=-\textstyle{{i\over 2}}F_3.\eqno (1.24)$$
The reason this is convenient is that the commutation relations for
$M_j$ are
$$\eqalign{[M_1,M_2]&=-2M_3\,,\cr
[M_1,M_3]&=M_1\,,\cr
[M_2,M_3]&=-M_2\,.\cr}
$$
The last two show that $M_1$ and $M_2$ are ladder operators on the
eigenspaces of $M_3$ and this will allow us to parametrize local
expansions of solutions to the Dirac equation in a sensible fashion.
\vfill\eject
\centerline{\S {\bf 2 Local Expansions}}
\vskip.2in \par\noindent
{\bf Eigenfunctions for infinitesimal rotations}.
We next wish to consider multivalued solutions to the Dirac
equation,
$$(m-D_k)\Psi =0,$$
which are branched at a point $a\in {\bf D}_R$ with fixed monodromy
about $a$.  Following SMJ [15] our principal tool in understanding
such solutions will be to look at expansions
in eigenfunctions for the infinitesimal ``k-rotations'' about the
point $a$.  To begin we consider the case $a=0$.  At $a=0$ the
infinitesimal generator of k-rotations is $M_3$.  In geodesic polar
coordinates $(r,\theta )$ centered at 0 we have
$$M_3=\left[\matrix{-i\partial_{\theta}+k+{1\over 2}&0\cr
0&-i\partial_{\theta}+k-{1\over 2}\cr}
\right],$$
where the relation between the complex Cartesian coordinate
$z=x_1+ix_2$ and geodesic polar coordinates is
$$z=Re^{i\theta}\hbox{th}\,{r\over 2}\,.\eqno (2.0)$$
Of course, geodesic polar coordinates are not uniquely determined
at $a=0$.  One must also specify the geodesic $\theta =0$.  Thus when
we speak of geodesic polar coordinates we will suppose that a
choice of the geodesic $\theta =0$ has been made.  This will be important
for us since the local constructions in this section will give rise to
multivalued functions.  These multivalued functions will have
different branches and in this section we will pick out such branches
by using the geodesic ray $\theta =0$ as a branch cut.

We turn now to the eigenvalue equation,
$$M_3\Psi (r,\theta )=(k+\lambda )\Psi (r,\theta ),\eqno (2.1)$$
whose solutions will have a monodromy multiplier,
$$e^{2\pi i(\lambda\pm{1\over 2})}=-e^{2\pi i\lambda},$$
after a counterclockwise circuit of 0.  We will now analyse the
multivalued solutions of the Dirac equation $(m-D_k)\Psi =0$  which
are also eigenfunctions (2.1) of $M_3.$  In studying solutions to the
Dirac equation $(m-D_k)\Psi =0$ it is very convenient to introduce the
associated Helmholtz equation
$$(m-D_k)^{*}(m-D_k)\Psi =(m^2-D_k^2)\Psi =0.\eqno (2.2)$$
In the notation of the previous section this is
$$\left[\matrix{m^2+K_{k-{1\over 2}}K_{k-{1\over 2}}^{*}&0\cr
0&m^2+K^{*}_{k-{1\over 2}}K_{k-{1\over 2}}\cr}
\right]\Psi =0.$$
We introduce
$$\Delta_{\kappa}=-K_{\kappa}^{*}K_{\kappa}+{{4\kappa (\kappa +1)}\over {
R^2}}\eqno (2.3)$$
and a little calculation shows
$$\Delta_{\kappa +1}=-K_{\kappa}K^{*}_{\kappa}+{{4\kappa (\kappa
+1)}\over {R^2}}.\eqno (2.4)$$
For reasons that will be clear later on we introduce a
parameter $s$ by
$$(Rs-1)^2=m^2R^2+4k^2.\eqno (2.5)$$
There are two roots to (2.5) and for definiteness we will in the future
restrict our attention to the solution $s$ of (2.5) with $Rs>1$.
Another form of (2.5) which is useful for us is
$$s\left(s-{2\over R}\right)=m^2+{{4k^2-1}\over {R^2}}\,.$$
If one rewrites the Helmholtz equation using (2.3)--(2.5) the
result is
$$\left[\matrix{\Delta_{k+{1\over 2}}&0\cr
0&\Delta_{k-{1\over 2}}\cr}
\right]\Psi =s\left(s-{2\over R}\right)\Psi .\eqno (2.6)$$
To analyze the solutions of the Dirac equation which
are eigenfunctions of $M_3$ above we will first consider the problem
of finding solutions to (see, e.g., [2])
$$\Delta_{\kappa}f=s\left(s-{2\over R}\right)f\eqno (2.7)$$
where
$$f(r,\theta )=e^{il\theta}\Phi (r).\eqno (2.8)$$
Note that we will have to set $\kappa =k\pm{1\over 2}$ and $l=\lambda
\pm{1\over 2}$ in appropriate
places to make use of (2.7) and (2.8) in the Dirac problem.   In
geodesic polar coordinates one finds
$$K_{\kappa}={2\over R}e^{-i\theta}\left\{\partial_r-{i\over {\hbox{sh}\,
r}}\partial_{\theta}-\kappa\hbox{th}\,{r\over 2}\right\},\eqno (2
.9)$$
$$K_{\kappa}^{*}=-{2\over R}e^{i\theta}\left\{\partial_r+{i\over {\hbox{sh}\,
r}}\partial_{\theta}+(\kappa +1)\hbox{th}\,{r\over 2}\right\},\eqno
(2.10)$$
and
$$\Delta_{\kappa}={4\over {R^2}}\left\{\partial_r^2+{{\hbox{ch}\,
r}\over {\hbox{sh}\,r}}\partial_r+{1\over {\hbox{sh}\,^2r}}\partial_{
\theta}^2-{{2i\kappa}\over {1+\hbox{ch}\,r}}\partial_{\theta}+{{2
\kappa^2}\over {1+\hbox{ch}\,r}}\right\}.\eqno (2.11)$$
We make a substitution to deal with (2.7) and the side
condition (2.8) [2,6].  Introduce the change of variables,
$$\Phi (r)=t^{l/2}(1-t)^{Rs/2}F(t),\eqno (2.12)$$
with
$$t={{\hbox{ch}\,r-1}\over {\hbox{ch}\,r+1}}={{|z|^2}\over {R^2}}\eqno
(2.13)$$
in (2.7) using (2.11).  One finds that $F(t)$ satisfies the hypergeometric
equation
$$t(1-t)F^{\prime\prime}+(c-(a+b+1)t)F'-abF=0\eqno (2.14)$$
with
$$\eqalign{&a={{Rs}\over 2}-\kappa ,\,\,\,b={{Rs}\over 2}+\kappa
+l,\,\,\,c=l+1\cr}
.\eqno (2.15)$$
Note that it is precisely to simplify the description of these
parameters that $s$ was introduced in (2.5) and (2.6) above.
For $c\ne\hbox{integer}$, two linearly independent solutions to the
hypergeometric equation are
$$F(a,b;c;t)\,\,\hbox{  and  }\,\,t^{1-c}F(a-c+1,b-c+1;2-c;t),$$
where $F(a,b;c;t)$ is the hypergeometric function of Gauss.  Putting
this together with (2.8) and (2.12) we see that two linearly
independent solutions to (2.7) satisfying (2.8) are given by
$$\eqalign{v_1(r,\theta ,l,\kappa ,s):=&e^{il\theta}P^{-l}_{Rs/2,
-\kappa}(r),\cr
v_2(r,\theta ,l,\kappa ,s):=&e^{il\theta}P^l_{Rs/2,\kappa}(r),\cr}
\eqno (2.16)$$
where
\def\tx{\textstyle}
$$P^l_{\alpha ,\kappa}(r):={{t^{-{l\over 2}}(1-t)^{\alpha}}\over {
\Gamma (1-l)}}F\left(\alpha +\kappa ,\alpha -\kappa -l,1-l,t\right
).$$
It will also be useful at this point to record the solution of (2.7) of
the form (2.8) which tends to 0 as $|z|\rightarrow R$ (or $r\rightarrow
\infty$).  Define
$$Q^l_{\alpha ,\kappa}(r):={{e^{il\pi}}\over {4\pi}}{{\Gamma (\alpha
-\kappa )\Gamma (\alpha +\kappa +l)}\over {\Gamma (2\alpha )}}t^{
-{l\over 2}}(1-t)^{\alpha}F(\alpha +\kappa ,\alpha -\kappa -l;2\alpha
;1-t)$$
and write
$$\hat {v}(r,\theta ,l,\kappa ,s)=e^{il\theta}Q^l_{Rs/2,\kappa}(r
).\eqno (2.17)$$
Then $\hat {v}(r,\theta ,l,\kappa ,s)$ is a solution to (2.7) which tends to
zero as
$r\rightarrow\infty$.  From the connection formula for the hypergeometric
equation one finds
$$\hat {v}(l,\kappa ,s)={{e^{i\pi l}}\over {4\sin(\pi l)}}\left\{
v_2(l,\kappa ,s)-{{\Gamma (\alpha -\kappa )\Gamma (\alpha +\kappa
-l)}\over {\Gamma (\alpha +\kappa )\Gamma (\alpha -\kappa -l)}}v_
1(l,\kappa ,s)\right\},$$
where $\alpha ={{Rs}\over 2}$ and we have written $\hat {v}(l,\kappa
,s)$ for $\hat {v}(r,\theta ,l,\kappa ,s)$.

Because we are interested in using the functions $v_j$ and $\hat {
v}$ to construct
solutions to the Dirac equation, it will be useful to compute the
action of the operators $K_{\kappa}$ and $K_{\kappa}^{*}$ on the functions $
v_j$ and $\hat {v}$.
Writing $v_j(l,\kappa ,s)$ for the function $v_j(r,\theta ,l,\kappa
,s)$ we now summarize the
action of $K_{\kappa}$ and $K_{\kappa}^{*}$ on these functions:
$$\eqalign{&K_{\kappa}v_1(l,\kappa ,s)=\tx{{2\over R}}v_1(l-1,\kappa
+1,s),\cr
&K_{\kappa}v_2(l,\kappa ,s)=\tx{{2\over R}}\left(\tx{{{Rs}\over 2}}
+\kappa\right)\left(\tx{{{Rs}\over 2}}-\kappa -1\right)v_2(l-1,\kappa
+1,s),\cr
&K_{\kappa}\hat {v}(l,\kappa ,s)=\tx{{2\over R}}\left(\tx{{{Rs}\over
2}}+\kappa\right)\left(\tx{{{Rs}\over 2}}-\kappa -1\right)\hat {v}
(l-1,\kappa +1,s),\cr
&K_{\kappa}^{*}v_1(l,\kappa +1,s)=-\tx{{2\over R}}\left(\tx{{{Rs}\over
2}}+\kappa\right)\left(\tx{{{Rs}\over 2}}-\kappa -1\right)v_1(l+1
,\kappa ,s),\cr
&K_{\kappa}^{*}v_2(l,\kappa +1,s)=-\tx{{2\over R}}v_2(l+1,\kappa
,s),\cr
&K_{\kappa}^{*}\hat {v}(l,\kappa +1,s)=-\tx{{2\over R}}\hat {v}(l
+1,\kappa ,s)\,.\cr}
$$
Finally we introduce the basic wave functions that we will use to
analyze the solutions to the Dirac equation:
$$\eqalign{w_{l,k}&:=\left[\matrix{{2\over {mR}}v_1\left(l-{1\over
2},k+{1\over 2},s\right)\cr
v_1\left(l+{1\over 2},k-{1\over 2},s\right)\cr}
\right],\cr
w_{l,k}^{*}&:=\left[\matrix{v_2\left(-l-{1\over 2},k+{1\over 2},s\right
)\cr
{2\over {mR}}v_2\left(-l+{1\over 2},k-{1\over 2},s\right)\cr}
\right],\cr}
\eqno (2.18)$$
and
$$\hat {w}_{l,k}:=\left[\matrix{\hat {v}\left(-l-{1\over 2},k+{1\over
2},s\right)\cr
{2\over {mR}}\hat {v}\left(-l+{1\over 2},k-{1\over 2},s\right)\cr}
\right],$$
where $s$ is the solution,
$$s=R^{-1}\left(1+\sqrt {m^2R^2+4k^2}\right),\eqno (2.19)$$
to (2.5).   The solution $\hat {w}_{l,k}$ which is well behaved at infinity
is
$$\hat {w}_{l,k}={{ie^{-i\pi l}}\over {4\cos(\pi l)}}\left\{w_{l,
k}^{*}-{{\Gamma (\alpha -k+\half)\Gamma (\alpha +k-l)}\over {\Gamma
(\alpha +k-\half)\Gamma (\alpha -k+l)}}{2\over {mR}}w_{-l,k}\right
\}.$$

In order to see that these are indeed solutions to the Dirac equation
it is useful to recompute the action of $K_{\kappa}$ and $K_{\kappa}^{
*}$ on $v_j$ and $\hat {v}$ in the
case where $s$ is given by (2.5).  Using
$$\tx{{2\over R}}\left(\tx{{{Rs}\over 2}}-\tx{{1\over 2}}+k\right
)\left(\tx{{{Rs}\over 2}}-\tx{{1\over 2}}-k\right)=\tx{{{m^2R}\over
2}}$$
one finds
$$\eqalign{&K_{k-{1\over 2}}v_1(l+\half,k-\half)={2\over R}v_1(l-\half
,k+\half),\cr
&K_{k-{1\over 2}}v_2(-l+\half,k-\half)={{m^2R}\over 2}v_2(-l-\half
,k+\half),\cr
&K_{k-{1\over 2}}\hat {v}(-l+\half,k-\half)={{m^2R}\over 2}\hat {
v}(-l-\half,k+\half),\cr
&K^{*}_{k-{1\over 2}}v_1(l-\half,k+\half)=-{{m^2R}\over 2}v_1(l+\half
,k-\half),\cr
&K^{*}_{k-{1\over 2}}v_2(-l-\half,k+\half)=-{2\over R}v_2(-l+\half
,k-\half),\cr
&K^{*}_{k-{1\over 2}}\hat {v}(-l-\half,k+\half)=-{2\over R}\hat {
v}(-l+\half,k-\half)\,.\cr}
$$
We have dropped the explicit dependence on $s$, it being
understood that $s$ is given by (2.19).

As mentioned in \S 1 the operators $M_1$ and $M_2$ are ladder operators
on the eigenvectors for $M_3.$ Next we record the action of these
infinitesimal symmetries of the Dirac operator on the wave
functions $w_{l,k}$ and $w_{l,k}^{*}$:
$$\eqalign{&M_3w_{l,k}=(k+l)w_{l,k},\cr
&M_3w_{l,k}^{*}=(k-l)w_{l,k}^{*},\cr
&M_1w_{l,k}=w_{l-1,k},\cr
&M_2w_{l,k}^{*}=w^{*}_{l-1,k},\cr
&M_1w^{*}_{l,k}=m_1(l,k)w^{*}_{l+1,k},\cr
&M_2w_{l,k}=m_2(l,k)w_{l+1,k},\cr}
\eqno (2.20)$$
where
$$\eqalign{m_1(l,k)&:=\left(\tx{{{m^2R^2}\over 4}}+k^2-\left(k-l-\tx{{
1\over 2}}\right)^2\right),\cr
m_2(l,k)&:=\left(\tx{{{m^2R^2}\over 4}}+k^2-\left(k+l+\tx{{1\over
2}}\right)^2\right).\cr}
$$
These results provide the rationale for the normalization of the
we have chosen for the wave functions.  The wave functions
$w_{l,k}$ and $w_{l,k}^{*}$ are parametrized so that each becomes less
singular at $z=0$  as $l$ increases. This is the reason for changing
the sign of $l$ in the definition of $w_{l,k}^{*}$.  The action of those
infinitesimal
symmetries that increase the local singularity of the wave functions
will play an important role in what follows and the normalization
we have adopted makes the action of $M_1$ on $w_{l,k}$ and of $M_
2$ on
$w_{l,k}^{*}$ as simple as possible.  Finally we have chosen the
normalization for $w_{l,k}^{*}$ so that the following conjugation
symmetry is valid,
$$w_{l,k}^{*}=C\bar {w}_{l,-k},$$
where $C=\left[\matrix{0&1\cr
1&0\cr}
\right].$  This conjugation will play a role later on
and it is especially important in the case $k=0$ where it
is the conjugation with respect to which the Dirac operator is a
{\it real} operator.

We will now make use of the covariance of the Dirac operator
under the action of $\tilde{\hbox{SU}}(1,1)$ to center these wave functions at
the
point $a\in {\bf D}_R$.  Before we do this we will establish a convention
for dealing with the branches of the multivalued functions $w_{l,
k}$
and $w_{l,k}^{*}$.  Suppose that one has chosen a geodesic ray, $
\ell ,$ at
the point $a=0$ in ${\bf D}_R$.  Then we fix a geodesic polar coordinate
system with $\{\theta =0\}=\ell$ and on ${\bf D}_R\backslash\ell$ we choose the
branches
of $w_{l,k}$ and $w^{*}_{l,k}$  that are associated with the determinations
(2.16) and (2.17) for $v_1$ and $v_2$ with $0\le\theta <2\pi$.
\vskip.1in \par\noindent
Now consider the translation to the point $a$ by the rigid
motion $T[a]$ on ${\bf D}_R$ given by
$$T[a]z=R{{{z\over R}+{a\over R}}\over {{{\bar {a}}\over R}{z\over
R}+1}}=R{{Rz+Ra}\over {\bar {a}z+R^2}}\,.\eqno (2.21)$$
Then one easily computes that
$$v(T[a],z)={{R^2-a\bar {z}}\over {R^2-\bar {a}z}}.$$
As a function of $z$ this clearly does not wind around the origin
and we can thus uniquely specify a branch of the logarithm by
$$\log v(T[a],0)=0$$
for all $a\in {\bf D}_R$.  Using this choice of the logarithm to determine
the fractional powers of $v$ in the following formulas,
we now define the local wave functions centered at $a$ by
$$w_{l,k}(z,a):=\left[\matrix{v(T[-a],z)^{k+{1\over 2}}&0\cr
0&v(T[-a],z)^{k-{1\over 2}}\cr}
\right]w_{l,k}(T[-a]z)\eqno (2.22)$$
and
$$w_{l,k}^{*}(z,a):=\left[\matrix{v(T[-a],z)^{k+{1\over 2}}&0\cr
0&v(T[-a],z)^{k-{1\over 2}}\cr}
\right]w_{l,k}^{*}(T[-a]z).\eqno (2.23)$$
Note that $T[-a]$ occurs in this formula not $T[a].$  Because the
functions $w_{l,k}$ and $w_{l,k}^{*}$ are multivalued the formulas (2.22) and
(2.23)
do not make sense as they stand.  We make sense of them in the
following manner.  Suppose that $\ell$ is a geodesic ray at $a$.  Then
$T[-a]^{-1}\ell =T[a]\ell$ is a geodesic ray at 0 and hence determines a branch
for $w_{l,k}$ and $w_{l,k}^{*}$.  With these branches for $w_{l,k}$ and $
w_{l,k}^{*}$ understood
in (2.22) and (2.23) the functions $w_{l,k}(z,a)$ and $w_{l,k}^{*}
(z,a)$ are well
defined for $z\notin\ell$ and they are branched along $\ell .$  We will think
of
$w_{l,k}(z,a)$ and $w_{l,k}^{*}(z,a)$ as multivalued functions with specific
branches
determined by the additional specification of a geodesic ray at $
z=a$.

The importance of
these wave functions is that the multivalued solutions
to the Dirac equation branched at ``$a$'' that are of interest to us
will have local expansions,
$$W(z)=\sum_{n\in {\bf Z}+{1\over 2}}\left\{a_nw_{n+\lambda}(z,a)
+b_nw^{*}_{n-\lambda}(z,a)\right\}.$$
\vskip.1in \par\noindent
{\bf Differentiating local expansions with respect to the branch points}.  In
this
subsection the dependence of the wave functions $w_{l,k}$ and $w_{
l,k}^{*}$ on
the parameter $k$ will not be very important and so we will
abbreviate $w_{l,k}=w_l$ and $w_{l,k}^{*}=w_l^{*}$.  In order to
differentiate local expansions in the $a$ parameter it will be useful
to develop {\it reciprocity relations} for the behavior of $w_l(z
,a)$ and
$w_l^{*}(z,a)$ as functions of $z$ and $a$.  Let $a(\epsilon )=T[
\epsilon R]a$ and $z(-\epsilon )=T[-\epsilon R]z$.
One easily computes that
$$v(T[-a(\epsilon )],z)={{R-\epsilon\bar {z}}\over {R-\bar{\epsilon }
z}}{{R+\epsilon\bar {a}}\over {R+\bar{\epsilon }a}}v(T[-a],z(-\epsilon
))$$
and
$$T[-a(\epsilon )]z={{R+\epsilon\bar {a}}\over {R+\bar{\epsilon }
a}}T[-a]z(-\epsilon ).$$
Observe also that under rotations $w_l$ and $w^{*}_l$ transform as
follows,
$$\left[\matrix{e^{i(k+{1\over 2})\theta}&0\cr
0&e^{i(k-{1\over 2})\theta}\cr}
\right]w_l(e^{i\theta}z)=e^{i(k+l)\theta}w_l(z)$$
and
$$\left[\matrix{e^{i(k+{1\over 2})\theta}&0\cr
0&e^{i(k-{1\over 2})\theta}\cr}
\right]w_l^{*}(e^{i\theta}z)=e^{i(k-l)\theta}w_l^{*}(z).$$
Of course, neither of these equations is correct without the proper
interpretation of the multivalued functions $w_l$ and $w_l^{*}$.  It will
suffice for our purposes to note that the equations are correct
provided that $w_l$ and $w_l^{*}$ are branched along the geodesic ray $
\ell$ at
0, we suppose that $z\notin\ell$, and we suppose that $\theta$ is restricted to
a
sufficiently
small neighborhood of $0\in {\bf R}$ so that the curve $\theta\rightarrow
e^{i\theta}z$ does not
intersect $\ell .$

Now we write
$$w_l(z,a(\epsilon ))=\left[\matrix{v(T[-a(\epsilon )],z)^{k+{1\over
2}}&0\cr
0&v(T[-a(\epsilon )],z)^{k-{1\over 2}}\cr}
\right]w_l(T[-a(\epsilon )]z)$$
and make use of the expressions for $v(T[-a(\epsilon )],z)$ and $
T[-a(\epsilon )]z$ above.
In the formula for $w_l(z,a(\epsilon ))$ that results, the transformation of
$w_l$ under rotations implies
$$\left[\matrix{\lambda^{k+{1\over 2}}&0\cr
0&\lambda^{k-{1\over 2}}\cr}
\right]w_l\left(\lambda T[-a]z(-\epsilon )\right)=\lambda^{k+l}w_
l(T[-a]z(-\epsilon ))$$
where
$$\lambda =\left({{R+\epsilon\bar {a}}\over {R+\bar{\epsilon }a}}\right
).$$
Furthermore one has
$$\left[\matrix{v(T[-a],z(-\epsilon ))^{k+{1\over 2}}&0\cr
0&v(T[-a],z(-\epsilon ))^{k-{1\over 2}}\cr}
\right]w_l(T[-a]z(-\epsilon ))=w_l(z(-\epsilon ),a).$$
The reader should understand these formulas in the following
manner.  A geodesic ray $\ell$ at $a$ has been chosen so that $z\notin
\ell$
and $\epsilon$ is chosen in a sufficiently small neighborhood, $U
,$ of $0\in {\bf C}$ so that
the range of
$$U\ni\epsilon\rightarrow{{R+\epsilon\bar {a}}\over {R+\bar{\epsilon }
a}}z(-\epsilon )$$
does not intersect $\ell .$  The functions $w_l(z,a)$ and $w_l^{*}
(z,a)$ are understood
to be branched along $\ell$ and the functions $w_l$ and $w_l^{*}$ are
understood
to be branched along $T[-a]\ell$.

Putting the preceeding results together one gets the
reciprocity formula,
$$w_l(z,a(\epsilon ))=\left({{R+\epsilon\bar {a}}\over {R+\bar{\epsilon }
a}}\right)^{k+l}\left[\matrix{v(T[-\epsilon R],z)^{k+{1\over 2}}&
0\cr
0&v(T[-\epsilon R],z)^{k-{1\over 2}}\cr}
\right]w_l(T[-\epsilon R]z,a)\eqno (2.24)$$
where we made use of the fact that
$$v(T[-\epsilon R],z)={{R-\epsilon\bar {z}}\over {R-\bar{\epsilon }
z}}\,.$$
In a precisely similar fashion one finds that
$$w_l^{*}(z,a(\epsilon ))=\left({{R+\epsilon\bar {a}}\over {R+\bar{
\epsilon }a}}\right)^{k-l}\left[\matrix{v(T[-\epsilon R],z)^{k+{1\over
2}}&0\cr
0&v(T[-\epsilon R],z)^{k-{1\over 2}}\cr}
\right]w_l^{*}(T[-\epsilon R]z,a).\eqno (2.25)$$
These formulae will allow us to compute the local expansions for the
derivatives of the wave functions $w_l(z,a)$ and $w^{*}_l(z,a)$ in the $
a$
variables.

One reason for the utility of these reciprocity relations for
differentiating the wave functions
$w_l(z,a)$ and $w^{*}_l(z,a)$ are the following easily
confirmed formulae,
$${{\partial}\over {\partial\epsilon}}\left[\matrix{v(T[\epsilon
R],z)^{k+{1\over 2}}&0\cr
0&v(T[\epsilon R],z)^{k-{1\over 2}}\cr}
\right]F(T[\epsilon R]z)|_{\epsilon =0}=M_1F(z)\eqno (2.26)$$
and
$${{\partial}\over {\partial\bar{\epsilon}}}\left[\matrix{v(T[\epsilon
R],z)^{k+{1\over 2}}&0\cr
0&v(T[\epsilon R],z)^{k-{1\over 2}}\cr}
\right]F(T[\epsilon R]z)|_{\epsilon =0}=M_2F(z).\eqno (2.27)$$
$$ $$

We will now use the reciprocity formulae above to provide local
expansions for the derivatives of the wave functions $w_l(z,a)$ and
$w^{*}_l(z,a)$ in the ``$a$'' variable.  The chain rule gives
$$\partial_{\epsilon}w_l(z,a(\epsilon ))|_{\epsilon =0}=R\left(\partial_
aw_l(z,a)-{{\bar {a}^2}\over {R^2}}\bar\partial_aw_l(z,a)\right),\eqno
(2.28)$$
and
$$\bar{\partial}_{\epsilon}w_l(z,a(\epsilon ))|_{\epsilon =0}=R\left
(\bar\partial_aw_l(z,a)-{{a^2}\over {R^2}}\partial_aw_l(z,a)\right
),\eqno (2.29)$$
with completely analogous formulas for $w^{*}_l(z,a)$.  Differentiating
the right hand side of the reciprocity formulas (2.24) and (2.25) with the
use of (2.26) and (2.27) one finds,
$$\partial_{\epsilon}w_l(z,a(\epsilon ))|_{\epsilon =0}=(k+l){{\bar {
a}}\over R}w_l(z,a)-M_1w_l(z,a),\eqno (2.30)$$
and
$$\bar{\partial}_{\epsilon}w_l(z,a(\epsilon ))|_{\epsilon =0}=-(k
+l){a\over R}w_l(z,a)-M_2w_l(z,a),\eqno (2.31)$$
with analogous results for $w_l^{*}(z,a)$.
To compute the action of $M_1$ and $M_2$ on $w_l(z,a)$ it is useful to
represent $M_j$ for $j=1,2,3$ in terms of the centered operators
$M_j^{(a)}$ obtained by conjugating $M_j$ with the action (2.23) of $
T[-a]$.
One finds that,
$$\eqalign{&M_1={1\over {1-|b|^2}}\left\{M^{(a)}_1-\bar b^2M^{(a)}_
2+2\bar bM^{(a)}_3\right\}\cr
&M_2={1\over {1-|b|^2}}\left\{M^{(a)}_2-b^2M^{(a)}_1-2bM^{(a)}_3\right
\}\cr
&M_3={1\over {1-|b|^2}}\left\{bM_1^{(a)}-\bar bM^{(a)}_2+(1+|b|^2
)M_3^{(a)}\right\},\cr}
\eqno (2.32)$$
where we have written
$$b={a\over R}$$
for the scaled variable.
Now substitute these results in (2.30) and (2.31) and equate the
results with (2.28) and (2.29); one can solve for $\partial_bw_l(
z,a)$ and
$\bar{\partial}_bw_l(z,a)$ to find
$$\partial_bw_l(z,a)={{(k+l)\bar {b}}\over {1+|b|^2}}w_l(z,a)-{1\over {
1-|b|^2}}\left(M_1^{(a)}w_l(z,a)+{{2\bar {b}}\over {1+|b|^2}}M_3^{
(a)}w_l(z,a)\right)$$
and
$$\bar{\partial}_bw_l(z,a)=-{{(k+l)b}\over {1+|b|^2}}w_l(z,a)-{1\over {
1-|b|^2}}\left(M_2^{(a)}w_l(z,a)-{{2b}\over {1+|b|^2}}M_3^{(a)}w_
l(z,a)\right).$$
The analogous results for $w^{*}_l(z,a)$ are
$$\partial_bw_l^{*}(z,a)={{(k-l)\bar {b}}\over {1+|b|^2}}w_l^{*}(
z,a)-{1\over {1-|b|^2}}\left(M_1^{(a)}w_l^{*}(z,a)+{{2\bar {b}}\over {
1+|b|^2}}M_3^{(a)}w_l^{*}(z,a)\right)$$
and
$$\bar{\partial}_bw_l^{*}(z,a)=-{{(k-l)b}\over {1+|b|^2}}w_l^{*}(
z,a)-{1\over {1-|b|^2}}\left(M_2^{(a)}w_l^{*}(z,a)-{{2b}\over {1+
|b|^2}}M_3^{(a)}w_l^{*}(z,a)\right).$$
It is now a simple matter to use the results (2.21)  for the wave
functions to get the local expansions,
$$\eqalign{\partial_bw_l(z,a)&=-{{(k+l)\bar {b}}\over {1-|b|^2}}w_
l(z,a)-{1\over {1-|b|^2}}w_{l-1}(z,a),\cr
\bar{\partial}_bw_l(z,a)&={{(k+l)b}\over {1-|b|^2}}w_l(z,a)-{{m_2
(l,k)}\over {1-|b|^2}}w_{l+1}(z,a),\cr
\partial_bw_l^{*}(z,a)&=-{{(k-l)\bar {b}}\over {1-|b|^2}}w_l^{*}(
z,a)-{{m_1(l,k)}\over {1-|b|^2}}w^{*}_{l+1}(z,a),\cr
\bar{\partial}_bw^{*}_l(z,a)&={{(k-l)b}\over {1-|b|^2}}w_l^{*}(z,
a)-{1\over {1-|b|^2}}w^{*}_{l-1}(z,a).\cr}
\eqno (2.33)$$

\vskip.1in\par\noindent
{\bf Estimates at infinity}.  We conclude this section with results
that we will use to justify some later applications of
Stokes' theorem in the non-compact domain ${\bf D}_R.$

If one considers (2.9) and (2.10) or (2.11) then one finds
$$K^{*}_{\kappa}K_{\kappa}=-{4\over {R^2}}\left\{\partial_r^2+{{\hbox{ch}\,
r}\over {\hbox{sh}\,r}}\partial_r+{1\over {\hbox{sh}^2r}}\partial_{
\theta}^2-{{i\kappa}\over {\hbox{ch}^2\left({r\over 2}\right)}}\partial_{
\theta}+{{\kappa^2}\over {\hbox{ch}^2\left({r\over 2}\right)}}-\kappa
(\kappa +1)\right\}.$$
Next we fix $s$ to the value given by (2.19).
Then we introduce
$$\Phi_{l,\kappa ,s}(r)=e^{-i\pi l}Q^l_{Rs/2,\kappa}(r)$$
for brevity and also because $\Phi_{l,k,s}(r)$ is {\it real valued}.  Then
$$\hat {v}(r,\theta ,l,k,s)=e^{il(\theta +\pi )}\Phi_{l,\kappa ,s}
(r).\eqno (2.34)$$
The differential equation $\left(K^{*}_{\kappa}K_{\kappa}+m^2\right
)\hat {v}=0$ becomes the ordinary
differential equation
$$\Phi_{l,\kappa ,s}^{\prime\prime}+{{\hbox{ch}r}\over {\hbox{sh}
r}}\Phi_{l,\kappa ,s}'-\left\{{{l^2}\over {\hbox{sh}^2r}}-{{\kappa
l}\over {\hbox{ch}^2\left({r\over 2}\right)}}-{{\kappa^2}\over {\hbox{ch}^
2\left({r\over 2}\right)}}+\kappa (\kappa +1)+{{m^2R^2}\over 4}\right
\}\Phi_{l,\kappa ,s}=0$$
for the radial wave function $\Phi_{l,\kappa ,s}.$
We eliminate the first derivative term in this last equation with
the substitution
$$u_{l,\kappa ,s}(r)=\sqrt {\hbox{sh}r}\,\,\,\Phi_{l,\kappa ,s}(r
).$$
After a little computation one finds that
$$\eqalign{u_{l,\kappa ,s}^{\prime\prime}&-\left({{m^2R^2}\over 4}
+\kappa (\kappa +1)+{1\over 4}\right)u_{l,\kappa ,s}+{{\kappa (l+
\kappa )}\over {\hbox{ch}^2\left({r\over 2}\right)}}u_{l,\kappa ,
s}-{{l^2-{1\over 4}}\over {\hbox{sh}^2r}}u_{l,\kappa ,s}=0\cr}
.\eqno (2.35)$$
As $r\rightarrow\infty$ the coefficients of $u_{l,\kappa ,s}$ in the third and
fourth terms on
the left hand side of (2.35) tend to 0 exponentially fast.  It is
not surprising then that the behavior of the solutions to (2.35) is
governed by the first two terms.  When
$${{m^2R^2}\over 4}+\kappa (\kappa +1)+{1\over 4}>0$$
there is one solution to (2.35) which is exponentially small at $
\infty$
(in Hartmann's terminology [3] this is the {\it principal solution}).
Suppose now that $\kappa =k-{1\over 2}$.  Then
$$M^2={{m^2R^2}\over 4}+\kappa (\kappa +1)+{1\over 4}={{m^2R^2}\over
4}+k^2\eqno (2.36)$$
is clearly positive and it follows from Theorem 8.1 in Hartmann
[3] that the principal solution $u_{l,k-{1\over 2},s}$ to (2.35) has the
asymptotics
$$u_{l,k-{1\over 2},s}(r)=O\left(e^{-Mr}\right)\hbox{  as  }r\rightarrow
\infty .$$
Thus
$$\Phi_{l,k-{1\over 2},s}(r)=O\left({{e^{-Mr}}\over {\sqrt {\hbox{sh}
r}}}\right)\hbox{  as  }r\rightarrow\infty .\eqno (2.37)$$
We are mostly interested in the consequences of (2.37) for the
solutions to the Dirac equation which are well behaved at $\infty
.$
Since the second component of $\hat {w}_{l,k}$, for which we write $\left
(\hat w_{l,k}\right)_2$,
is just $\hat {v}(-l+{1\over 2},k-{1\over 2},s)$ it follows that
$$|\left(w_{l,k}(r)\right)_2|=|\Phi_{-l+{1\over 2},k-{1\over 2},s}
(r)|=O\left({{e^{-Mr}}\over {\sqrt {\hbox{sh}r}}}\right)\hbox{  as  }
r\rightarrow\infty .\eqno (2.38)$$
We would like to have something similar for the first component
$\left(w_{l,k}\right)_1$.  This component is a solution to the Helmholtz
equation
$$\left(K_{k-{1\over 2}}K_{k-{1\over 2}}^{*}+m^2\right)w=0.\eqno
(2.39)$$
Since,
$$K_{k-{1\over 2}}K_{k-{1\over 2}}^{*}=K_{k+{1\over 2}}^{*}K_{k+{
1\over 2}}-{{8(k+{1\over 2})}\over {R^2}},\eqno (2.40)$$
it follows from (2.35) that the leading behavior (at $\infty$) of the
second order ODE that
governs the radial part of the wave function $\hbox{$\sqrt {\hbox{sh}
r}$}\left(w_{l,k}\right)_1$ is
$$u^{\prime\prime}-M^2u=0.$$
Then one has the following asymptotic estimate for the first
component of $w_{l,k}$,
$$|\left(w_{l,k}(r)\right)_1|={{mR}\over 2}|\Phi_{-l-{1\over 2},k
+{1\over 2},s}(r)|=O\left({{e^{-Mr}}\over {\sqrt {\hbox{sh}(r)}}}\right
)\hbox{  as  }r\rightarrow\infty .\eqno (2.41)$$
We will now use the estimates (2.38) and (2.41) to establish the
following representation for the $L^2$ norm,
$$\int_{D_{r_0,\infty}}|\hat {w}_{l,k}(z)|^2d\mu (z)=c_{l,k}(r_0)
,\eqno (2.42)$$
where
$$c_{l,k}(r)=2\pi R^2\hbox{sh}(r)\,\Phi_{-l-{1\over 2},k+{1\over
2},s}(r)\,\Phi_{-l+{1\over 2},k-{1\over 2},s}(r)\eqno (2.43)$$
and we write
$$D_{a,b}=\left\{z:R\tanh{a\over 2}\le |z|<R\tanh{b\over 2}\right
\}.$$
To prove (2.42) suppose first that $f$ and $g$ are multivalued solutions
to the Dirac equation in ${\bf D}_R\backslash \{0\}$ with the same unitary
monodromy
about 0.  Then a simple calculation shows that
$$m(\bar {f}\cdot g)\,d\mu (z)=2d\left\{{{\bar {f}_1g_2}\over {1-{{
|z|^2}\over {R^2}}}}id\bar z\right\}.$$
Since $\bar {f}_ig_j$ is a smooth single valued function away from the origin
Stokes' theorem implies
$${m\over 2}\int_{D_{r_0,r_1}}\bar {f}\cdot g\,(z)d\mu (z)=\int_{
C_{r_1}}{{\bar {f}_1g_2}\over {1-{{|z|^2}\over {R^2}}}}id\bar {z}
-\int_{C_{r_0}}{{\bar {f}_1g_2}\over {1-{{|z|^2}\over {R^2}}}}id\bar {
z},\eqno (2.44)$$
where we write
$$C_r=\left\{z:z=Re^{i\theta}\tanh{r\over 2}\right\}.$$
In (2.44) let $f=g=w_{l,k}.$ Then substitute
geodesic polar coordinates (2.0) in (2.44).
Finally use the asymptotics (2.38) and (2.41) in the resulting
equation to evaluate the limit $r_1\rightarrow\infty$.  One obtains (2.42).
The
reader should note that the sign change due to the factor
$e^{i\pi l}$ by which $Q_{Rs/2,\kappa}^l$ differs from $\Phi_{l,\kappa
,s}$ is important in getting
the signs to work out properly in (2.42).

Now suppose that
in the exterior of the the disk $|z|<r_0-\epsilon$ the function $
f$ is a
multivalued solution to the Dirac equation with monodromy multiplier
$e^{2\pi i\lambda}$.  If $f$ is square integrable in the exterior of $
|z|<r_0-\epsilon$ with
respect to the measure $d\mu$ then it has an expansion
$$f=\sum_{n\in {\bf Z}+{1\over 2}}c_n(f)\hat {w}_{n+\lambda ,k}\,
.\eqno (2.45)$$
This expansion gives the Fourier series representation for $f$ in
the angular variable $\theta$.  If one makes use of this to do the circuit
integrals that appear on the right hand side (2.44) then one
finds,
$${m\over 2}\int_{C_r}{{\bar {f}_1f_2}\over {1-{{|z|^2}\over {R^2}}}}
id\bar {z}=\sum_{n\in {\bf Z}+{1\over 2}}|c_n(f)|^2c_{n+\lambda ,
k}(r).\eqno (2.46)$$
 From (2.42) it is clear that $c_{n+\lambda ,k}(r)$ is a positive, monotone
decreasing function of $r$ which tends to 0 as $r\rightarrow\infty
.$  Thus the
monotone convergence theorem (for sums) and (2.46) imply that
$$\matrix{\cr
\hbox{lim}\cr
r\rightarrow\infty\cr}
\int_{C_r}{{\bar {f}_1f_2}\over {1-{{|z|^2}\over {R^2}}}}id\bar {
z}=0\eqno (2.47)$$
when $f$ is a multivalued solution to the Dirac equation which is
square integrable in a neighborhood of infinity.   Finally, we want
to know that something similar is true if $f$ and $g$ are two
multivalued solutions to the Dirac equation both square integrable
in a neighborhood of $\infty .$ Consider the Hermitian form,
$$(f,g)_r=\int_{C_r}{{\bar {f}_1g_2}\over {1-{{|z|^2}\over {R^2}}}}
id\bar {z}\,.$$
Then (2.45) shows that this is positive definite and so the
Schwarz inequality,
$$|(f,g)_r|^2\le (f,f)_r(g,g)_r,$$
is valid.  It follows immediately from (2.47) that
$$\matrix{\cr
\hbox{lim}\cr
r\rightarrow\infty\cr}
\int_{C_r}{{\bar {f}_1g_2}\over {1-{{|z|^2}\over {R^2}}}}id\bar {
z}=0.\eqno (2.48)$$
We summarize these developments in the following proposition:
\vskip.1in \par\noindent
{\bf Proposition} 2.0 {\sl Suppose that $f$ and $g$ are two
solutions to the Dirac equation which have expansions of type
(2.45) in the region $D_{r_0,\infty}$.  Then }
$$\matrix{\cr
\hbox{lim}\cr
r\rightarrow\infty\cr}
\int_{C_r}{{\bar {f}_1g_2}\over {1-{{|z|^2}\over {R^2}}}}id\bar {
z}=0.$$

\vskip.2in \hfill\break
\centerline{\S 3 ${\bf L}^2$ {\bf Existence Results}}
\vskip.2in \par\noindent {\bf A model for the simply connected covering $
\tilde {{\bf D}}_R(a)$}.
To understand the asymptotics of the Green function for a Dirac
operator with vertex insertions it will be important to have an
understanding of certain spaces of multivalued solutions to the Dirac
equation.  We will follow SMJ [15] here in establishing an $L^2$ existence
theory for wave functions with specified branching and restricted
singularities.  Let $\{a_1,a_2,\ldots ,a_n\}$ denote $n$ distinct points in the
hyperbolic disk ${\bf D}_R.$ It will be useful for us to use complex variables
writing $a_j=\alpha_j+i\beta_j$ when $a_j=(\alpha_j,\beta_j).$ We will also
write
$$a=\{a_1,a_2,\ldots ,a_n\}.$$
Following SMJ [15] we will now show that the space of solutions, $
w$, to
the Dirac equation,
$$(m-D_k)w=0,$$
which are branched at $\{a_1,a_2,\ldots ,a_n\}$ with monodromy $e^{
2\pi i\lambda_j}$ at $a_j$
(with $0<|\lambda_j|<{1\over 2}$ ) and {\it globally} in $L^2({\bf D}_
R,d\mu )$ is an $n$ dimensional
space.

It will be convenient (particularly in discussing smoothness questions)
to work in the simply connected covering space of ${\bf D}_R\backslash
\{a_1,a_2,\ldots ,a_n\}$
and we will now specify our conventions for such considerations.
Write ${\bf D}_R(a)$ for ${\bf D}_R\backslash \{a_1,a_2,\ldots ,a_
n\}$ and $\tilde {{\bf D}}_R(a)$ for the simply connected
covering space of the punctured disk ${\bf D}_R(a)$.  Multivalued functions
on ${\bf D}_R(a)$ can, of course, be regarded as functions on $\tilde {
{\bf D}}_R(a).$
Introducing appropriate branch cuts, $\ell_j$, at $a_j$ a multivalued function
is also specified by a particular branch defined on ${\bf D}_R\backslash
\{\ell_1,\ell_2,\ldots ,\ell_n\}$.
Both descriptions will be important for us and to pass easily from
one to the other we will always work in a specific model for
$\tilde {{\bf D}}_R(a)$.
Fix a base point $a_0$ in ${\bf D}_R$ different from any of the points
$a_1,a_2,\ldots ,a_n$.  We take our model of $\tilde {{\bf D}}_R(
a)$ to be the homotopy classes
of paths in ${\bf D}_R(a)$ which start at $a_0$.  The projection, pr, from $
\tilde {{\bf D}}_R(a)$
onto ${\bf D}_R(a)$ maps a path $\gamma$ onto its endpoint $\gamma
(1).$ This model of $\tilde {{\bf D}}_R(a)$
comes with a distinguished point, $\tilde {a}_0$, in the fiber over $
a_0$ which is
the class of the constant path $t\rightarrow a_0.$

For simplicity we fix the convention that our branch cuts $\ell_j$ are
always geodesic rays joining the point $a_j$ to a point on the boundary
of ${\bf D}_R.$ If $\{\ell_1,\ell_2,\ldots ,\ell_n\}$ is a pairwise disjoint
collection of branch cuts
then the inverse image of
$${\bf D}_R\backslash \{\ell_1,\ell_2,\ldots ,\ell_n\}$$
under the projection,
$$\hbox{pr}:\tilde {{\bf D}}_R(a)\rightarrow {\bf D}_R,$$
splits into path components on which the projection is a
diffeomorphism.  Let $C_0(\ell )$ denote the path component of this inverse
image which contains the point $\tilde {a}_0$.  If $f$ is a map defined on $
\tilde {{\bf D}}_R(a)$
then we will refer to the restriction of $f$ to the path component
$C_0(\ell )$ as the principal branch of $f$.  The principal branch of such a
function $f$ can also be regarded as a function defined on
${\bf D}_R\backslash \{\ell_1,\ell_2,\ldots ,\ell_n\}$ and we will do this
without further comment.
Conversely if $f$ is a function defined on ${\bf D}_R\backslash \{
\ell_1,\ell_2,\ldots ,\ell_n\}$ we can
regard it as a function on $C_0(\ell ).$ For multivalued functions that have
a natural extension to maps on $\tilde {{\bf D}}_R(a)$ this gives a unique way
to go
from a branch for such a function to a function defined on $\tilde {
D}_R(a)$.
For example, we have already described how to fix a branch of the
multivalued function $w_l(z,a_j)$ when we are given a branch cut, $
\ell_j,$ at
$a_j$.  We can regard this as a function on $C_0(\ell )$ which has an obvious
extension to a smooth function defined on $\tilde {{\bf D}}_R(a)$.  When we
wish to
regard $w_l(\cdot ,a_j)$ as a function on $\tilde {{\bf D}}_R(a)$ we will write
$
w_l(\tilde {z},a_j)$ with
the understanding that $\tilde {z}$ is a point in $\tilde {{\bf D}}_
R(a)$ which projects onto
$z\in {\bf D}_R(a)$.  It is understood that a branch cut $\ell_j$ must be
chosen
before the function $w_l(\tilde {z},a_j)$ is uniquely determined.

Let $[\alpha_j]$ denote the homotopy class of the simple $a_0$ based loop, $
\alpha_j,$
about $a_j$ in ${\bf D}_R(a)$ which circles $a_j$ in a counter-clockwise
fashion
and does not wind around any of the other points $a_i$ with $i\ne
j$.  Let
$R_j$ denote the deck transformation on $\tilde {{\bf D}}_R(a)$ which maps $
\tilde {a}_0$ to the
point in $\tilde {{\bf D}}_R(a)$ which corresponds to the homotopy class $
[\alpha_j].$ To
simplify notation in what follows we will, as above, write $\tilde {
z}$ for
points in $\tilde {{\bf D}}_R(a)$ and $z=\hbox{pr}(\tilde {z})$ for their
projections in $
{\bf D}_R(a).$ Suppose
that $f$ is a smooth function on $\tilde {{\bf D}}_R(a)$ for which
$$f(R_j\tilde {z})=e^{2\pi i\lambda_j}f(\tilde {z}).$$
If $F$ is the principal branch of such a function $f$ for some choice of
branch cuts $\{\ell_1,\ell_2,\ldots ,\ell_n\}$ with $a_0\notin\ell_
j$ for $j=1,2,\ldots ,n$ then we will
say that $F$ {\it is branched at $a_j$ with monodromy multiplier $
e^{2\pi i\lambda_j}$ for
$j=1,2,\ldots ,n$.
}\vskip.1in \par\noindent
{\bf Multivalued solutions with specified branching}.  We now turn to the
consideration of multivalued solutions, $w,$ to the Dirac equation with
specified monodromy at the branch points $a_j$.  We can locally expand
such functions near the branch points in Fourier series in the
angular variable $\theta$ of geodesic polar coordinates suitably modified to
reflect the appropriate monodromy.  As in SMJ [15] these Fourier series
translate into local expansions
$$w(z)=\sum_{n\in {\bf Z}+{1\over 2}}a_n^j(w)w_{n+\lambda_j}(z,a_
j)+b^j_n(w)w^{*}_{n-\lambda_j}(z,a_j)\hbox{  for  }z\hbox{ near }
a_j,\eqno (3.0)$$
where (3.0) is given a precise
meaning when the choice of a pairwise disjoint collection of branch
cuts $\{\ell_1,\ell_2,\ldots ,\ell_n\}$ has been made with $a_0\notin
\ell_j$ for $j=1,2,\ldots ,n.$

The following local asymptotics for the wave functions may be
deduced from the formulas for $w_l(z)$ and $w_l^{*}(z)$ in the preceeding
section:
$$w_l(z)=\left[\matrix{{{2e^{i(l-{1\over 2})\theta}\left({r\over
2}\right)^{l-{1\over 2}}}\over {mR\Gamma (l+{1\over 2})}}+O(r^{l+{
1\over 2}})\cr
{{e^{i(l+{1\over 2})\theta}\left({r\over 2}\right)^{l+{1\over 2}}}\over {
\Gamma (l+{3\over 2})}}+O(r^{l+{3\over 2}})\cr}
\right]\eqno (3.1)$$
and
$$w_l^{*}(z)=\left[\matrix{{{e^{-i(l+{1\over 2})\theta}\left({r\over
2}\right)^{l+{1\over 2}}}\over {\Gamma (l+{3\over 2})}}+O(r^{l+{3\over
2}})\cr
{{2e^{-i(l-{1\over 2})\theta}\left({r\over 2}\right)^{l-{1\over 2}}}\over {
mR\Gamma (l+{1\over 2})}}+O(r^{l+{1\over 2}})\cr}
\right]\eqno (3.2)$$
 From these asymptotics we deduce that $w_l$ and $w_l^{*}$ are locally in
$L^2(d\mu )$ provided that $l>-{1\over 2}$.  Thus in the expansion (3.0) of an
$
L^2$
wave function we expect that the coefficients $a_n^j(w)$ will vanish
unless $n+\lambda_j>-{1\over 2}$ and the coefficient $b_n^j(w)$ will vanish
unless
$n-\lambda_j>-{1\over 2}.$ This leads us to define
$$c_j(w)=\left\{\matrix{a_{{1\over 2}}^j(w)\hbox{  if  }\lambda_j
<0\cr
a_{-{1\over 2}}^j(w)\hbox{  if  }\lambda_j>0\cr}
\right.\eqno (3.3)$$
and
$$c^{*}_j(w)=\left\{\matrix{b_{{1\over 2}}^j(w)\hbox{  if  }\lambda_
j>0\cr
b_{-{1\over 2}}^j(w)\hbox{  if  }\lambda_j<0\,.\cr}
\right.\eqno (3.4)$$
The coefficients $c_j(w)$ and $c_j^{*}(w)$ are the lowest order coefficients
that can occur (i.e.,  be non zero) in an expansion of type (3.0) for a
function which is locally in $L^2(d\mu ).$ We will now make this more
precise with a formula for the $L^2$ inner product of two multivalued
solutions to the Dirac equation with the same monodromy.
The infinitesimal version of formal
skew symmetry for $D_k$ is
$$\eqalign{\left(\bar f\cdot D_kg+\overline {D_kf}\cdot g\right)d
\mu (z)=2d\left\{{{\overline {f_1}g_2}\over {1-{{|z|^2}\over {R^2}}}}
id\bar z-{{\overline {f_2}g_1}\over {1-{{|z|^2}\over {R^2}}}}idz\right
\}\cr}
.\eqno (3.5)$$
The $d$ on the right hand side is the usual exterior derivative.  This
is true if $f$ and $g$ are smooth functions on ${\bf D}_R$ and it will remain
true for functions $f$ and $g$ that are smooth branched functions with
the same {\it unitary} monodromy at the points $a_j$ for $j=1,2,\ldots
,n$ ---at
least if one stays away from the branch points.  In this case each
side of (3.5) will be single valued on ${\bf D}_R\backslash \{a_1
,a_2,\ldots ,a_n\}$.  \vskip.1in
\par\noindent {\bf Theorem} 3.0{\sl\ Suppose that $f$ and $g$ are multivalued
solutions to the Dirac equation, }
$$(m-D_k)f=(m-D_k)g=0,$$
{\sl which are branched at $a_1,a_2,\ldots ,a_n$ with monodromy $
e^{2\pi i\lambda_j}$ at $a_j$.
If $f$ and $g$ are in $L^2({\bf D}_R,d\mu )$ then}
$${{m^3R}\over {16}}\int_{{\bf D}_R}\overline {f(z)}\cdot g(z)\,d
\mu (z)=-\sum_{j=1}^n|s_j|\overline {c_j(f)}c^{*}_j(g)=-\sum_{j=1}^
n|s_j|\overline {c_j^{*}(f)}c_j(g)\eqno (3.6)$$
{\sl where $s_j=\sin\pi\lambda_j.$ } \vskip.1in \par\noindent Proof.
Substituting
$D_kf=mf$ and $D_kg=mg$ in equation (3.5) one finds
$$\eqalign{m\left(\overline f\cdot g\right)d\mu (z)=d\left\{{{\overline {
f_1}g_2}\over {1-{{|z|^2}\over {R^2}}}}id\bar z-{{\overline {f_2}
g_1}\over {1-{{|z|^2}\over {R^2}}}}idz\right\}\cr}
.$$
In fact, calculating the exterior derivatives of each of the terms on
the right hand side of this last equation one finds that more is true
in case $f$ and $g$ satisfy the Dirac equation, namely,
$$m\left(\bar f\cdot g\right)d\mu (z)=2d\left\{{{\overline {f_1}g_
2}\over {1-{{|z|^2}\over {R^2}}}}id\bar z\right\}=-2d\left\{{{\overline {
f_2}g_1}\over {1-{{|z|^2}\over {R^2}}}}idz\right\}.\eqno (3.7)$$

Now we wish to calculate the $L^2$ inner product of $f$ and $g$ by first
integrating (3.7)  over the complement of the union of the disks of
radius $\epsilon$ about $a_j$ for $j=1,\ldots ,n$ inside a disk of radius $
\rho$.
Then we use Stokes' theorem to
reduce the integrals on the right to the sum of integrals over the
circles, $C_j(\epsilon )$, of radius $\epsilon$ about the points $
a_j$ with the standard
counter-clockwise orientation (this will account for some minus signs
showing up below) and an integral over the circle of radius $\rho
.$
It follows from Proposition 2.0 that the
boundary integral over the circle $C_{\rho}$ is 0 in the limit $\rho
\rightarrow\infty$.
Finally we recover the $L^2$ inner product from the remaining circuit
integrals in the
limit $\epsilon\rightarrow 0.$ The result is
$$\eqalign{{m\over 2}\int_{{\bf D}_R}\bar {f}\cdot g\,d\mu (z)=&-\matrix{\cr
\hbox{lim}\cr
\epsilon\rightarrow 0\cr}
\sum_{j=1}^n\int_{C_j(\epsilon )}{{\overline {f_1}g_2}\over {1-{{
|z|^2}\over {R^2}}}}id\bar {z}\cr
=&\matrix{\cr
\hbox{lim}\cr
\epsilon\rightarrow 0\cr}
\sum_{j=1}^n\int_{C_j(\epsilon )}{{\overline {f_2}g_1}\over {1-{{
|z|^2}\over {R^2}}}}idz.\cr}
$$
Now introduce the local variables $u_j=T[-a_j]z$ in the $C_j(\epsilon
)$ integrals.
The change of variables gives
$$id\bar {z}={{1-{{|a_j|^2}\over {R^2}}}\over {\left(1+{{\bar {a}_
ju_j}\over {R^2}}\right)}}id\bar {u}_j$$
and
$$idz={{1-{{|a_j|^2}\over {R^2}}}\over {\left(1+{{a_j\bar {u}_j}\over {
R^2}}\right)}}idu_j.$$
Substituting these results in the appropriate contour integrals above
and taking limits as $\epsilon\rightarrow 0$ of some of the multiplicative
factors that
appear inside the integrands one finds,
$$\eqalign{{m\over 2}\int_{{\bf D}_R}\bar {f}\cdot g\,d\mu (z)=&-\matrix{\cr
\hbox{lim}\cr
\epsilon\rightarrow 0\cr}
\sum_{j=1}^n\int_{C_j(\epsilon )}\overline {f_1}g_2(u_j)id\bar {u}_
j\cr
=&\matrix{\cr
\hbox{lim}\cr
\epsilon\rightarrow 0\cr}
\sum_{j=1}^n\int_{C_j(\epsilon )}\overline {f_2}g_1(u_j)idu_j,\cr}
$$
where the notation $\bar {f}_jg_k(u)$ is meant to remind the reader that it is
the product $\bar {f}_jg_k$ that is a function on ${\bf D}_R(a)$.  Geodesic
polar
coordinates $(r_j,\theta_j)$ about $a_j$ (determined by the branch cut $
\ell_j$) give
$$u_j=Re^{i\theta_j}\hbox{th}\,{{r_j}\over 2}\sim{{Rr_j}\over 2}e^{
i\theta_j}.$$
We can use this asymptotic formula in the contour integrals to get
$$\eqalign{{m\over 2}\int_{{\bf D}_R}\bar {f}\cdot g\,d\mu (z)=&-\matrix{\cr
\hbox{lim}\cr
\epsilon\rightarrow 0\cr}
\sum_{j=1}^n{{\epsilon R}\over 2}\int_0^{2\pi}\overline {f_1}g_2(
u_j)e^{-i\theta_j}d\theta_j\cr
=&-\matrix{\cr
\hbox{lim}\cr
\epsilon\rightarrow 0\cr}
\sum_{j=1}^n{{\epsilon R}\over 2}\int_0^{2\pi}\overline {f_2}g_1(
u_j)e^{i\theta_j}d\theta_j.\cr}
$$
Finally if we use the asymptotics of the local wave functions, given
by (3.1) and (3.2), in these last two formulas and the identity,
$${{\pi}\over {\Gamma ({1\over 2}-l)\Gamma ({1\over 2}+l)}}=\cos\pi
l,$$
then one obtains the two local expansion results for the $L^2$ inner
product given in (3.6) above.  To see this it is also useful to note
that the multiplicative factor $v(T[-a_j],z)$, which appears in the wave
functions, $w_l(z,a_j)$ and $w_l^{*}(z,a_j)$ tends to 1 as $z\rightarrow
a_j$.  QED \vskip.1in
\par\noindent {\bf Definition}.  {\sl Let $l$ denote the $n\times
n$ diagonal matrix with $jj$
entry $l_j$ with $l_j$ chosen so that $0<l_j<1$.  Define ${\cal W}
(a,l)$ to be the
space of multivalued solutions, $w,$ to the Dirac equation $(m-D_
k)w=0$
which are branched at $\{a_1,a_2,\ldots ,a_n\}$ with monodromy $e^{
2\pi il_j}$ in a
counter-clockwise circuit of $a_j$ and which are globally in $L^2
({\bf D}_R,d\mu )$.}
\vskip.1in \par\noindent Remark:  We parametrize the monodromy in this
definition by $l_j$ chosen between 0 and 1 rather than $\lambda_j$ chosen
between $-{1\over 2}$ and ${1\over 2}$ since the local expansions of functions
in $
{\cal W}(a,l)$
will be slightly simpler to characterize in terms of the parameters
$l_j$.  \vskip.1in \par\noindent Theorem 3.0 shows that the elements, $
w,$ in
${\cal W}(a,l)$ are completely determined by the coefficients $c_
j(w)$ for
$j=1,2,\ldots ,n$ or by the coefficients $c_j^{*}(w)$ for $j=1,2,
\ldots ,n$.  Thus the
dimension of ${\cal W}(a,l)$ is less than or equal to $n.$
Following SMJ [15] we will now give a
functional analytic proof that the dimension of ${\cal W}(a,l$)  is exactly
$n$ by constructing a canonical basis.  The functional analytic part
of the proof requires that we first consider an associated problem
for the Helmholtz operator $m^2+K_{\kappa}K_{\kappa}^{*}$ where
$\kappa =k-{1\over 2}$.  Recall from (2.2) that
$$(m-D_k)^{*}(m-D_k)=\left[\matrix{m^2+K_{\kappa}K_{\kappa}^{*}&0\cr
0&m^2+K_{\kappa}^{*}K_{\kappa}\cr}
\right].$$
As a start towards constructing solutions to the Dirac equation
with prescribed branching at $\{a_1,a_2,\ldots ,a_n\}$ we will first consider
the problem of finding multivalued solutions, $v,$ to the Helmholtz
equation
$$(m^2+K_{\kappa}K_{\kappa}^{*})v=0.\eqno (3.8)$$
We will construct solutions of the desired type by fixing the
leading singularity of $v$ at one branch point.  Then we will get
a solution to the Dirac equation of the appropriate type by
considering
$$w=\left[\matrix{v\cr
m^{-1}K_{k-{1\over 2}}^{*}v\cr}
\right].\eqno (3.9)$$
The reader might observe in what follows that using (3.8) instead of
$$(m^2+K_{\kappa}^{*}K_{\kappa})v=0,$$
and (3.9) instead of
$$w=\left[\matrix{-m^{-1}K_{k-{1\over 2}}v\cr
v\cr}
\right],$$
is dictated by the desire to produce canonical solutions satifying
$c_j(W_i)=\delta_{ij}$ rather than the dual canonical basis $c_j^{
*}(W_i)=\delta_{ij}.$

To formulate local expansions for the Helmholtz equation in a
convenient form
we wish to introduce here the analogue of the wave functions
$w_l(z,a_j)$ and $w_l^{*}(z,a_j)$ for $v_j(l,\kappa ,s)$ introduced in (2.16)
above.
Recall from \S 2 that (3.8) is equivalent to,
$$\left\{s\left(s-{2\over R}\right)-\Delta_{\kappa +1}\right\}v=0
,$$
with $\kappa =k-{1\over 2}$ and $s$ given by (2.5).  Now we write
$$v_{l,\kappa}(z)=v_1(z,l,\kappa ,s),\eqno (3.10)$$
and
$$v_{l,\kappa}^{*}(z)=v_2(z,-l,\kappa ,s),\eqno (3.11)$$
with $s$ given by (2.5).
Then $v_{l,\kappa +1}$ and $v_{l,\kappa +1}^{*}$ are both multivalued solutions
to (3.8) above
with monodromy $e^{2\pi il}$ and $e^{-2\pi il}$ respectively, in a
counter-clockwise
circuit of 0.  We
translate $v_{l,\kappa}$ and $v_{l,\kappa}^{*}$ so that they are ``centered at
$
a\in {\bf D}_R$'' in
the following manner,
$$v_{l,\kappa}(z,a):=v(T[-a],z)^{\kappa}v_{l,\kappa}(T[-a]z)\eqno
(3.12)$$
and
$$v_{l,\kappa}^{*}(z,a):=v(T[-a],z)^{\kappa}v_{l,\kappa}^{*}(T[-a
]z).\eqno (3.13)$$
Precisely as in \S 2 a branch cut $\ell$ at $a$ is used to fix a unique branch
for the functions $v_{l,k}(z,a)$ and $v_{l,k}^{*}(z,a)$.
The local singularity in both $v_{l,\kappa}(z)$ and $v_{l,\kappa}^{
*}(z)$ goes like $r^l$ where
$r$ is the hyperbolic distance between 0 and $z$.  To make use of the
formula (3.9) to get an $L^2$ solution to the Dirac equation we will
require that the solutions we consider for (3.8) have both $v$ and
$K_{\kappa}^{*}v$ in $L^2$.  It is now a simple matter to check (using the
formulas
for the action of $K_{\kappa}^{*}$ on $v_l$ and $v_l^{*}$ given in the
preceeding section)
that if $v$ is a multivalued solution to (3.8) with monodromy
multiplier $e^{2\pi il_j}$ in a counter-clockwise circuit of $a_j$ and if
$v$ and $K_{\kappa}^{*}v$ are in $L^2$ then provided $l_j$ is chosen so that $
0<l_j<1$ the
function $v$ will have restricted local expansions
$$v(z)=\sum_{n=-1}^{\infty}\alpha_n^j(v)v_{n+l_j,\kappa +1}(z,a_j
)+\sum_{n=1}^{\infty}\beta_n^j(v)v_{n-l_j,\kappa +1}^{*}(z,a_j)\eqno
(3.14)$$
for $z$ near $a_j.$
In each case the sum is now over integer values of $n$ rather than
half integers.  Our strategy in finding such solutions will be to
seek a canonical basis $\{V_i\}$ with
$$\alpha_{-1}^j(V_i)=\delta_{ij}.\eqno (3.15)$$
\vskip.1in \par\noindent
{\bf Existence for a canonical ${\bf L}^2$ basis}.
Let $\varphi_j(z)$ be a $C_0^{\infty}$ function which is identically 1 in a
neighborhood
of the point $z=a_j$  and which vanishes in the complement of an
open ball about $z=a_j$.  This open ball should be small enough so
that $\varphi_j$ vanishes in a neighborhood of each of the points $
a_i$ with
$i\ne j$.  Observe next that if $V_j$ is a solution to (3.8) with local
expansions of type (3.14) and which further satisfies (3.15)
then,
$$\eqalign{\left(m^2+K_{\kappa}K_{\kappa}^{*}\right)\left(V_j-\varphi_
jv_{l_j-1,\kappa +1}(z,a_j)\right)=&\cr
-\left(m^2+K_{\kappa}K_{\kappa}^{*}\right)\varphi_jv_{l_j-1,\kappa
+1}(z,a_j):=&f_j,\cr}
\eqno (3.16)$$
where $f_j$ is a smooth multivalued function which vanishes in a
neighborhood of the point $a_j$ (and in a neighborhood of all the
other branch points as well).  For simplicity we will refer to
branched functions that have smooth extensions (from some
$C_0(\ell )$) to functions on $\tilde {{\bf D}}_R(a)$ as {\it smooth
multivalued functions.}
We will construct
$V_j$ by finding a suitably regular solution to (3.16)---one whose local
expansion coefficients $\alpha_{-1}^i$ vanish for $i=1,2,\ldots ,
n$ and which is
in the Sobolev space $H^1$.

Motivated by (3.16) we are interested in the existence question for
solutions to
$$\left(m^2+K_{\kappa}K_{\kappa}^{*}\right)u_j=f_j\eqno (3.17)$$
for $u_j$ in a completion of
$$C^{\infty}_0(\tilde {{\bf D}}_R(a))$$
with specified branching at $a_i$ for $i=1,2,\ldots ,n.$  Here
$C_0^{\infty}(\tilde {{\bf D}}_R(a))$ denotes the space of $C^{\infty}$
functions on $
\tilde {{\bf D}}_R(a)$ whose
supports project onto compact subsets of ${\bf D}_R(a)$.  In particular
such functions vanish in a neighborhood of the points $a_j$ and in
a neighborhood of the boundary of ${\bf D}_R$. Let
$C_{a,l}^{\infty}$ denote the subset of $C_0^{\infty}(\tilde {{\bf D}}_
R(a))$ consisting of those functions
$f$ which transform under the deck transformation $R_j$ by,
$$f(R_j\tilde {z})=e^{2\pi il_j}f(\tilde {z}),$$
where $0\le l_j<1$ for $j=1,2,\ldots ,n$ and we write
$$l=(l_1,l_2,\ldots ,l_n).$$
Now let $H_{a,l}$ denote the Hilbert space
completion of $C_{a,l}^{\infty}$ with respect to the norm derived from the
inner product,
$$(F,G)=\int_{{\bf D}_R}\left\{\overline {K_{\kappa}^{*}F(\tilde {
z})}K_{\kappa}^{*}G(\tilde z)+m^2\overline {F(\tilde {z})}G(\tilde
z)\right\}d\mu (z).$$
Note that the integrand descends to a function on ${\bf D}_R$ because
the monodromy multipliers of $\bar {F}$ and $G$ cancel.  Now consider the
element $f_j\in C_{a,l}$ defined above in (3.16).
  For $v\in C_{a,l}^{\infty}$ consider the linear functional,
$$v\rightarrow\int_{{\bf D}_R}\overline {f_j(\tilde {z})}v(\tilde {
z})\,d\mu (z)=<f_j,v>.\eqno (3.18)$$
Since $(h,h)\ge m^2<h,h>$, for all $h\in C_{a,l}^{\infty}$, it is clear that
the
linear functional (3.18) is continuous on the Hilbert space $H_{a
,l}$.
Thus by the Riesz representation theorem there exists an
element $u_j\in H_{a,l}$ so that
$$<f_j,v>=(u_j,v).$$
Since the formal adjoint of $K_{\kappa}^{*}$ is $K_{\kappa}$ one sees from this
last
equation that $u_j$ is a distribution solution to (3.17).  We wish to be a
little more explicit about the local regularity of this solution near
the branch point points $a_j$ for $j=1,\ldots ,n$ and for this purpose it is
useful to consider in more detail functions $F\in H_{a,l}$.  Such a function
$F$, is the limit a sequence  of functions $F_m\in C^{\infty}_{a,
l}$ in the $H_{a,l}$ norm.
Recall that functions in $C^{\infty}_{a,l}$ have ``compact supports'' so that
integration by parts in the formula for the $H_{a,l}$ norm of $F_
m$ is
justified and one finds that $(F_m,F_m)$ is,
$$\int_{{\bf D}_R}\overline {F_m(\tilde {z})}K_{\kappa}K_{\kappa}^{
*}F_m(\tilde {z})\,d\mu (z)+m^2\int_{{\bf D}_R}\overline {F_m(\tilde {
z})}F_m(\tilde {z})\,d\mu (z).\eqno (3.19)$$
Now a simple computation shows that
$$K^{*}_{\kappa +1}K_{\kappa +1}-K_{\kappa}K^{*}_{\kappa}={{8(\kappa
+1)}\over {R^2}}.$$
Substituting $K_{\kappa +1}^{*}K_{\kappa +1}-{{8(\kappa +1)}\over {
R^2}}$ for $K_{\kappa}K_{\kappa}^{*}$ in (3.19) and integrating by
parts one finds that the $L^2(d\mu )$ norm of $K_{\kappa +1}F_m$ can be
expressed in
terms of the $L^2(d\mu )$ norm of $K_{\kappa}^{*}F_m$ and the $L^
2(d\mu )$ norm of $F_m$.  The
analogous result for the difference $F_m-F_{m'}$ shows that $K_{\kappa
+1}F_m$ is a
Cauchy sequence in $L^2(d\mu )$.  Thus we find that $K_{\kappa +1}
F$ is in $L^2(d\mu )$.

Next observe that if $F,$ $K_{\kappa}^{*}F$, and $K_{\kappa +1}F$ are in $
L^2(d\mu )$ it follows that
$F$ is locally in the Sobolev space $H^1$ (at least away from the branch
points).  Returning now to the distribution solution $u_j\in H_{a
,l}$ of (3.17)
above one can now conclude from the standard regularity results
for solutions to elliptic equations that since $u_j$ is locally in the
Sobolev space $H^1$ it must actually be $C^{\infty}$ away from the branch
points $a_i$ for $i=1,2,\ldots ,n$.

Using the fact that $u_j$, $K_{\kappa}^{*}u_j$ and $K_{\kappa +1}
u_j$ must all be locally in $L^2$ it
is not difficult to see that the local expansions for $u_j$ must have
the form
$$u_j(\tilde {z})=\sum_{n=0}^{\infty}\alpha_n^i(u_j)v_{n+l_i}(\tilde {
z},a_i)+\sum_{n=1}^{\infty}\beta_n^i(u_j)v_{n-l_i}^{*}(\tilde {z}
,a_i)$$
for $z$ near to $a_i$.  The reader should note that it is precisely
the additional condition that $K_{\kappa +1}u_j$ must be in $L^2$ that forces
$\alpha_{-1}^i(u_j)=0$ for all $i$.
Thus if we define,
$$V_j(\tilde {z})=u_j(\tilde {z})+\varphi_j(z)v_{l_j-1,\kappa +1}
(\tilde {z},a_j),$$
then it is clear that $V_j$ is a solution to (3.9) which is appropriately
branched at $a_1,a_2,\ldots ,a_n$, and it is in $L^2({\bf D}_R,d\mu
)$ together with $K_{\kappa}^{*}V_j$
by construction, since $K_{\kappa}^{*}v_{l_j-1}$ is locally less singular than
$
v_{l_j-1}$.
Finally
$$\alpha_{-1}^i(V_j)=\delta_{ij}$$
since $u_j$ does not contribute to the local expansion coefficients
$\alpha_{-1}^i(V_j).$  To exhibit the dependence of $V_j$ on the various
parameters
we will write,
$$V_j=V_j(l,\kappa ),$$
where $l=(l_1,l_2,\ldots ,l_n).$

We now exhibit the canonical $L^2$ basis of
Dirac wave functions.  Define,
$$w_j(l,k)=\left[\matrix{{2\over {mR}}V_j(l,k-{1\over 2})\cr
{2\over {m^2R}}K_{k-{1\over 2}}^{*}V_j(l,k-{1\over 2})\cr}
\right],\eqno (3.20)$$
where $0<l_j<1$ for each $j=1,2,\ldots ,n$.
Then $w_j(l,k)$ is an $L^2$ wave
function for the Dirac  equation with monodromy $e^{2\pi il_j}$ about
the point $a_j$ and with
$$c_i(w_j(l,k))=\delta_{ij}.\eqno (3.21)$$
\vskip.1in \par\noindent
{\bf The response functions}.
The canonical wave functions are very useful but they are not the
wave functions that are fundamental
in our treatment of $\tau -$functions.  We turn now to the
consideration of the Dirac wave functions $W_j(\tilde {z},\lambda
,k)$ and $W_j^{*}(\tilde {z},\lambda ,k)$
that will be central for us.  Because of the role they play in
the formula (4.50) for the derivative of the Green function we will
refer to these wave functions as the response functions.  We start
with a characterization of the desired response function $W_j$ and
then we use the $L^2$ wave functions to prove the existence of $W_
j$ and
$W_j^{*}.$

The response function $W_j$ whose existence we wish to establish is
characterized by the following three conditions:
\vskip.1in \par\noindent
(1) $W_j(\tilde {z},\lambda ,k)$ is a multivalued solution to the Dirac
equation which is
in $L^2(d\mu )$ in the complement of some compact neighborhood of
$\{a_1,\ldots ,a_n\}$.
\vskip.1in \par\noindent
(2) $W_j(\tilde {z},\lambda ,k)$  has monodromy multiplier $e^{2\pi
i\lambda_j}$ in a counter-clockwise
circuit of $a_j$ (with $\lambda_j$ now chosen so that $0<|\lambda_
j|<{1\over 2})$.
\vskip.1in \par\noindent
(3) $W_j(\tilde {z},\lambda ,k)$ has local expansions at each of the points $
a_j$ of the
following type,

$$W_j(\tilde {z},\lambda ,k)=\delta_{ij}w_{-{1\over 2}+\lambda_i}
+\sum_{n>0}\left\{a_{nj}^i(\lambda ,k)w_{n+\lambda_i}+b_{nj}^i(\lambda
,k)w^{*}_{n-\lambda_i}\right\},\eqno (3.22)$$
where $n\in {\bf Z}+{1\over 2}$ and
we abbreviate $w_{n+\lambda_i,k}(\tilde {z},a_i)$ by $w_{n+\lambda_
i}$ and $w_{n-\lambda_i,k}^{*}(\tilde {z},a_i)$ by
$w_{n-\lambda_i}^{*}.$  The first thing to observe about (3.22) is that it
characterizes $W_j(\tilde {z},\lambda ,k).$  Any two functions with local
expansions of
type (3.22) differ by a wave function whose local expansion has
only terms involving $w_{n+\lambda_i}$ and $w_{n-\lambda_i}^{*}$ for $
n\ge{1\over 2}.$  A little computation
using Theorem 3.0 shows that any such wave function, $w$, has  $L^
2$
norm 0 since either $c_i(w)=0$ or $c_i^{*}(w)=0$ for each $i=1,2,
\ldots ,n.$
\vskip.1in \par\noindent\
{\bf Remark}. {\it Because
$W_j(\tilde {z},\lambda ,k)$ depends on many parameters we will often omit
those which
are of less interest in the current calculations and rely on the
characteristic names for the variables to save the reader from
confusion.  Thus in what follows we will write $W_j(\lambda ,k)$ for the
function $W_j(\tilde {z},\lambda ,k).$

}\vskip.1in \par\noindent\
To prove the existence of $W_j(\lambda ,k)$ we will consider two cases. Let
${\bf P}$ denote the subset of $\{1,2,\ldots ,n\}$ which consists of those $
j$ for
which $0<\lambda_j<{1\over 2}$ and let ${\bf P}'$ denote the complementary
subset of
$\{1,2,\ldots ,n\}$ consisting of those $j$ for which $-{1\over 2}
<\lambda_j<0.$  If $j\in {\bf P}$ then
it is not hard to see that any function $W_j(\lambda ,k)$ with local expansions
of type (3.22) is locally in $L^2.$  In fact, one can see that if
$W_j(\lambda ,k)$ satisfies (3.22) and $j\in {\bf P}$ then
$$c_i(W_j(\lambda ,k))=\delta_{ij}\hbox{  for  }i,j\in {\bf P}\eqno
(3.23)$$
and
$$c_i^{*}(W_j(\lambda ,k))=0\hbox{  for  }i\in {\bf P}'.\eqno (3.
24)$$
But Theorem 3.0 shows that if $w\in {\cal W}(a,l)$ and $c_i(w)=0$ for
$i\in {\bf P}$ and $c_i^{*}(w)=0$ for $i\in {\bf P}'$ then $w=0.$ Since the
dimension of
${\cal W}(a,l)$ is $n$ it follows that $\{c_i\}_{i\in {\bf P}}\cup
\{c_i^{*}\}_{i\in {\bf P}'}$ is a collection of
independent linear functionals on ${\cal W}(a,l)$.  It follows that there
exists
an element $W_j(\lambda ,k)$ of ${\cal W}(a,l)$ which satisfies the conditions
(3.23) and
(3.24).  Without difficulty one can see that this implies $W_j(\lambda
,k)$ has
local expansions of type (3.22) and finishes the existence proof when
$j\in {\bf P}$.

When $j\in {\bf P}'$ the leading singularity in (3.22) at $a_j$ is no longer
locally
in $L^2$.  To prove the existence of $W_j(\lambda ,k)$ in this case it is
probably
simplest to return to the construction of the function $V_j$.  Instead
of forcing the singularity of $V_j$ to agree with $v_{l_j-1,\kappa
+1}(\tilde {z},a_j)$ near
$z=a_j$ one can construct an analogous function $\tilde {V}_j$ whose
singularity
at $z=a_j$ is matched with $v_{\lambda_j-1,\kappa +1}(\tilde {z},
a_j)$ where $\lambda_j=l_j-1$.  From
$\tilde {V}_j$ we construct
$$\tilde {W}_j=\left[\matrix{{2\over {mR}}\tilde {V}_j\cr
{2\over {m^2R}}\tilde {V}_j\cr}
\right]$$
as before.  Again using the independence of the collection
of $c_i$ for $i\in {\bf P}$ and the collection of $c_i^{*}$ for $
i\in {\bf P}'$ we see that
there exists an element $G_j\in {\cal W}(a,l)$ so that
$$c_i(G_j)=c_i(\tilde {W}_j)\hbox{  for  }i\in {\bf P}$$
and
$$c_i^{*}(G_j)=c_i^{*}(\tilde {W}_j)\hbox{  for  }i\in {\bf P}'.$$
If we now define
$$W_j(\lambda ,k)=\tilde {W}_j-G_j\hbox{  for  }\kappa =k-\half\,
,$$
then it is easy to check that $W_j(\lambda ,k)$ has local expansions of type
(3.22).  This finishes our existence proof for the response functions
satisfying (3.22).

The response functions $W^{*}_j(\tilde {z},\lambda ,k)$ are characterized by
conditions
precisely analogous to 1--3 above.  In particular $W_j^{*}$ is a multivalued
solution to the Dirac equation with monodromy multiplier $e^{2\pi
i\lambda_j}$
in a counter-clockwise circuit about $a_j$.  Instead of (3.22), however,
the local expansion for $W_j^{*}$ is,
$$W_j^{*}(\tilde {z},\lambda ,k)=\delta_{ij}w^{*}_{-{1\over 2}-\lambda_
i}+\sum_{n>0}\left\{c_{nj}^i(\lambda ,k)w_{n+\lambda_i}+d_{nj}^i(
\lambda ,k)w^{*}_{n-\lambda_i}\right\}.\eqno (3.25)$$
The proof of the existence of the response functions $W_j^{*}$ is precisely
analogous to the existence proof for $W_j$ and we leave this to the
reader.  The definition of $W_j^{*}$ we have given here differs by
the change $\lambda\rightarrow -\lambda$ compared to the analogous response
function
in [12].  In the present context the response functions $W_j$ and
$W_j^{*}$ are both solutions to the same Dirac equation with the
{\it same monodromy}.
\vskip.1in \par\noindent
{\bf Remark}.  {\it We would like to point out one possible source of confusion
for readers of this paper and the original papers of SMJ.  In [12] there
are two different bases for the same space ${\cal W}(a,l)$ referred to as the
canonical basis $w_j$ and the dual canonical basis $w_j^{*}$.  The response
functions $W_j^{*}$ above should not be confused with a dual basis of
any sort.  As the reader can easily see, the span of the response
functions $W_j$ and $W_j^{*}$  for $j=1,2,\ldots ,n$ is $2n$ dimensional as the
local
expansions (3.22) and (3.25) show that these functions are linearly
independent.   }

\vskip.2in \par\noindent
\centerline{\S 4 {\bf The Green function ${\bf G}^{a,\lambda}$ for the
Hyperbolic Dirac operator  }}
\vskip.2in \par\noindent
The goal of this section is to construct a Green function for the
Dirac operator in the presence of branch type singularities at the
points $a_j$ for $j=1,2,\ldots ,n$.  Before we attempt this there are a
number of preliminary considerations.
\vskip.1in \par\noindent
{\bf Green functions in the absence of branch points}.
We begin by identifying the Green functions for the
Helmholtz and Dirac operators in the absence of any singularities.
To start we require Green's identity.
First observe that the formula which
identifies $K^{*}_{\kappa}$ with the formal adjoint of $K_{\kappa}$ may be
written,
$$(\overline {K_{\kappa}f(z)}g(z)-\overline {f(z)}K^{*}_{\kappa}g
(z))d\mu (z)=d\varphi (\bar {f},g),\eqno (4.1)$$
where the one-form $\varphi$ is,
$$\varphi (f,g)=-2i\left({{f(z)g(z)}\over {1-{{|z|^2}\over {R^2}}}}
dz\right),\eqno (4.2)$$
and $d$ denotes the exterior derivative.
Taking complex conjugates of (4.1) one finds
$$(\overline {K^{*}_{\kappa}f}g-\bar {f}K_{\kappa}g)d\mu =-d\overline {
\varphi (f,\bar {g})}.\eqno (4.3)$$
 From (4.1) and (4.3) one may deduce
$$(\overline {K_{\kappa}f}K_{\kappa}g-\bar {f}K^{*}_{\kappa}K_{\kappa}
g)d\mu =d\varphi (\bar {f},K_{\kappa}g)\eqno (4.4)$$
and
$$(\overline {K_{\kappa}^{*}K_{\kappa}f}g-\overline {K_{\kappa}f}
K_{\kappa}g)d\mu =-d\overline {\varphi (K_{\kappa}f,\bar {g})}.\eqno
(4.5)$$
Adding (4.4) and (4.5) we find the infinitesimal version of Green's
theorem
$$(\overline {\Delta_{\kappa}f}g-\bar {f}\Delta_{\kappa}g)d\mu =d
(\overline {\varphi (K_{\kappa}f,\bar {g})}-\varphi (\bar {f},K_{
\kappa}g))\,,\eqno (4.6)$$
where we recall (2.3).
Finally if we write,
$$\lambda (s)=s\left(s-{2\over R}\right)=m^2+{{4\kappa (\kappa +1
)}\over {R^2}},$$
then we can rewrite (4.6) as
$$(\overline {(\Delta_{\kappa}-\lambda (s))f}g-\bar {f}(\Delta_{\kappa}
-\lambda (s))g)d\mu =d(\overline {\varphi (K_{\kappa}f,\bar {g})}
-\varphi (\bar {f},K_{\kappa}g)).\eqno (4.7)$$
We are interested in the Green function for $\lambda (s)-\Delta_{
\kappa}$ and so we
will look for radially symmetric solutions to
$$(\lambda (s)-\Delta_{\kappa})g=0$$
which have the appropriate singularity at $z=0$ and which are
in $L^2(d\mu )$ in a neighborhood of $|z|=R.$  Recall now from
(2.7)--(2.15) that a solution to the Helmholtz equation
can be written in the form,
$$g_{\kappa ,s}(z)=\left(1-{{|z|^2}\over {R^2}}\right)^{{{Rs}\over
2}}F_{\kappa ,s}\left({{|z|^2}\over {R^2}}\right),\eqno (4.8)$$
where $F_{\kappa ,s}(t)$ satisfies the hypergeometric equation (2.14) with,
$$\eqalign{a=&\textstyle{{{Rs}\over 2}}-\kappa ,\,\,\,b=\textstyle{{{
Rs}\over 2}}+\kappa ,\,\,\,c=1,\cr}
$$
and where $s$ is again given by (2.19).

The case $c=1$ is special for the hypergeometric equation.  One
solution, $F(a,b;1;t)$ is regular at $t=0$, while the second independent
solution has a logarthmic singularity
at $t=0.$  A local analysis near $t=1$ shows that when $Rs=a+b>1$
only one solution, $F_{\kappa ,s}(t),$ determines a solution, $g_{
\kappa ,s}(z)$, via (4.8) which
is square integrable near $|z|=R$ (square integrable with respect to
$d\mu$ that is).  Of course this solution is only
determined up to a constant multiple by this requirement.  We
fix the normalization of $F_{\kappa ,s}(t)$ by requiring that
$$g_{\kappa ,s}(z)\sim -{1\over {4\pi}}\log|z|+O(|z|\log|z|)\eqno
(4.9)$$
for $z$ near 0.  For this to work we must know that the solution
$F_{\kappa ,s}(t)$ which produces the solution $g_{\kappa ,s}(z)$ which is
square integrable
near $|z|=R$ is not also the solution of the hypergeometric equation
which is analytic at $t=0.$  We will sketch how one can see this.
Suppose that $v$ is the radial solution of
$$(K_{\kappa}^{*}K_{\kappa}+m^2)v=0$$
which tends to 0 as $|z|\rightarrow R$,  then for $\kappa =k-{1\over
2}$
$$w=\left[\matrix{-K_{k-{1\over 2}}v\cr
mv\cr}
\right]$$
is a solution to the Dirac equation which is square integrable
at $\infty .$ If $v$ is also well behaved near $|z|=0$, then $w$ will be
square integrable in a disk,
$$D_{\rho}=\left\{z:|z|\le R\hbox{th}\,{{\rho}\over 2}\right\},$$
of radius $\rho >0$ about 0.  A calculation
similar to the calculation that produced (2.42) and (2.43) shows
that the $L^2$ norm of $w$ on $D_{\rho}$ is expressible in a form precisely
analogous to (2.43).  The estimates (2.38) and (2.41) for $w$
show that this expression for the $L^2$ norm tends to 0 as $\rho\rightarrow
\infty .$
This contradiction shows that the solutions well behaved at $\infty$ and
at 0 cannot match up.

We next consider the sense in which $g_{\kappa ,s}$ is a Green function for
$\lambda (s)-\Delta_{\kappa}$. Let $f$ denote a smooth function of compact
support in ${\bf D}_R.$  Integrate both sides of (4.7) with
$f=f$ and $g=g_{\kappa ,s}$ over the complement of a disk of radius $
\epsilon$
about $z=0$.
Use Stokes' theorem to reduce the
result to a boundary integral over $C_{\epsilon},$ the circle of radius $
\epsilon$
about 0, and then use the asymptotics
(4.9) to evaluate the boundary integrals in the limit $\epsilon\rightarrow
0$.  We
find
$$\int_{{\bf D}_R}(\lambda (s)-\Delta_{\kappa})f(z)\overline {g_{
\kappa ,s}(z)}\,d\mu (z)=f(0).\eqno (4.11)$$
This shows that $g_{\kappa ,s}(z)$ is the solution appropriate to determine the
Green function for $\lambda (s)-\Delta_{\kappa}$.  To obtain this Green
function we must
translate $g_{\kappa ,s}$ in the appropriate fashion.  Introduce the variable
$w,$
$$w:=T[-a]z=R{{Rz-Ra}\over {-\bar {a}z+R^2}},$$
and define
$$v(w):={{R^2+\bar {a}w}\over {R^2+a\bar {w}}}.$$
Let $v(w)^{\kappa}$ denote the fractional power of $v(w)$ normalized so that
$v(0)^{\kappa}=1.$  Then the transformation property of $\Delta_{
\kappa}$ which is the
analogue of the transformation property (1.18) of the Dirac operator
is
$$v(w)^{-\kappa}\Delta_{\kappa}^{(w)}v(w)^{\kappa}=-K_{\kappa}^{*}
(w)K_{\kappa}(w)+{{4\kappa (\kappa +1)}\over {R^2}}:=\Delta_{\kappa}
(w).\eqno (4.12)$$
This should be understood in the following fashion.  The operator
$\Delta_{\kappa}^{(w)}$ is the expression for $\Delta_{\kappa}$ that one
obtains by the change of
variables $z=z(w)$.  The operators $K_{\kappa}(w)$ and $K_{\kappa}^{
*}(w)$ are the analogues
of $K_{\kappa}$ and $K_{\kappa}^{*}$ in the $w$ coordinates,
$$K_{\kappa}(w)=2\left\{\left(1-{{|w|^2}\over {R^2}}\right)\partial_
w-{{\kappa}\over {R^2}}\bar w\right\},$$
and
$$K_{\kappa}^{*}(w)=-2\left\{\left(1-{{|w|^2}\over {R^2}}\right)\bar
\partial_w+{{\kappa +1}\over {R^2}}w\right\},$$
so that $\Delta_{\kappa}(w)$ is the same operator in the $w-$coordinates that
$\Delta_{\kappa}$ is in the $z$ coordinates.
Now we introduce
$$g_{\kappa ,s}(y,z)=v(T[-y]z)^{\kappa}g_{\kappa ,s}(T[-y]z)=\left
({{R^2-y\bar {z}}\over {R^2-\bar {y}z}}\right)^{\kappa}g_{\kappa
,s}(T[-y]z).\eqno (4.13)$$
To see that this is the appropriate Green function suppose that
$f$ is a smooth function of compact support in ${\bf D}_R$ as above and
consider the integral
$$\eqalign{&\int_{{\bf D}_R}(\lambda (s)-\Delta_{\kappa})f(z)\overline {
g_{\kappa ,s}(y,z)\,}d\mu (z)\cr
&=\int_{{\bf D}_R}(\lambda (s)-\Delta_{\kappa}^{(w)})f(T[y]w)\overline {
v(w)^{\kappa}g_{\kappa ,s}(w)}\,d\mu (w)\cr
&=\int_{{\bf D}_R}(\lambda (s)-\Delta_{\kappa}(w))v(w)^{-\kappa}f
(T[y]w)\overline {g_{\kappa ,s}(w)\,}d\mu (w)\cr
&=v(w)^{-\kappa}f(T[y]w)_{w=0}=f(y),\cr}
\eqno (4.14)$$
where we made the substitution $z\leftarrow w$ to obtain the transformed
integral in the second line then used (4.13), (4.12) and finally (4.11) to
obtain
the third and fourth lines.

Next we construct the Green function for the Dirac operator, $m-D_
k,$
using the Green function $g_{\kappa ,s}(y,z).$
Recall now (2.2) and (2.6) from which it follows that
$$(m+D_k)(m-D_k)=\left[\matrix{\lambda (s)-\Delta_{k+{1\over 2}}&
0\cr
0&\lambda (s)-\Delta_{k-{1\over 2}}\cr}
\right].\eqno (4.15)$$
Now let $f(z)$ denote a smooth function of compact support on
${\bf D}_R$ with values in ${\bf C}^2$.  Then from (4.15) and (4.14) we have
$$\int_{{\bf D}_R}(m+D_k)(m-D_k)f(z)\cdot\left[\matrix{\overline {
g_{k+{1\over 2},s}(y,z)}\cr
0\cr}
\right]\,d\mu (z)=f_1(y).\eqno (4.16)$$
If we exclude an $\epsilon$ disk about $z=y$ in (4.16) and use Stokes'
theorem to ``integrate by parts'' once in (4.16) then we find
$$\int_{{\bf D}_R}(m-D_k)f(z)\cdot\overline {G_1(y,z)}\,d\mu (z)=
f_1(y),\eqno (4.17)$$
where
$$G_1(y,z)=(m+D_k^{*})\left[\matrix{g_{k+{1\over 2},s}(y,z)\cr
0\cr}
\right]=(m-D_k)\left[\matrix{g_{k+{1\over 2},s}(y,z)\cr
0\cr}
\right].\eqno (4.18)$$
The boundary of the $\epsilon$ disk about $z=y$ makes a contribution to the
Stokes' theorem calculation that vanishes in the limit $\epsilon\rightarrow
0$ due to
the weak logarithmic singularity in $g_{k+{1\over 2},s}(y,z)$ at $
z=y$.  In a
precisely similar fashion one finds that for
$$G_2(y,z):=(m-D_k)\left[\matrix{0\cr
g_{k-{1\over 2},s}(y,z)\cr}
\right],\eqno (4.19)$$
one has
$$\int_{{\bf D}_R}(m-D_k)f(z)\cdot\overline {G_2(y,z)}\,d\mu (z)=
f_2(y).\eqno (4.20)$$
\vskip.2in \par\noindent
{\bf The Green function for the Helmholtz operator in the presence of branch
points}.
Before tackling the existence question for a Green function
associated with the branched version of $m-D$ we will first consider
the analogous problem for the ``Helmholtz'' operator
$$\lambda (s)-\Delta_{\kappa}=m^2+K_{\kappa}^{*}K_{\kappa},$$
with
$$\lambda (s)=m^2+{{4\kappa (\kappa +1)}\over {R^2}}.$$
We wish to consider the action of $\Delta_{\kappa}$ on functions
on the unit disk with specified branching at a finite collection
of points $\{a_1,a_2,\ldots ,a_n\}$ in the disk ${\bf D}_R$.  It will be useful
for us to
recall the conventions of \S 3.  Thus we
identify points $a_j=(\alpha_j,\beta_j)$ in the disk with their standard
representation as complex numbers $a_j=\alpha_j+i\beta_j$.
Write ${\bf D}_R(a)$ for ${\bf D}_R\backslash \{a_1,a_2,\ldots ,a_
n\}$ and
$\tilde {{\bf D}}_R(a)$ for the simply connected covering space of
the punctured disk ${\bf D}_R(a)$.
Let $C_0^{\infty}(\tilde {{\bf D}}_R(a))$ denote the smooth
functions on $\tilde {{\bf D}}_R(a)$ whose supports project onto compact
subsets of
${\bf D}_R(a)$.  Fix a base point $a_0$ in ${\bf D}_R$ different from any
of the points $a_1,a_2,\ldots ,a_n$ and let $[\alpha_j]$ denote the homotopy
class of the simple
$a_0$ based loop, $\alpha_j,$ about $a_j$ in ${\bf D}_R(a)$ which circles $
a_j$ in a
counterclockwise fashion and does not wind around any of the
other points $a_i$ with $i\ne j$.  Points in $\tilde {{\bf D}}_R(
a)$ over $a_0$ correspond to
homotopy classes of $a_0$ based loops in ${\bf D}_R(a)$ and as above
we fix our model of $\tilde {{\bf D}}_R(a)$ to be the collection of homotopy
classes of paths in ${\bf D}_R(a)$ which start at $a_0$.  The canonical
projection,
$$\hbox{pr}:\tilde {{\bf D}}_R(a)\rightarrow {\bf D}_R,$$
maps each class $[\gamma ]$ of paths that start at $a_0$ into the endpoint $
\gamma (1)$.
This model of $\tilde {{\bf D}}_R(a)$ comes with a distinguished base point.
Let
$\tilde {a}_0$ denote the point in $\tilde {{\bf D}}_R(a)$ which corresponds to
the constant path
$[0,1]\ni t\rightarrow a_0$.
Let $R_j$ denote the deck transformation
on $\tilde {{\bf D}}_R(a)$ which maps $\tilde {a}_0$ to the point in $
\tilde {{\bf D}}_R(a)$ which corresponds to
the homotopy class $[\alpha_j].$  Although it is not completely precise it
will be convenient to adopt the following notation to avoid clumsy
expressions.  We will write $\tilde {z}$ for a typical point in $
\tilde {{\bf D}}_R(a)$ and
$$\hbox{pr}(\tilde {z})=z$$
for the projection on ${\bf D}_R.$

We wish to consider the action of $\Delta_{\kappa}$ on a subset of
$C^{\infty}_0(\tilde {{\bf D}}_R(a))$ with specified branching at $
a_j$ for $j=1,2,\ldots ,n.$
Let
$C_{a,l}^{\infty}$ denote the subset of $C_0^{\infty}(\tilde {{\bf D}}_
R(a))$ consisting of those functions
$f$ which transform under $R_j$ by
$$f(R_j\tilde {z})=e^{-2\pi il_j}f(\tilde {z}),$$
where $0\le l_j<1$ for $j=1,2,\ldots ,n$.  Note that since $C_{a,
l}^{\infty}$ is
a module for the multiplicative action of smooth functions on
${\bf D}_R$ and since the differential operators $\partial_z$ and $
\bar{\partial}_z$ lift in an obvious
way to differential operators on $C_{a,l}^{\infty}$ there is no difficulty in
defining the action of $\Delta_{\kappa}$ on $C_{a,l}^{\infty}.$
We will now construct a Green function for $\lambda (s)-\Delta_{\kappa}$ acting
on
$C^{\infty}_{a,l}$.  This is not completely accurate since we will implicitly
limit the strength of the singularities at the points $a_j$ in a
completion of $C^{\infty}_{a,l}$ but it will prove simpler to construct the
Green function than it will to give a complete discussion of the
domain of the corresponding operator.  By analogy with (4.14)
above we seek a function $g^{a,l}(\tilde {y},\tilde {z})$ so that for any $
f\in C^{\infty}_{a,l}$ we
have
$$\int_{{\bf D}_R}(\lambda (s)-\Delta_{\kappa})f(\tilde {z})\overline {
g^{a,l}(\tilde {y},\tilde {z})\,}d\mu (z)=f(\tilde {y}).\eqno (4.
21)$$
Of course, for this integral to make
sense it is necessary for the integrand,
$$\tilde {{\bf D}}_R(a)\ni\tilde {z}\rightarrow (\lambda (s)-\Delta_{
\kappa})f(\tilde {z})\overline {g^{a,l}(\tilde {y},\tilde {z})},$$
to descend to a function on ${\bf D}_R$.  This will happen if $g^{
a,l}(\tilde {y},\tilde {z})$
transforms as follows,
$$g^{a,l}(\tilde {y},R_j\tilde {z})=e^{-2\pi il_j}g^{a,l}(\tilde {
a},\tilde {z}),\eqno (4.22)$$
under the deck transformations $R_j$.  Equation (4.21) suggests that
$g^{a,l}(\tilde {y},\tilde {z})$ should transform in the first variable
$$g^{a,l}(R_j\tilde {y},\tilde {z})=e^{2\pi il_j}g^{a,l}(\tilde {
y},\tilde {z}).\eqno (4.23)$$

We will now give a functional analytic proof for the existence of
a suitable version of $g^{a,l}(\tilde {y},\tilde {z})$.  To begin we will first
concentrate
on the behavior of $g^{a,l}(\tilde {y},\tilde {z})$ as a function of the second
variable
$\tilde {z}.$ To obtain a Green function $g^{a,l}(\tilde {y},\tilde {
z})$ for $\lambda (s)-\Delta_{\kappa}$ satisfying (4.21)
above we seek a function
$$\tilde {{\bf D}}_R(a)\ni\tilde {z}\rightarrow g^{a,l}(\tilde {y}
,\tilde {z})$$
which is a solution to the Helmholtz equation,
$$(\lambda (s)-\Delta_{\kappa})g^{a,l}(\cdot ,\tilde {z})=0,$$
defined for $\tilde {z}\in\tilde {{\bf D}}_R(a)$ with $z\in {\bf D}_
R(a)\backslash \{y\}$ (recall that $\hbox{pr}(\tilde {z})=z$ and
$\hbox{pr}(\tilde {y})=y)$, with monodromy multiplier $e^{-2\pi i
l_j}$
in a counterclockwise circuit of
the branch point $a_j$.  Also, to obtain a Green function we want
$g^{a,l}(\tilde {y},\tilde {z})$ to differ from $g_{\kappa ,s}(y,
z)$ by a smooth function when
$\tilde {y}$ is close to $\tilde {z}$.  Note that, because of the
transformation properties we want for $g^{a,l}(\tilde {y},\tilde {
z})$ it will
have singularities on the set
$${\cal D}:=\hbox{$\{$ $\hbox{pr}^{-1}(x)\times\hbox{pr}^{-1}(x):
x\in {\bf D}_R(a)\}$}$$
as well as on the diagonal of $\tilde {{\bf D}}_R(a)\times\tilde {
{\bf D}}_R(a)$.

To make this precise we let $\phi (z)$ denote a $C^{\infty}_0$ function on $
{\bf R}^2$ with
$$\phi (z)=1\hbox{  for  }|z|\le 1$$
and
$$\phi (z)=0\hbox{  for  }|z|\ge 2,$$
and define
$$\phi_{\epsilon}(z):=\phi\left({z\over {\epsilon}}\right).$$
Then for $y$ fixed and different from $a_j$ for $j=1,\ldots ,n$ and $
\epsilon$ small
enough the support of $z\rightarrow\phi_{\epsilon}\left(z-y\right
)$ will
not contain any of the branch points $\{a_1,a_2,\ldots ,a_n\}.$  In such
circumstances it is easy to see that that the function,
$$z\rightarrow\phi_{\epsilon}\left(z-y\right)g_{\kappa ,s}(y,z),$$
has support in an {\it elementary neighborhood of the point
$z=y.$}  An
elementary neighborhood, $U,$ of a point $y\in {\bf D}_R(a)$ is one
for which the inverse image of $U$ under the projection,
$$\hbox{pr}:\tilde {{\bf D}}_R(a)\rightarrow {\bf D}_R(a),$$
splits into components $V_{\alpha}$ with the property that
$$\hbox{pr}:V_{\alpha}\rightarrow U$$
is a diffeomorphism.  The sets $V_{\alpha}$ are indexed by
homotopy classes, $\alpha ,$ of paths that join the base point $a_
0$ to $y$.
Using this property we will now indicate how to
obtain a lift of the function,
$$U\times U\ni (x,y)\rightarrow\phi_{\epsilon}(y-x)g_{\kappa ,s}(
x,y),$$
to a function defined on
$$\hbox{pr}^{-1}(U)\times\hbox{pr}^{-1}(U)\backslash {\cal D}$$
with prescribed multipliers.  Let $a_0$ denote the base point in
${\bf D}_R(a)$ as above.  Let $\alpha_0$ denote a path
joining $a_0$ to $y$ representing fixed homotopy class $[\alpha_0
].$
Once [$\alpha_0]$ is fixed we can index the components of $\hbox{pr}^{
-1}(U)$ by homotopy
classes of loops based at $y$.  Furthermore the choice of $\alpha_
0$ allows
us to identify the homotopy group of loops based at $a_0\in {\bf D}_
R(a)$  with the
homotopy group of loops based at $y\in {\bf D}_R(a)$.  Thus if we let
$\pi$ denote the representation of the fundamental group of (${\bf D}_
R(a),a_0)$
which results from assigning the number $e^{-2\pi il_j}$ to the generator
which starts at $a_0$ and
makes a simple counterclockwise circuit of $a_j$ not winding around
any of the other points $a_k$ for $k\ne j$, then this naturally gives a
representation of the fundamental group of $({\bf D}_R(a),y)$ as well;
namely,
$$\pi [\gamma ]:=\pi [\alpha_0^{-1}\gamma\alpha_0]\hbox{  for  }[
\gamma ]\in\pi_1({\bf D}_R(a),y).$$
Now we define a lift of
$$g_{\kappa ,s}^{\epsilon}(x,y):=\phi_{\epsilon}(y-x)g_{\kappa ,s}
(x,y)$$
which is defined on $U\times U\backslash \{(x,x):x\in {\bf D}_R(a
)\}$ by
$$\tilde {g}_{\kappa ,s}^{\epsilon}(\tilde {x},\tilde {y}):=\pi [
\gamma_2\gamma_1^{-1}]g_{\kappa ,s}^{\epsilon}(x,y)\hbox{  for  }
\tilde {x}\in V_{[\gamma_1\alpha_0]}\hbox{  and  }\tilde {y}\in V_{
[\gamma_2\alpha_0]}.$$
The function $\tilde {g}_{\kappa ,s}^{\epsilon}(\tilde {x},\tilde {
y})$ is defined on $\hbox{pr}^{-1}(U)\times\hbox{pr}^{-1}(U)\backslash
{\cal D}$, and $[\gamma_k]\in\pi_1({\bf D}_R(a),y)$ for
$k=1,2.$  It is not hard to see that this
extension is independent of the choice of $[\alpha_0]$, for if $\alpha_
0'$ is another
path from $a_0$ to $y$ one has
$$\pi [\gamma_2\gamma_1^{-1}]=\pi [\alpha_0^{-1}\gamma_2\gamma_1^{
-1}\alpha_0]=\pi [\alpha_0^{-1}\alpha_0']\pi [\alpha_0^{\prime -1}
\gamma_2\gamma_1^{-1}\alpha_0']\pi [\alpha_{0^{}}^{\prime -1}\alpha_
0]$$
and
$$\pi [\alpha_0^{-1}\alpha_0']\pi [\alpha_{0^{}}^{\prime -1}\alpha_
0]=1.$$
It is now natural to extend $\tilde {g}_{\kappa ,s}^{\epsilon}(\tilde {
y},\tilde {z})$ to all of $\tilde {{\bf D}}_R(a)\times\tilde {{\bf D}}_
R(a)\backslash\!{\cal D}$ by
setting it equal to 0 on the complement of $\hbox{pr}^{-1}(U)\times\hbox{pr}^{
-1}(U)\backslash {\cal D}$.
Now fix $\tilde {y}\in\tilde {{\bf D}}_R(a)$.  We will construct $
g^{a,l}(\tilde {y},\tilde {z})$ so that up to
a smooth error it looks like $\tilde {g}_{\kappa ,s}^{\epsilon}(\tilde {
y},\tilde {z})$ whenever $z,$ the projection
of $\tilde {z}$, is close to $y$, the projection of $\tilde {y}$.   More
precisely, we
require that
$$(\lambda (s)-\Delta_{\kappa})\left\{g^{a,l}(\tilde y,\tilde z)-
\tilde g_{\kappa ,s}^{\epsilon}(\tilde y,\tilde z)\right\}\eqno (
4.24)$$
be free of singularities as a function of  $\tilde {z}\in$ $\tilde {
D}_R(a)$.  When
$\tilde {z}$ is not in the fiber over $y=\hbox{pr}(\tilde {y})$ we want $
(\lambda (s)-\Delta_{\kappa})g^{a,l}(\tilde {y},\tilde {z})=0$.
Thus with the formal cancellation of the delta functions in (4.24) the
only remaining contribution comes when the differential operator
part of $\Delta_{\kappa}$ hits the function $\phi_{\epsilon}(z-y)$ in $
\tilde {g}_{\kappa ,s}^{\epsilon}(\tilde {y},\tilde {z})$ (note that the
multiplicative part of $\Delta_{\kappa}$ which does not contain any derivatives
is absorbed in the delta function that arises when $\Delta_{\kappa}$ hits $
g_{\kappa ,s}(y,z)$).
Now choose $\epsilon$ small enough so that $2\epsilon$ will produce an
elementary
neighborhood $U$ above.  Then $\phi_{2\epsilon}(z-y)$ is identically 1 on the
support of $\phi_{\epsilon}(z-y)$. Using this one can easily see that there is
an
explicit second order differential operator, $p(\partial_z,\bar{\partial}_
z$),
whose lowest order terms are first order in
$\partial_z$ and $\bar{\partial}_z$ and for which we want
$$(\lambda (s)-\Delta_{\kappa})\left\{g^{a,l}(\tilde y,\tilde z)-
\tilde g_{\kappa ,s}^{\epsilon}(\tilde y,\tilde z)\right\}=\left\{
p(\partial_z,\bar\partial_z)\phi_{\epsilon}(z-y)\right\}\tilde {g}_{
\kappa ,s}^{2\epsilon}(\tilde {y},\tilde {z}).\eqno (4.25)$$
Since $p(\partial_z,\bar{\partial}_z)$ kills $\phi_{\epsilon}(z-y
)$ near $z=y$ it follows that the right
hand side of (4.25) is a smooth function of $\tilde {z}$ in $C^{\infty}_{
a,l}$.
We will now show that for any $f\in C_{a,l}^{\infty}$ whose
support projects onto an elementary neighborhood  it is
possible to find a multivalued function $F$, transforming as $f$ does
under deck transformations, so that
$$(\lambda (s)-\Delta_{\kappa})F=f.\eqno (4.26)$$
Once we have this existence result
we can then recover $g^{a,l}(\tilde {y},\tilde {z})$ from the solution to
(4.25) above by
adding in $\tilde {g}_{\kappa ,s}^{\epsilon}(\tilde {y},\tilde {z}
)$.

Now we will recall the Hilbert space technique of \S 3 for
finding a solution to (4.26).  Let $H_{a,l}$ denote the Hilbert space
completion of $C_{a,l}^{\infty}$ with respect to the norm derived from the
inner product
$$(F,G)=\int_{{\bf D}_R}\left\{\overline {K_kF(\tilde {z})}K_kG(\tilde
z)+m^2\overline {F(\tilde {z})}G(\tilde z)\right\}d\mu (z).$$
Note that the integrand descends to a function on ${\bf D}_R$ because
the monodromy multipliers of $\bar {F}$ and $G$ cancel.  Now suppose that
$f\in C^{\infty}_{a,l}$ and the support of $f$ projects onto an elementary
neighborhood.  For $v\in C_{a,l}^{\infty}$ consider the linear functional
$$v\rightarrow\int_{{\bf D}_R}\overline {f(\tilde {z})}v(\tilde {
z})\,d\mu (z):=<f,v>.\eqno (4.27)$$
Since $(h,h)\ge m^2<h,h>$, for all $h\in C_{a,l}^{\infty}$, it is clear that
the
linear functional (4.27) is continuous on the Hilbert space $H_{a
,l}$.
Thus by the Riesz representation theorem there exists an
element $F\in H_{a,l}$ so that
$$<f,v>=(F,v).$$
Since the formal adjoint of $K_k$ is $K_k^{*}$ one sees from this last
equation that $F$ is a distribution solution to (4.26).  Recall from the
discussion in \S 3 that $F$ will be locally in the Sobolev space $
H^1$
(at least away from the branch points) and so by standard regularity
results for elliptic equations, one can conclude that $F$
must actually be $C^{\infty}$ away from the branch points $a_j$ for $
j=1,2,\ldots ,n$.

We now use this result to solve the equation
$$(\lambda (s)-\Delta_{\kappa})F=\left\{p(\partial_z,\bar\partial_
z)\phi_{\epsilon}(z-y)\right\}\tilde {g}_{\kappa ,s}^{2\epsilon}(
\tilde {y},\tilde {z}).\eqno (4.28)$$
Let $g^{a,l}(\tilde {y},\tilde {z})$ denote the sum of the solution $
F(\tilde {y},\tilde {z})$ to (4.28) and
$\tilde {g}_{\kappa ,s}^{\epsilon}(\tilde {y},\tilde {z})$.
Then the reader can check that $g^{a,l}$
satisfies the following three conditions:
\vskip.1in \par\noindent {\bf G}1: The map $\tilde {z}\rightarrow
g^{a,l}(\tilde {y},\tilde {z})$ is a $C^{\infty}(\tilde {{\bf D}}_
R(a))$ solution
of the Helmholtz equation,
$$(\lambda (s)-\Delta_{\kappa})g^{a,l}(\tilde {y},\tilde {z})=0,$$
for $\hbox{$z\ne y$}$ and ~$z\ne a_j$ for $j=1,2,\ldots ,n$ which transforms
under the
deck transformations $R_j$ as follows,
$$g^{a,l}(\tilde {y},R_j\tilde {z})=e^{-2\pi il_j}g^{a,l}(\tilde {
y},\tilde {z}).$$
\vskip.1in \par\noindent {\bf G}2: If $U$ is an elementary neighborhood of the
point $y$ then in the complement $\tilde {{\bf D}}_R(a)\backslash\hbox{pr}^{
-1}(U)$ the functions,
$$\eqalign{&\tilde {z}\rightarrow g^{a,l}(\tilde {y},\tilde {z}),\cr
&\tilde {z}\rightarrow K_kg^{a,l}(\tilde {y},\tilde {z}),\cr
&\tilde {z}\rightarrow K^{*}_{k-1}g^{a,l}(\tilde {y},\tilde {z}),\cr}
$$
have absolute squares that are integrable with
respect to the natural hyperbolic measure, $d\mu (z),$ on ${\bf D}_
R.$
\vskip.1in \par\noindent {\bf G}3: For some elementary neighborhood, $
U,$ of the
point $y$ the difference
$$g^{a,l}(\tilde {y},\tilde {z})-\tilde {g}_{\kappa ,s}^{\epsilon}
(\tilde {y},\tilde {z})$$
is smooth as a function of $\tilde {z}\in\hbox{pr}^{-1}(U)$.
\vskip.2in \par\noindent\
Next we want to recall the consequences of {\bf G}2 for the local
expansions of $\tilde {z}\rightarrow g^{a,l}(\tilde {y},\tilde {z}
)$ about the points $a_j$ for $j=1,2,\ldots ,n$.
Because $g^{a,l}(\tilde {y},\tilde {z})$ is a multivalued solution to the
Helmholtz
equation with monodromy $e^{-2\pi il_j}$ about the point $a_j$ it follows that
$g^{a,l}(\tilde {y},\tilde {z})$ has a local expansion,
$$g^{a,l}(\tilde {y},\tilde {z})=\sum_{n\in {\bf Z}}\{\alpha_n^j(
\tilde {y})v_{n-l_j,\kappa}(\tilde {z},a_j)+\beta_n^j(\tilde {y})
v^{*}_{n+l_j,\kappa}(\tilde {z},a_j)\},$$
valid for $z$ near $a_j.$  Recall now that $l_j$ is chosen so that
$0\le l_j<1$ and that the leading singularity in $v_{l,\kappa}(\tilde {
z},a)$ and $v^{*}_{l,\kappa}(\tilde {z},a)$ goes
like $r^l$ in both cases, where $r$ is the distance from $z$ to $
a$.  It is now
a straightforward exercise to verify that the condition {\bf G}2 implies
the following restriction on the local expansion for $g^{a,l}(\tilde {
y},\tilde {z}),$
$$g^{a,l}(\tilde {y},\tilde {z})=\sum_{n=1}^{\infty}\alpha_n^j(\tilde {
y})v_{n-l_j,\kappa}(\tilde {z},a_j)+\sum_{n=0}^{\infty}\beta_n^j(
\tilde {y})v^{*}_{n+l_j,\kappa}(\tilde {z},a_j),\eqno (4.29)$$
for $z$ near $a_j$.  We will see that this restriction on the local
expansions of $g^{a,l}(\tilde {y},\tilde {z})$ together with {\bf G}1 and {\bf
G}3 and the fact that
$\tilde {z}\rightarrow g^{a,l}(\tilde {y},\tilde {z})$ is square integrable on
$
{\bf D}_R$ suffice to characterize
$g^{a,l}(\tilde {y},\tilde {z})$.  We also wish to characterize $
g^{a,l}(\tilde {y},\tilde {z})$ as a function of
the first variable $\tilde {y}$.  Since $\Delta_{\kappa}$ is Hermitian
symmetric one
might expect that
$$g^{a,l}(\tilde {y},\tilde {z})=\overline {g^{a,l}(\tilde {z},\tilde {
y})}.\eqno (4.30)$$
Observe that the right hand side of (4.29) transforms under
$\tilde {y}\rightarrow R_j\tilde {y}$ by the multiplier $e^{2\pi
il_j}$ which is appropriate for $g^{a,l}(\tilde {y},\tilde {z}).$
Also since $g_{\kappa ,s}(y,z)=\overline {g_{\kappa ,s}(z,y)}$ it follows that
the right hand side of (4.30)
differs from the Green function $g_{\kappa ,s}(y,z)$ along the diagonal by a
smooth function---this is additional evidence that the right hand
side of (4.30) should agree with $g^{a,l}(\tilde {y},\tilde {z})$.  We will now
confirm (4.30)
by a
Stokes' theorem calculation along the lines of the standard argument
for such symmetry.
Now choose $\tilde {x}$ and $\tilde {y}$ two {\it distinct} points in $
\tilde {{\bf D}}_R(a)$ and let
$$g(\tilde {z})=g^{a,l}(\tilde {z},\tilde {y})$$
and
$$f(\tilde {z})=g^{a,l}(\tilde {z},\tilde {x}).$$
Substitute these last two functions in (4.7) and integrate the resulting
equality over the complement of the union of the balls of radius $
\epsilon$ about
each of the points $a_j$ and the balls of radius $\epsilon$ about both $
x$ and
$y$.
The left hand side gives 0 since
$$\hbox{$(\lambda (s)-\Delta_{\kappa})f=(\lambda (s)-\Delta_{\kappa}
)g=0$}$$
on this set and
Stokes' theorem reduces the integral on the right to an integral
over the circles of radius $\epsilon$ about the points $a_j$ for $
j=1,\ldots ,n$ and
$x$ and $y$.
Using the results for the local expansion of $g^{a,l}(\tilde {z},
\tilde {x})$ given in (4.28)
above one finds that each of the integrals,
$$\int_{\partial B_{\epsilon}(a_j)}\varphi (\bar {f},K_{\kappa}g)
-\overline {\varphi (K_{\kappa}f,\bar {g})},$$
vanishes in the limit $\epsilon\rightarrow 0$.  Finally, using the fact that $
g^{a,l}(\tilde {x},\tilde {y})$
differs from the Green function for $\lambda (s)-\Delta_{\kappa}$ by a smooth
function
near the diagonal one finds that the contribution to the integral
of
$$\varphi (\bar {f},K_{\kappa}g)-\overline {\varphi (K_{\kappa}f,
\bar {g})}$$
over the circles of radius $\epsilon$ about $x$ and $y$ is just
$$g(\tilde {x})-\overline {f(\tilde {y})}=g^{a,l}(\tilde {x},\tilde {
y})-\overline {g^{a,l}(\tilde {y},\tilde {x})}$$
in the limit $\epsilon\rightarrow 0$.
This last quantity must vanish and this establishes the
Hermitian symmetry of the Green function we are considering.
\vskip.1in\par\noindent
{\bf The Green function for the Dirac operator in the presence of branch
points}.
Next we turn to the construction of a Green function for the
Dirac operator acting on $C_{a,l}^{\infty}$ by using formulas (4.18) and
(4.19) with $g_{\kappa ,s}$ replaced by $g^{a,l}.$  This will not give us the
Green function we want because the local expansions will not
have the ``minimal singularity'' type that we desire.  However, by
subtracting suitable wave functions from the results we are
able to pass to the desired Green function in much the same fashion
that we constructed the wave functions $W_j$ from the canonical
$L^2$ wave functions $w_j.$  In analogy with (4.18) and (4.19) we now
define ${\cal G}^{a,l}$ by
$${\cal G}^{a,l}_1(\tilde {y},\tilde {z})=(m-D_k)\left[\matrix{g_{
k+{1\over 2},s}^{a,l}(\tilde {y},\tilde {z})\cr
0\cr}
\right]=\left[\matrix{mg_{k+{1\over 2},s}^{a,l}(\tilde {y},\tilde {
z})\cr
K_{k-{1\over 2}}^{*}g_{k+{1\over 2},s}^{a,l}(\tilde {y},\tilde {z}
)\cr}
\right],\eqno (4.31)$$
$${\cal G}_2^{a,l}(\tilde {y},\tilde {z})=(m-D_k)\left[\matrix{0\cr
g_{k-{1\over 2},s}^{a,l}(\tilde {y},\tilde {z})\cr}
\right]=\left[\matrix{-K_{k-{1\over 2}}g_{k-{1\over 2},s}^{a,l}(\tilde {
y},\tilde {z})\cr
mg_{k-{1\over 2},s}^{a,l}(\tilde {y},\tilde {z})\cr}
\right],\eqno (4.32)$$
where we have explicitly noted the dependence of $g^{a,l}_{\kappa
,s}$ on $(\kappa ,s)$ to
avoid confusion.  We now record the local expansion results for (4.31)
and (4.32) which are obtained from (4.29).  There are two different
cases to consider depending on whether $0<l_j<{1\over 2}$  or ${1\over
2}\le l_j<1$.  In
case $0<l_j<{1\over 2}$ we write $\lambda_j=l_j$ and the local expansions for
${\cal G}^{a,l}$ are about $z=a_j$ are
$$\eqalign{{\cal G}^{a,l}_1(\tilde {y},\tilde {z})&=\sum_{n\ge{3\over
2}}\alpha_{n,1}^jw_{n-\lambda_j,k}(\tilde {z},a_j;-m)+\sum_{n\ge
-{1\over 2}}\beta_{n,1}^jw_{n+\lambda_j,k}^{*}(\tilde {z},a_j;-m)
,\cr
{\cal G}_2^{a,l}(\tilde {y},\tilde {z})&=\sum_{n\ge{1\over 2}}\alpha_{
n,2}^jw_{n-\lambda_j,k}(\tilde {z},a_j;-m)+\sum_{n\ge{1\over 2}}\beta_{
n,2}^jw_{n+\lambda_j,k}^{*}(\tilde {z},a_j;-m),\cr}
\eqno (4.33)$$
where all the sums are over elements of ${\bf Z}+{1\over 2}$ and we have
written
$w_{l,k}(\tilde {z},a;m)$ and $w_{l,k}^{*}(\tilde {z},a;m)$ for (2.22) and
(2.23) in order to
recognize explicitly the dependence on the parameters $k$ and $m$ (it is $
-m$ which
occurs in (4.33)).

In case ${1\over 2}\le l_j<1$ we write $\lambda_j=l_j-1$ and the local
expansion results
about $z=a_j$ are
$$\eqalign{{\cal G}^{a,l}_1(\tilde {y},\tilde {z})&=\sum_{n\ge{1\over
2}}\alpha_{n,1}^jw_{n-\lambda_j,k}(\tilde {z},a_j;-m)+\sum_{n\ge{
1\over 2}}\beta_{n,1}^jw_{n+\lambda_j,k}^{*}(\tilde {z},a_j;-m),\cr
{\cal G}_2^{a,l}(\tilde {y},\tilde {z})&=\sum_{n\ge -{1\over 2}}\alpha_{
n,2}^jw_{n-\lambda_j,k}(\tilde {z},a_j;-m)+\sum_{n\ge{3\over 2}}\beta_{
n,2}^jw_{n+\lambda_j,k}^{*}(\tilde {z},a_j;-m).\cr}
\eqno (4.34)$$
In (4.33) and (4.34) we have suppressed the dependence on parameters
$\tilde {y}$, $k$, $m$, and $\lambda_j$
for the coefficients $\alpha_{n,i}^j$ and $\beta_{n,i}^j$ to avoid further
complicating
expressions that are already notationally overburdened.

To obtain the Green function we desire our object will be to
eliminate those terms in the local expansions (4.33) and (4.34)
with $n<{1\over 2}$.  We can do this by subtracting suitable
multiples of the wave functions $W_j(-m)$ and $W_j^{*}(-m)$ without introducing
additional singularities on the ``diagonal'' $y=z$.  As is the case in
(4.33) and (4.34) it is now convenient to explicitly identify the
$m-$dependence of the wave functions $W_j$ and $W_j^{*}$ since the Green
functions ${\cal G}_j^{a,l}(\tilde {y},\tilde {z})$ are solutions to the
equation $
(m+D_k)\psi =0$ in
the $\tilde {z}$ variable instead of the original Dirac equation $
(m-D_k)\psi =0$.
As always our notation for the response functions will suppress
dependence on parameters that do not play a role in the current
calculations and will rely on characteristic names for the parameters
to make our intentions clear to the reader.  Thus $W_j(\tilde {z}
,-m)$ is our short
hand notation for $W_j(\tilde {z},k,\lambda ,-m)$.
Define
$$\eqalign{G_1^{a,\lambda}(\tilde {y},\tilde {z})&={\cal G}_1^{a,
l}(\tilde {y},\tilde {z})-\sum_{j:\lambda_j>0}\beta_{-{1\over 2},
1}^jW_j^{*}(\tilde {z},-m),\cr
G_2^{a,\lambda}(\tilde {y},\tilde {z})&={\cal G}_2^{a,l}(\tilde {
y},\tilde {z})-\sum_{j:\lambda_j<0}\alpha_{-{1\over 2},2}^jW_j(\tilde {
z},-m).\cr}
\eqno (4.37)$$
As the reader can easily verify the subtractions in (4.37) remove all
terms in the local expansions on the right hand side with $n<{1\over
2}$ and
we find
$$G^{a,\lambda}_i(\tilde {y},\tilde {z})=\sum_{n\ge{1\over 2}}\left
\{e_{n,i}^j(\tilde y)w_{n-\lambda_j}(\tilde z,a_j;-m)+f_{n,i}^j(\tilde
y)w^{*}_{n+\lambda_j}(\tilde z,a_j;-m)\right\},\eqno (4.38)$$
where we have once again omitted the $k$ dependence of $w_{l,k}$ and $
w_{l,k}^{*}$
for brevity.

We turn next to a characterization of $G_i^{a,\lambda}(\tilde {y}
,\tilde {z})$ along the lines of
the characterization {\bf G}1, {\bf G}2, and {\bf G}3 of $g^{a,l}
(\tilde {y},\tilde {z}).$
\vskip.1in\par\noindent
{\bf Proposition} 4.0 {\sl The components $G_i^{a,\lambda}(\tilde {
y},\tilde {z})$ of the Green function
constructed above for $(m-D_k)$ acting on $C_{a,l}^{\infty}$ are uniquely
characterized by
the following three properties: }
\vskip.1in\par\noindent
{\sl(1) For $i=1,2$ the vector valued functions $\tilde {z}\rightarrow
G_i^{a,\lambda}(\tilde {y},\tilde {z})$ are
solutions to the equation,}
$$(m+D_k)G_i^{a,\lambda}(\tilde {y},\tilde {z})=0,$$
{\sl when $z$ is not equal to $a_j$ for $j=1,\ldots ,n$ and not equal to $
y$.  Under
the deck transformation $R_j$, $G^{a,\lambda}_i(\tilde {y},\tilde {
z})$ transforms as follows,}
$$G^{a,\lambda}_i(\tilde {y},R_j\tilde {z})=e^{-2\pi i\lambda_j}G^{
a,\lambda}_i(\tilde {y},\tilde {z}).$$
\vskip.1in\par\noindent
{\sl
(2) About each of the points $a_j$ for $j=1,2,\ldots ,n$ the functions
$G_i^{a,\lambda}(\tilde {y},\tilde {z})$ have restricted local expansions given
by (4.38) above.
In the complement of a compact neighborhood of the set
$\{y,a_1,a_2,\ldots ,a_n\}$ the function $\tilde {z}\rightarrow G^{
a,\lambda}_i(\tilde {y},\tilde {z})$ has a norm which is
square integrable.}
\vskip.1in\par\noindent
{\sl(3) For some elementary neighborhood $U,$ of $y$ the
difference,}
$$G_i^{a,\lambda}(\tilde {y},\tilde {z})-\tilde {G}^{\epsilon}_i(
\tilde {y},\tilde {z}),$$
{\sl is smooth as a function of $\tilde {z}\in\hbox{pr}^{-1}(U)$ and $
\tilde {G}_i^{\epsilon}(\tilde {y},\tilde {z})$ is defined
by,}
$$\eqalign{\tilde {G}^{\epsilon}_1(\tilde {y},\tilde {z})&=\left[\matrix{
m\tilde {g}^{\epsilon}_{k+{1\over 2},s}(\tilde {y},\tilde {z})\cr
K_{k-{1\over 2}}^{*}\tilde {g}^{\epsilon}_{k+{1\over 2},s}(\tilde {
y},\tilde {z})\cr}
\right]\cr
\tilde {G}_2^{\epsilon}(\tilde {y},\tilde {z})&=\left[\matrix{-K_{
k-{1\over 2}}\tilde {g}^{\epsilon}_{k-{1\over 2},s}(\tilde {y},\tilde {
z})\cr
m\tilde {g}_{k-{1\over 2},s}^{\epsilon}(\tilde {y},\tilde {z})\cr}
\right].\cr}
$$
\vskip.1in\par\noindent
Proof.
These three properties of $G^{a,\lambda}_i(\tilde {y},\tilde {z})$ are simple
translations of
the properties {\bf G}1, {\bf G}2, and {\bf G}3 for the Green function $
g^{a,l}(\tilde {y},\tilde {z}).$
The fact that these three conditions characterize $G^{a,\lambda}_
i(\tilde {y},\tilde {z})$ can
be understood as follows.  Suppose that $G_i^{a,\lambda}(\tilde {
y},\tilde {z})$ and $H_i^{a,\lambda}(\tilde {y},\tilde {z})$
both satisfy all the conditions in Proposition 4.0 and consider the
difference
$$\Delta (\tilde {y},\tilde {z})=G_i^{a,\lambda}(\tilde {y},\tilde {
z})-H_i^{a,\lambda}(\tilde {y},\tilde {z}).$$
Then (3) of Proposition 4.0  implies that $\Delta (\tilde {y},\tilde {
z})$ is free of
singularities near $z=y$ and (1) and (2) imply that $\tilde {z}\rightarrow
\Delta (\tilde {y},\tilde {z})$
is a solution to $(m+D_k)\Delta (\tilde {y},\tilde {z})=0$ with monodromy
multipliers
$e^{-2\pi i\lambda_j}$ about $a_j$ and which is globally in $L^2(
d\mu ).$
However, the restricted local expansion (4.38) implies that for
each $j=1,2,\ldots ,n$ either $c_j(\Delta )=0$ or $c_j^{*}(\Delta
)=0$ and this together with
Theorem 3.0 implies that $\Delta =0$.  QED
\vskip.2in \par\noindent
{\bf The derivative of the Green function $G^{a,\lambda}$}.
Next we will identify the low order expansion coefficients $e_{{1\over
2},i}^j(\tilde {y})$
and $f_{{1\over 2},i}^j(\tilde {y})$ in (4.38) as the components of wave
functions for the Dirac
equation.  Suppose that $f(\tilde {z})$ is a solution to the Dirac equation
$(m-D_k)f=0$ which transforms under the deck transformations
$R_j$ as follows,
$$f(R_j\tilde {z})=e^{-2\pi i\lambda_j}f(\tilde {z}),$$
and which has restricted local expansions,
$$f(\tilde {z})=\sum_{n\ge -{1\over 2}}\left\{a_n^jw_{n-\lambda_j}
(\tilde z,a_j;m)+b_n^jw^{*}_{n+\lambda_j}(\tilde z,a_j;m)\right\}
.\eqno (4.39)$$
To lighten the notation for the next calculation we write
$$g(\tilde {z})=G^{a,\lambda}_i(\tilde {y},\tilde {z}).$$
Now let $D_{\epsilon}(y)$ denote the disk of radius $\epsilon$ about $
y$, and let $C_{\epsilon}(y)$
denote the circle of radius $\epsilon$ about $y$.  Then for $z$ in the
complement
of the union,
$$D_{\epsilon}(y)\cup D_{\epsilon}(a_1)\cup\cdots\cup D_{\epsilon}
(a_n),\eqno (4.40)$$
we have
$$\eqalign{D_kf(\tilde {z})&=mf(\tilde {z}),\cr
D_kg(\tilde {z})&=-mg(\tilde {z}).\cr
\cr}
\eqno (4.41)$$
If we now make use of (4.41) in (3.5) we find that
$$d\left\{{{\bar {f}_1g_2}\over {1-{{|z|^2}\over {R^2}}}}id\bar z
-{{\bar {f}_2g_1}\over {1-{{|z|^2}\over {R^2}}}}idz\right\}=0$$
in the complement of the set (4.40).  If we now integrate this
last equality in the complement of the set (4.40) and use Stokes'
theorem to rewrite the result as a circuit integral on the boundary,
then we find
$$\sum_{j=1}^n\int_{C_{\epsilon}(a_j)}\left\{{{\bar {f}_1g_2}\over {
1-{{|z|^2}\over {R^2}}}}id\bar z-{{\bar {f}_2g_1}\over {1-{{|z|^2}\over {
R^2}}}}idz\right\}=-\int_{C_{\epsilon}(y)}\left\{{{\bar {f}_1g_2}\over {
1-{{|z|^2}\over {R^2}}}}id\bar z-{{\bar {f}_2g_1}\over {1-{{|z|^2}\over {
R^2}}}}idz\right\}.\eqno (4.42)$$
Of course, this makes sense since $\bar {f}_ig_j$ is a well defined function
on ${\bf D}_R(a)\backslash \{y\}$.  We now wish to evaluate both sides of
(4.42) in the
limit $\epsilon\rightarrow 0.$ As in Theorem 3.0 it is useful at this point to
introduce
the local coordinate $u_j=T[-a_j]z$ in the $C_{\epsilon}(a_j)$ integral and the
local coordinate $u=T[-y]z$ in the $C_{\epsilon}(y)$ integral.  Then in
geodesic
polar coordinates $(r_j,\theta_j)$ and $(r,\theta )$ we have
$$\eqalign{u_j&=Re^{i\theta_j}\hbox{th}\,{{r_j}\over 2}\sim{{Rr_j}\over
2}e^{i\theta_j},\cr
u&=Re^{i\theta}\hbox{th}\,{r\over 2}\sim{{Rr}\over 2}e^{i\theta},\cr}
$$
and the asymptotics are given for small $r_j$ and small $r$.
Making use of this asymptotic parametrization in (4.42) and
replacing some
factors arising from the change of variables by their limits as
$\epsilon\rightarrow 0$, one finds
$$\eqalign{\matrix{\cr
\hbox{lim}\cr
\epsilon\rightarrow 0\cr}
\sum_{j=1}^n\epsilon R\int_0^{2\pi}\left\{\bar f_1g_2(u_j)e^{-i\theta_
j}+\bar f_2g_1(u_j)e^{i\theta_j}\right\}\,d\theta_j\cr
=-\matrix{\cr
\hbox{lim}\cr
\epsilon\rightarrow 0\cr}
\epsilon R\int_0^{2\pi}\left\{\bar f_1g_2(u)e^{-i\theta}+\bar f_2
g_1(u)e^{i\theta}\right\}\,d\theta .\cr}
\eqno (4.43)$$
We have taken the liberty of scaling the result by a factor
of 2.  Next we use (4.38), the asymptotics of the wave functions $
w_l$ and
$w_l^{*}$ given in (3.1) and (3.2), and the local asymptotics of the Green
function for $\tilde {z}$ near $\tilde {y}$ to evaluate both sides of (4.43) in
the limit
$\epsilon\rightarrow 0.$  We find
$$\sum_{j=1}^n{{8\sin(\pi\lambda_j)}\over {m^2R}}\left\{\overline {
b_{-{1\over 2}}^j}e^j_{{1\over 2},i}(\tilde y)+\overline {a_{-{1\over
2}}^j}f_{{1\over 2},i}^j(\tilde y)\right\}=-\overline {f_i(\tilde {
y})}\,,\eqno (4.44)$$
where we used
$${{\pi}\over {\Gamma (\lambda )\Gamma (1-\lambda )}}=\sin(\pi\lambda
).$$
We now use (4.44) to evaluate the low order expansion coefficients
$e_{{1\over 2},i}^j(\tilde {y})$  and $f_{{1\over 2},i}^j(\tilde {
y})$ in terms of the wave functions $W_j$ and $W^{*}_j.$
First choose
$$f(\tilde {z})=W_{\nu}(\tilde {z},m)\,,$$
then
$$\eqalign{a_{-{1\over 2}}^j&=\delta_{j\nu},\cr
b^j_{-{1\over 2}}&=0,\cr}
$$
and (4.44) specializes to
$$f_{{1\over 2},i}^{\nu}(\tilde {y})=-{{m^2R}\over {8s_{\nu}}}\overline {
W_{\nu}(\tilde {y},m)_i},\eqno (4.45)$$
where $W_{\nu}(\cdots )_i$ is the $i^{th}$ component of $W_{\nu}(
\cdots ).$
Next choose
$$f(\tilde {z})=W^{*}_{\nu}(\tilde {z},m)\,,$$
then
$$\eqalign{a_{-{1\over 2}}^j&=0,\cr
b^j_{-{1\over 2}}&=\delta_{j\nu},\cr}
$$
and (4.44) specializes to
$$e^{\nu}_{{1\over 2},i}(\tilde {y})=-{{m^2R}\over {8s_{\nu}}}\overline {
W^{*}_{\nu}(\tilde {y},m)_i}.\eqno (4.46)$$
Now we are ready for the principal result of this section.  We
write $G_{ij}^{a,\lambda}(\tilde {y},\tilde {z})$ for the $j^{th}$ component of
the vector $
G^{a,\lambda}_i(\tilde {y},\tilde {z}).$
We are interested in calculating the derivative of $G^{a,\lambda}_{
ij}(\tilde {y},\tilde {z})$
with respect to the $a$ variables.  As in \S 2 it is
convenient to introduce the scaled variables
$$b_{\nu}={{a_{\nu}}\over R}.$$
Taking the derivative of $G^{a,\lambda}_i(\tilde {y},\tilde {z})$ in the
parameter $
b_{\nu}$ kills the
singularity on the diagonal $y=z$.  Thus the function
$$\tilde {z}\rightarrow\partial_{b_{\nu}}G^{a,\lambda}_i(\tilde {
y},\tilde {z})$$
will be a solution to $(m+D_k)\psi =0$ with monodromy multiplier
$e^{-2\pi i\lambda_j}$ associated with the deck transformation $R_
j$.  This function
will be determined uniquely by its local expansions at $a_j$ for
$j=1,2,\ldots ,n.$  Using (2.32) and (2.34)  to differentiate the local
expansions (4.38) we find that
$$\partial_{b_{\nu}}G_i^{a,\lambda}(\tilde {y},\tilde {z})=-(1-|b_{
\nu}|^2)^{-1}e^{\nu}_{{1\over 2},i}(\tilde {y})W_{\nu}(\tilde {z}
,-m).\eqno (4.48)$$
Using (2.33) and (2.35) to differentiate the local expansions (4.38) we
find that
$$\bar{\partial}_{b_{\nu}}G_i^{a,\lambda}(\tilde {y},\tilde {z})=
-(1-|b_{\nu}|^2)^{-1}f^{\nu}_{{1\over 2},i}(\tilde {y})W^{*}_{\nu}
(\tilde {z},-m).\eqno (4.49)$$
Finally substituting (4.45) and (4.46) in (4.48) and (4.49) we find
$$\eqalign{\partial_{b_{\nu}}G_{ij}^{a,\lambda}(\tilde {y},\tilde {
z})&={{m^2R}\over {8s_{\nu}(1-|b_{\nu}|^2)}}\overline {W^{*}_{\nu}
(\tilde {y},m)_i}W_{\nu}(\tilde {z},-m)_j,\cr
\bar{\partial}_{b_{\nu}}G_{ij}^{a,\lambda}(\tilde {y},\tilde {z})&
={{m^2R}\over {8s_{\nu}(1-|b_{\nu}|^2)}}\overline {W_{\nu}(\tilde {
y},m)_i}W^{*}_{\nu}(\tilde {z},-m)_j.\cr}
\eqno (4.50)$$

\vfill\eject

\centerline{\S 5 {\bf Deformation Equations}}
\vskip.2in \par\noindent
{\bf A holonomic system}.
In this section we will find deformation equations for the
low order expansion coefficients of the response functions,
$$W_{\nu}=\delta_{j\nu}w_{-{1\over 2}+\lambda_j}+\sum_{n>0}\left\{
a_{n\nu}^jw_{n+\lambda_j}+b_{n\nu}^jw^{*}_{n-\lambda_j}\right\}\eqno
(5.1)$$
and
$$W_{\nu}^{*}=\delta_{j\nu}w^{*}_{-{1\over 2}-\lambda_j}+\sum_{n>
0}\left\{c_{n\nu}^jw_{n+\lambda_j}+d_{n\nu}^jw^{*}_{n-\lambda_j}\right
\}.\eqno (5.2)$$
It might be helpful for the reader to observe a parallel between the
developments here and one way of understanding the
{\it Schlesinger equations}
from the theory of monodromy preserving deformations of
linear equations in the complex plane.  In the Schlesinger theory the
fundamental object is a multivalued analytic function $z\rightarrow
Y(z,a)$ which
takes values in the general linear group.  The matrix valued function
$Y(z,a)$ has fixed (i.e., independent of the $a_j$) monodromy matrices $
M_j$
in simple circuits of the
branch points $a_j$ and satisfies an ordinary differential equation,
$${{dY}\over {dz}}=\sum_{j=1}^n{{A_j}\over {z-a_j}}Y,$$
in the $z$ variable.  This equation is the analogue of (5.3) below.
The fixed monodromy condition allows one to extend this differential
equation to a differential equation,
$$d_aY=-\sum_{j=1}^n{{A_jda_j}\over {(z-a_j)}}Y,$$
in the {\it a} variables.  This is the analogue of (5.20) below.  The
compatibility conditions for the $z$ and $a$ equations gives the
Schlesinger equations,
$$d_aA_{\nu}=-\sum_{\mu\ne\nu}{{[A_{\nu},A_{\mu}]}\over {a_{\nu}-
a_{\mu}}}d(a_{\nu}-a_{\mu}),$$
which is the analogue of (5.26) below.  We derive the deformation
equations (5.26) in a manner precisely analogous to this.  One
difference is that the internal compatibility conditions for the $
z$ and
$a$ equations do not give anything new in the Schlesinger case.  In
our case there is some interesting information that needs to be
extracted from the internal consistency of (5.3).

Our first tool will be the following linear relations, that arise when
the infinitesimal symmetries $M_j$ are applied to $W_{\nu}$ and $
W_{\nu}^{*}:$

$$\eqalign{M_3W_{\nu}-b_{\nu}M_1W_{\nu}&=-\half W_{\nu}+\sum_{\mu}\left
\{e_{\nu\mu}W_{\mu}+f_{\nu\mu}\bar b_{\mu}W_{\mu}^{*}\right\}\,,\cr
M_3W^{*}_{\nu}+\bar {b}_{\nu}M_2W_{\nu}^{*}&=\half W_{\nu}^{*}+\sum_{
\mu}\left\{g_{\nu\mu}b_{\mu}W_{\mu}+h_{\nu\mu}W_{\mu}^{*}\right\}\,
,\cr
b_{\nu}^2M_1W_{\nu}+M_2W_{\nu}&=\sum_{\mu}\left\{\alpha_{\nu\mu}W_{
\mu}+\beta_{\nu\mu}W_{\mu}^{*}\right\}\,,\cr
M_1W_{\nu}^{*}+\bar {b}_{\nu}^2M_2W_{\nu}^{*}&=\sum_{\mu}\left\{\gamma_{
\nu\mu}W_{\mu}+\delta_{\nu\mu}W_{\mu}^{*}\right\}\,.\cr}
\eqno (5.3)$$
In these relations the matrices $e$, $f,$ $g,$ $h$, $\alpha$, $\beta
,$ $\gamma ,$ and $\delta$ are matrices
that depend on the branch points $a_j$ for $j=1,\ldots ,n$ but not on $
\tilde {z}$.  In
the case of $e,f,g,$ and $h$ we have also partly anticipated the form
that the relations will take.  This will simplify some later results
and will avoid the necessity of introducing further definitions.  The
terms $-{1\over 2}W_{\nu}$ and ${1\over 2}W_{\nu}^{*}$ are stuck out in front
of the first two
equations instead of being combined with $e_{\nu\mu}$ and $h_{\nu
\mu}$ to avoid
introducing the Kronecker $\delta_{\nu\mu}$.  The $\delta_{\nu\mu}$ that
appears in the last
equation is not the Kronecker delta---the possible confusion with this
will shortly be addressed when $\alpha ,\beta ,\gamma ,$ and $\delta$ are all
eliminated in
favor of $e,f,g,$ and $h.$

These relations (5.3) will be deduced using a combination of (2.20) and (2.32)
---the results that describe the action of the infinitesimal symmetries
$M_j$ on local expansions and the results for centering the operators $
M_j$
at different points in the disk.
As the reader can easily check using (2.20) and (2.32) the linear
combinations on the left
hand side of (5.3) are all chosen so that the leading singularities
in the local expansions cancel at level $n=-{3\over 2}$.  Since $
W_{\mu}$ and $W_{\mu}^{*}$ for
$\mu =1,\ldots ,n$ are a basis for the wave functions whose local expansions
contain terms no lower than $n=-{1\over 2}$ and which are in $L^2$ near $
|z|=R$
relations of the form (5.3) follow immediately.  To find the
coefficient matrices $e,$ $f,$ and etc.  one need only compare the local
expansions on both sides of (5.3) at level $n=-{1\over 2}.$  One finds,
$$\eqalign{e&=(k+\lambda )+[{\bf a}_1,B]\,,\cr
f&=(B{\bf b}_1\bar {B}-{\bf b}_1)\,,\cr
g&=({\bf c}_1-\bar {B}{\bf c}_1B)\,,\cr
h&=(k+\lambda )-[{\bf d}_1,\bar {B}],\cr}
\eqno (5.4)$$
where we used the notation
$$\eqalign{({\bf a}_n)_{\nu\mu}&=a_{n-{1\over 2},\nu}^{\mu}(1-|b_{
\mu}|^2)^{-1}\,,\cr
({\bf b}_n)_{\nu\mu}&=b_{n-{1\over 2},\nu}^{\mu}(1-|b_{\mu}|^2)^{
-1}\,,\cr
({\bf c}_n)_{\nu\mu}&=c_{n-{1\over 2},\nu}(1-|b_{\mu}|^2)^{-1}\,,\cr
({\bf d}_n)_{\nu\mu}&=d_{n-{1\over 2},\nu}^{\mu}(1-|b_{\mu}|^2)^{
-1}.\cr}
\eqno (5.5)$$
Also we have written $B$ for the diagonal matrix with $\nu\nu^{th}$ entry
given by $b_{\nu}$ and $(k+\lambda$) for the diagonal matrix with entry
$(k+\lambda_{\nu})$ on the diagonal.  The analogue of the display (5.4) for the
coefficients $\alpha ,$ $\beta ,$ etc. is
$$\eqalign{\alpha&=-2B(k+\lambda -\half)-[{\bf a}_1,B^2]\,,\cr
\beta&=({\bf b}_1-B^2{\bf b}_1\bar {B}^2)\,,\cr
\gamma&=({\bf c}_1-\bar {B}^2{\bf c}_1B^2)\,,\cr
\delta&=2\bar {B}(k+\lambda +\half)-[{\bf d}_1,\bar {B}^2].\cr}
\eqno (5.6)$$
 From (5.4) and (5.6) one may deduce some simple relations among
the matrix coefficients,
$$\eqalign{\alpha&=B-eB-Be\,,\cr
\beta&=-f-Bf\bar {B}\,,\cr
\gamma&=g+\bar {B}gB\,,\cr
\delta&=\bar {B}+h\bar {B}+\bar {B}h\,.\cr}
\eqno (5.7)$$
Next we will deduce some less obvious relations from the
compatibility requirements of (5.3).  Consider the Lie algebra
relation,
$$(M_3M_2-M_2M_3-M_2)W=0,\eqno (5.8)$$
where we've written $W$ for the column vector,
$$W=\left[\matrix{W_1\cr
\vdots\cr
W_n\cr}
\right].$$
We now treat (5.3) as equations for $M_2W$ and $M_3W$ in terms of
$M_1W$ and $W$ and $W^{*}$ and as equations for $M_1W^{*}$ and $M_
3W^{*}$ in terms
of $M_2W^{*}$ and $W$ and $W^{*}.$  If we now use (5.3) to
eliminate the appearance of $M_3W,$ $M_2W,$ $M_3W^{*},$ and $M_1W^{
*}$ in favor
of $M_1W,$ $M_2W^{*}$, $W$ and $W^{*}$ one finds that the coefficient of $
M_1W$ in
the resulting equation is,
$$[e,B^2]+[\alpha ,B],$$
which vanishes as a consequence of (5.7).  The coefficient of $M_
2W^{*}$
in (5.8) is,
$$(B^2f\bar {B}^2-f+B\beta\bar {B}^{}-\beta )\bar {B},$$
which also vanishes as a consequence of (5.7).  Since $W$ and $W^{
*}$ are
linearly independent we can equate the remaining coefficients of
$W$ and $W^{*}$ to 0 in (5.8) and one finds,
$$\eqalign{\alpha +[e,\alpha ]+2Be+B\beta\gamma +B^2f\bar {B}\gamma
-\beta gB-B&=0\,,\cr
e\beta +2Bf\bar {B}+B\beta\delta +B^2f\bar {B}\delta -\alpha f\bar {
B}-\beta h&=0\,.\cr}
\eqno (5.9)$$
In a similar fashion if one starts with the Lie algebra relation,
$$(M_3M_1-M_1M_3+M_1)W^{*}=0,\eqno (5.10)$$
and uses (5.3) to eliminate $M_1W^{*},$ $M_3W^{*},$ $M_2W,$
and $M_3W$ in favor of $M_2W^{*}$, $M_1W,$ $W$ and $W^{*}$ then one finds
the coefficient of $M_1W$ is,
$$(\bar {B}^2gB^2-g+\gamma -\bar {B}\gamma B)B,$$
which vanishes as a consequence of (5.7) and the coefficient of
$M_2W^{*}$ is,
$$[h,\bar {B}^2]+[\bar {B},\delta ],$$
which also vanishes as a consequence of (5.7).  Equating the
coefficients of $W$ and $W^{*}$ to 0 in what remains of (5.10)  one
finds,
$$\eqalign{h\gamma +\delta gB+\gamma e+\bar {B}\gamma\alpha -2\bar {
B}gB-\bar {B}^2gB\alpha&=0\,,\cr
\delta -[h,\delta ]+\gamma f\bar {B}+\bar {B}\gamma\beta -2\bar {
B}h-\bar {B}^2gB\beta -\bar {B}&=0\,.\cr}
\eqno (5.11)$$
Both (5.9) and (5.11) simplify if one uses (5.7) to eliminate $\alpha
,\beta ,\gamma ,$ and
$\delta$ from these equations.
Substituting (5.7) in (5.9) one finds
$$\eqalign{[B,e^2-fg]=0\cr}
\eqno (5.12)$$
and
$$X-BX\bar {B}=0$$
where
$$X=ef-fh.$$
However, $X-BX\bar {B}=0$, implies $X=0$ so the second consequence of
(5.9) can be written
$$ef-fh=0.\eqno (5.13)$$
Observe also that (5.12) implies that the matrix $e^2-fg$ is diagonal.
In a precisely analogous way one can substitute (5.7) in (5.11) to
get
$$[h^2-gf,\bar {B}]=0\eqno (5.14)$$
and
$$X-\bar {B}XB=0$$
where
$$X=ge-hg,$$
and as before this implies
$$ge-hg=0,\eqno (5.15)$$
and (5.14) implies that the matrix $h^2-gf$ is diagonal.  The equations
we've found in this fashion do not determine the diagonal parts
for $e^2-fg$ and $h^2-gf.$  However, we can determine these by looking
at the coefficients of $w_{{1\over 2}+\lambda_j}$ in the local expansions of
the
first and third equations in (5.3).  One finds for the first
equation,
$$\eqalign{{{b_j-b_{\nu}}\over {1-|b_j|^2}}a_{{3\over 2}\nu}^j+{{
b_{\nu}\bar {b}_j^2-\bar {b}_j}\over {1-|b_j|^2}}m_2(-\half+\lambda_
j)\delta_{j\nu}+{{1+|b_j|^2-2b_{\nu}\bar {b}_j}\over {1-|b_j|^2}}
(k+\lambda_j+\half)a_{{1\over 2}\nu}^j\cr
=\sum_{\mu =1}^n(e_{\nu\mu}-\half\delta_{\nu\mu})a_{{1\over 2}\mu}^
j+\sum_{\mu =1}^nf_{\nu\mu}\bar {b}_{\mu}c_{{1\over 2}\mu}^j.\cr}
$$
If we set $\nu =j$ this simplifies to,
$$-\bar {b}_jm_2(-\half+\lambda_j)+(k+\lambda_j+1)a_{{1\over 2}j^{}}^
j=\sum_{\mu =1}^ne_{j\mu}a_{{1\over 2}\mu}^j+\sum_{\mu =1}^nf_{j\mu}
\bar {b}_{\mu}c_{{1\over 2}\mu}^j,\eqno (5.16)$$
where we have taken the liberty of collecting some of the terms involving
$a_{{1\over 2}j}^j$ on the left hand side.  If we look at the coefficient of $
w_{{1\over 2}+\lambda_j}$
in the third equation of (5.3) then set $\nu =j$ and collect some of the terms
involving $a_{{1\over 2}j}^j$ on one side then we find,
$$\eqalign{(1+|b_j|^2)m_2(\lambda_j-\half)-2b_j(k+\lambda_j+1)a_{{
1\over 2}j}^j\cr
=-\sum_{\mu =1}^n(b_{\mu}+b_j)e_{j\mu}a_{{1\over 2}\mu}^j-\sum_{\mu
=1}^n(1+b_j\bar {b}_{\mu})f_{j\mu}c_{{1\over 2}\mu}^j.\cr}
\eqno (5.17)$$
If one now multiplies (5.16) by $2b_j$ and adds the result to (5.17) then
one finds (after a little simplification),
$$(e^2-fg)_{jj}=k^2+{{m^2R^2}\over 4}.$$
Since we already know that $e^2-fg$ is diagonal it follows that
$$e^2-fg=\left(k^2+{{m^2R^2}\over 4}\right)I.\eqno (5.18)$$
In a precisely analogous fashion one can use the second and fourth
equations in (5.3) to show that
$$h^2-gf=\left(k^2+{{m^2R^2}\over 4}\right)I.\eqno (5.19)$$
\vskip.2in \par\noindent
{\bf The Holonomic Extension.}
Next we consider extending the differential equations (5.3) to the
$a$ variables.  To do this it is convenient to use (2.32)--(2.35) to
compute the local expansions for the exterior derivatives
$d_bW_{\nu}$ and $d_bW_{\nu}^{*}$.  One finds for the local expansion of $
d_bW_{\nu}$
about the point $a_j$,
$$\eqalign{\left(1-|b_j|^2\right)d_bW_{\nu}=&w_{-{3\over 2}+\lambda_
j}\left[-\delta_{j\nu}db_j\right]\cr
&+w_{-{1\over 2}+\lambda_j}\left[\delta_{j\nu}k_{-}(\lambda_j,b_j
)-a_{{1\over 2}\nu}^jdb_j\right]\cr
&+w_{-{1\over 2}-\lambda_j}^{*}\left[-b_{{1\over 2}\nu}^jd\bar b_
j\right]\cr
&+w_{{1\over 2}+\lambda_j}\left[da_{{1\over 2}\nu}^j+a_{{1\over 2}
\nu}^jk_{+}(\lambda_j,b_j)-a_{{3\over 2}\nu}^jdb_j-\delta_{j\nu}m_
2(\lambda_j-\half)d\bar b_j\right]\cr
&+w^{*}_{{1\over 2}-\lambda_j}\left[db_{{1\over 2}\nu}^j+b_{{1\over
2}\nu}^jk_{-}(\lambda_j,b_j)-b_{{3\over 2}\nu}^jd\bar b_j\right]+
\cdots\,\,\,\,\,\,.\cr}
$$
where we've introduced the abreviation
$$k_{\pm}(\lambda_j,b_j)=(k+\lambda_j\pm\half)(b_jd\bar {b}_j-\bar {
b}_jdb_j).$$
For the local expansion of $d_bW_{\nu}^{*}$ about the point $a_j$ one finds,
$$\eqalign{\left(1-|b_j|^2\right)d_bW^{*}_{\nu}=&w_{-{3\over 2}-\lambda_
j}^{*}\left[-\delta_{j\nu}d\bar b_j\right]\cr
&+w_{-{1\over 2}-\lambda_j}^{*}\left[\delta_{j\nu}k_{+}(\lambda_j
,b_j)-d_{{1\over 2}\nu}^jd\bar b_j\right]\cr
&+w_{-{1\over 2}+\lambda_j}\left[-c_{{1\over 2}\nu}^jdb_j\right]\cr
&+w^{*}_{{1\over 2}-\lambda_j}\left[d(d^j_{{1\over 2}\nu})+d_{{1\over
2}\nu}^jk_{-}(\lambda_j,b_j)-d_{{3\over 2}\nu}^jd\bar b_j-\delta_{
j\nu}m_1(-\lambda_j-\half)\right]\cr
&+w_{{1\over 2}+\lambda_j}\left[dc_{{1\over 2}\nu}^j+c_{{1\over 2}
\nu}^jk_{-}(\lambda_j,b_j)-c_{{3\over 2}\nu}^jdb_j\right]+\cdots\cr}
$$
Using these results one can easily check that the terms
at level $n=-{1\over 2}$ cancel in the local expansions of
$$d_bW_{\nu}+db_{\nu}M_1W_{\nu}$$
and
$$d_bW_{\nu}^{*}+d\bar {b}_{\nu}M_2W^{*}_{\nu}.$$
Thus one has relations of the form,
$$\eqalign{d_bW_{\nu}+db_{\nu}M_1W_{\nu}&=\sum_{\mu}\left(E_{\nu\mu}
W_{\mu}+F_{\nu\mu}W^{*}_{\mu}\right)\,,\cr
d_bW^{*}_{\nu}+d\bar {b}_{\nu}M_2W_{\nu}^{*}&=\sum_{\mu}\left(G_{
\nu\mu}W_{\mu}+H_{\nu\mu}W_{\mu}^{*}\right).\cr}
\eqno (5.20)$$
By comparing terms in the local expansions of both sides of (5.20)
at level $n={1\over 2}$ one finds that,
$$\eqalign{E=&{{(k+\lambda -\half)(Bd\bar {B}+\bar {B}dB)}\over {
1-|B|^2}}+[dB,{\bf a}_1]\,,\cr
F=&-dB{\bf b}_1\bar {B}^2-{\bf b}_1d\bar {B}\,,\cr
G=&-d\bar {B}{\bf c}_1B^2-{\bf c}_1dB\,,\cr
H=&-{{(k+\lambda +\half)(Bd\bar {B}+\bar {B}dB)}\over {1-|B|^2}}+
[d\bar {B},{\bf d}_1].\cr}
\eqno (5.21)$$
{\bf The deformation equations}.
The combination of the Dirac equation, the first two equations of
(5.3) and (5.21) determine an extended holonomic system for the
response functions $W$ and $W^{*}$.  We will deduce deformation equations
for the coefficients $e,f,g,$ and $h$ by examining the compatibility
conditions between (5.20) and the first two equations in (5.3).
The two compatibility conditions we will examine are
$$\eqalign{\left(M_3d_b-d_bM_3\right)W=0,\cr
\left(M_3d_b-d_bM_3\right)W^{*}=0.\cr}
$$
We use the  equations in (5.3) and (5.20) to express all
the terms in the preceeding equations in terms of $M_jW$, $M_jW^{
*}$,
$W$ and $W^{*}$ for $j=1,2$ and $\mu =1,2,\ldots ,n$.  In a single term in
each of the equations
this also requires the use the commutation relations for the
Lie algebra of $M_1,$ $M_2$, and $M_3$ and one should also make the
observation that the most singular terms containing $M^2_1W$ and
$M_2^2W^{*}$ manifestly cancel.  The condition that the lowest
coefficients in the local expansions of the resulting expressions must
vanish leads to the following identities:
$$\eqalign{&[E,B]-[dB,e]=0\,,\cr
&dBf\bar {B}^2+fd\bar {B}+BF\bar {B}-F=0\,,\cr
&[\bar {B},H]-[d\bar {B},h]=0\,,\cr
&d\bar {B}gB^2+gdB-\bar {B}GB+G=0\,.\cr}
\eqno (5.22)$$
All these identities are simple consequences of (5.21) and (5.4) but it
will be useful to record them here for future use.  Once one has
identified these relations all the remaining terms involving $M_j
W$
and $M_jW^{*}$ can be written as linear combinations of $W$ and $
W^{*}$
using (5.3).  Since $W$ and $W^{*}$ are linearly independent we can
equate the coefficients of the resulting equations to 0.  We find
the first form of the deformation equations,
$$\eqalign{de&=-(dBf\bar {B}+BF)\gamma +[E,e]+FgB-f\bar {B}G\,,\cr
d(f\bar {B})&=-(dBf\bar {B}+BF)\delta +(Ef\bar {B}-eF)+(Fh-f\bar {
B}H)+F\,,\cr
d(gB)&=(\bar {B}G-d\bar {B}gB)\alpha +(Ge-gBE)+(HgB-hG)-G\,,\cr
dh&=(\bar {B}G-d\bar {B}gB)\beta +[H,h]+(Gf\bar {B}-gBF)\,.\cr}
\eqno (5.23)$$
We will first eliminate the appearance of $\alpha ,\beta ,\gamma
,$ and $\delta$ from these
equations using (5.7) and then we use the identities (5.22) to rework
the equations as explicit equations for $df$ and $dg$.  We will illustrate
the calculations involved for $e$ and $f$ and leave the analogous
reductions for $g$ and $h$ to the reader.

First substitute $\gamma =g+\bar {B}gB$ into (5.23) to get
$$de=-(dBf\bar {B}+BF)g-(dBf\bar {B}^2+BF\bar {B})gB+[E,e]+FgB-f\bar {
B}G.$$
Now use (5.22) to substitute $F-fd\bar {B}$ for $dBf\bar {B}^2+BF
\bar {B}$ on the right
hand side of the last equation.  After some cancellation one finds
$$de=f(d\bar {B}gB-\bar {B}G)-(dBf\bar {B}+BF)g+[E,e].$$
It will prove useful to introduce,
$$\hat {F}=-(dBf\bar {B}+BF)=dB{\bf b}_1\bar {B}+B{\bf b}_1d\bar {
B}\eqno (5.24)$$
and
$$\hat {G}=d\bar {B}gB-\bar {B}G=d\bar {B}{\bf c}_1B+\bar {B}{\bf c}_
1dB.\eqno (5.25)$$
The equation for $de$ becomes
$$de=f\hat {G}+\hat {F}g+[E,e].$$
Now we work on the equation for $df$.  Again we start by
replacing $\delta$ on the right hand side by $\bar {B}+h\bar {B}+
\bar {B}h$.  One finds
that,
$$(df)\bar {B}+fd\bar {B}=\hat {F}(\bar {B}+h\bar {B}+\bar {B}h)+
Ef\bar {B}-eF+Fh-f\bar {B}H+F.$$
Next we use (5.22) to replace $\hat {F}\bar {B}$ by $fd\bar {B}-F$ in this last
equation
and make some cancellations to find,
$$(df)\bar {B}=\hat {F}h\bar {B}+fd\bar {B}h+Ef\bar {B}-eF-f\bar {
B}H+fhd\bar {B}.$$
Using (5.22) we can replace the term $f(d\bar {B}h-\bar {B}H)$ that appears on
the
right of this equation by
$$f(hd\bar {B}-H\bar {B})=efd\bar {B}-fH\bar {B},$$
where we used $fh=ef.$  One finds,
$$(df)\bar {B}=\hat {F}h\bar {B}+Ef\bar {B}-fH\bar {B}+e(fd\bar {
B}-F).$$
But using (5.22) we can replace $fd\bar {B}-F$ in this last result by
$\hat {F}\bar {B}$ and then cancel the common factor of $\bar {B}$ on the right
to obtain
$$df=\hat {F}h+e\hat {F}+Ef-fH.$$
Doing precisely analogous calculations for $g$ and $h$ one finds,
$$\eqalign{de=&f\hat {G}+\hat {F}g+[E,e]\,,\cr
df=&e\hat {F}+\hat {F}h+Ef-fH\,,\cr
dg=&h\hat {G}+\hat {G}e+Hg-gE\,,\cr
dh=&g\hat {F}+\hat {G}f+[H,h].\cr}
\eqno (5.26)$$
\vskip.2in \par\noindent\
{\bf Further identities.}  There are a number of identities among the
low order expansion coefficients of the response functions $W$ and
$W^{*}$ that are associated with the transformations $m\leftarrow
-m$ and
$(k,\lambda )\leftarrow (-k,-\lambda )$ which we will now record.  First
observe that,
$$\eqalign{\left[\matrix{-1&0\cr
0&1\cr}
\right]w_{l,k}(\cdot ,m)&=w_{l,k}(\cdot ,-m)\,,\cr
\left[\matrix{1&0\cr
0&-1\cr}
\right]w_{l,k}^{*}(\cdot ,m)&=w_{l,k}^{*}(\cdot ,-m).\cr}
\eqno (5.27)$$
Because we are interested in the dependence on the parameter $m$
we have made it explicit in the preceeding equation when we did
not always do so before.  From these relations it follows by
comparing local expansions at level $n=-{1\over 2}$ that,
$$\eqalign{\left[\matrix{-1&0\cr
0&1\cr}
\right]W_{\nu}(k,\lambda ,m)&=W_{\nu}(k,\lambda ,-m)\,,\cr
\left[\matrix{1&0\cr
0&-1\cr}
\right]W_{\nu}^{*}(k,\lambda ,m)&=W_{\nu}^{*}(k,\lambda ,-m).\cr}
\eqno (5.28)$$
Now comparing local expansions in (5.28) at level $n={1\over 2}$ one
finds,
$$\eqalign{a_{{1\over 2}\nu}^{\mu}(k,\lambda ,m)&=a_{{1\over 2}\nu}^{
\mu}(k,\lambda ,-m)\,,\cr
b_{{1\over 2}\nu}^{\mu}(k,\lambda ,m)&=-b_{{1\over 2}\nu}^{\mu}(k
,\lambda ,-m)\,,\cr
c_{{1\over 2}\nu}^{\mu}(k,\lambda ,m)&=-c_{{1\over 2}\nu}^{\mu}(k
,\lambda ,-m)\,,\cr
d_{{1\over 2}\nu}^{\mu}(k,\lambda ,m)&=d_{{1\over 2}\nu}^{\mu}(k,
\lambda ,-m)\,.\cr}
\eqno (5.29)$$
Since the dependence of the low order coefficients on the
parameter $m$ is so simple we will surpress the explicit
notational dependence on $m$ with the convention that
$$x^{\mu}_{j\nu}(k,\lambda )=x^{\mu}_{j\nu}(k,\lambda ,m)$$
for $x=a,b,c$ and $d$.  If $b^{\mu}_{j\nu}(k,\lambda ,-m)$ (or $c^{
\mu}_{j\nu}(k,\lambda ,-m)$) occurs we will replace it
with $-b^{\mu}_{j\nu}(k,\lambda )$ (or $-c^{\mu}_{j\nu}(k,\lambda
)$) without further comment.

For $k=0$ the Dirac operator is a real operator.  One
consequence of this is that there is a natural conjugation that acts
on the $k=0$ solutions to the Dirac equation.  This conjugation
is,
$$w\rightarrow\left[\matrix{0&1\cr
1&0\cr}
\right]\bar {w}.\eqno (5.30)$$
It is not difficult to check that this conjugation acts on the
local wave functions (2.22) and (2.23) as follows,
$$\eqalign{\left[\matrix{0&1\cr
1&0\cr}
\right]\overline {w_{l,k}(\cdot ,a)}&=w_{l,-k}^{*}(\cdot ,a)\,,\cr
\left[\matrix{0&1\cr
1&0\cr}
\right]\overline {w_{l,k}^{*}(\cdot ,a)}&=w_{l,-k}(\cdot ,a).\cr}
\eqno (5.31)$$
Using (5.31) to compare local expansions at level $n=-{1\over 2}$ one
finds,
$$\eqalign{\left[\matrix{0&1\cr
1&0\cr}
\right]\overline {W_{\nu}(k,\lambda )}&=W_{\nu}^{*}(-k,-\lambda )\,
,\cr
\left[\matrix{0&1\cr
1&0\cr}
\right]\overline {W_{\nu}^{*}(k,\lambda )}&=W_{\nu}(-k,-\lambda )\,
.\cr}
\eqno (5.32)$$
Comparing local expansions at level $n={1\over 2}$ in (5.32) one finds,
$$\eqalign{d_{{1\over 2}\nu}^{\mu}(k,\lambda )&=\overline {a_{{1\over
2}\nu}^{\mu}(-k,-\lambda )}\,,\cr
c_{{1\over 2}\nu}^{\mu}(k,\lambda )&=\overline {b_{{1\over 2}\nu}^{
\mu}(-k,-\lambda )}.\cr}
\eqno (5.33)$$
Finally we will note some identities that follow from the derivative
formula for the Green function (4.50).  To make use of this formula
it is convenient to modify the Green function with (5.32) in mind.
We introduce
$$\hat {G}^{a,\lambda}={8\over {m^2R}}\left[\matrix{0&1\cr
1&0\cr}
\right]G^{a,\lambda}$$
and (4.50) translates into
$$\eqalign{\partial_{b_{\nu}}\hat {G}^{a,\lambda}=&{1\over {s_{\nu}
(1-|b_{\nu}|^2)}}W_{\nu}(-k,-\lambda ,m)\otimes W_{\nu}(k,\lambda
,-m)\,,\cr
\bar{\partial}_{b_{\nu}}\hat {G}^{a,\lambda}=&{1\over {s_{\nu}(1-
|b_{\nu}|^2)}}W^{*}_{\nu}(-k,-\lambda ,m)\otimes W_{\nu}^{*}(k,\lambda
,-m)\,.\cr}
\eqno (5.34)$$
If we now use this equation and the local expansion formulas for
$d_bW_{\nu}$ and $d_bW_{\nu}^{*}$ given above to compute the local expansion
coefficient of
$$w_{-{1\over 2}-\lambda_i}(\cdot ,a_i)\otimes w_{-{1\over 2}+\lambda_
j}(\cdot ,a_j)$$
in the equation,
$$\left(\partial_{b_{\nu}}\partial_{b_{\mu}}-\partial_{b_{\mu}}\partial_{
b_{\nu}}\right)\hat {G}^{a,\lambda}=0,$$
one finds after some simplification
$$s^{-1}a_{{1\over 2}}(-k,-\lambda )-a_{{1\over 2}}(k,\lambda )^{
\tau}s^{-1}=\hbox{diagonal},\eqno (5.35)$$
where $a_{{1\over 2}}$ is the matrix with $\nu\mu$ matrix element $
a_{{1\over 2}\nu}^{\mu}$ and $X^{\tau}$ denotes
the transpose of the matrix $X$ and $s$ denotes the diagonal matrix
with $\nu\nu$ element $s_{\nu}=\sin(\pi\lambda_{\nu}).$

In a similar fashion if one calculates
the local expansion coefficient of
$$w^{*}_{-{1\over 2}+\lambda_i}(\cdot ,a_i)\otimes w_{-{1\over 2}
+\lambda_j}(\cdot ,a_j)$$
in
the identity,
$$\left(\bar\partial_{b_{\nu}}\partial_{b_{\mu}}-\partial_{b_{\mu}}
\bar\partial_{b_{\nu}}\right)\hat {G}^{a,\lambda}=0,$$
then one finds
$$s^{-1}b_{{1\over 2}}(-k,-\lambda )-c_{{1\over 2}}(k,\lambda )^{
\tau}s^{-1}=0.\eqno (5.36)$$
Now the combination of (5.35) and (5.33) produces the following
relation between the deformation variables $h$ and $e$ above
(in which the parameters are understood to be $(k,\lambda ,m)$),
$$h=s(1-|B|^2)e^{*}s^{-1}(1-|B|^2)^{-1},\eqno (5.37)$$
where $e^{*}$ is the conjugate transpose of $e.$

The combination of (5.36) and (5.33) produces the
relation,
$$\left(s^{-1}b_{{1\over 2}}(k,\lambda )\right)^{*}=s^{-1}b_{{1\over
2}}(k,\lambda ),\eqno (5.38)$$
from which it follows that
$$\left(fs(1-|B|^2)\right)^{*}=fs(1-|B|^2).\eqno (5.39)$$
One also has
$$\left(s^{-1}c_{{1\over 2}}(k,\lambda )\right)^{*}=s^{-1}c_{{1\over
2}}(k,\lambda ),\eqno (5.40)$$
from which it follows that
$$\left(s^{-1}(1-|B|^2)^{-1}g\right)^{*}=s^{-1}(1-|B|^2)^{-1}g.\eqno
(5.41)$$
We summarize these developments in the following theorem:
\vskip.1in \par\noindent
{\bf Theorem} 5.0. {\sl The low order expansion coefficients $e$, $
f$, $g$, and $h$
defined by (5.4) above satisfy the non-linear deformation
equations, }
$$\eqalign{de=&f\hat {G}+\hat {F}g+[E,e]\,,\cr
df=&e\hat {F}+\hat {F}h+Ef-fH\,,\cr
dg=&h\hat {G}+\hat {G}e+Hg-gE\,,\cr
dh=&g\hat {F}+\hat {G}f+[H,h]\,.\cr}
$$
{\sl The coefficients $e,$ $f,$ $g,$ and $h$ also satisfy the relations, }
$$\eqalign{ef-fh&=0\,,\cr
ge-hg&=0\,,\cr
e^2-fg&=M^2\,,\cr
h^2-gf&=M^2\,,\cr}
$$
{\sl where $M^2=k^2+{{m^2R^2}\over 4}$.  If we further introduce, }
$$\eqalign{{\bf f}&=fs(1-|B|^2)\,,\cr
{\bf g}&=s^{-1}(1-|B|^2)^{-1}g,\cr}
$$
{\sl then we have the additional symmetry relations, }
$$\eqalign{{\bf f}^{*}&={\bf f}\,,\cr
{\bf g}^{*}&={\bf g}\,,\cr
h&=s(1-|B|^2)e^{*}s^{-1}(1-|B|^2)^{-1}.\cr}
$$
\vskip.2in \par\noindent\
We have one final observation to make concerning these deformation
equations.  If we combine the commutator equation $ef-fh=0$ with
the final expression for $h$ we find,
$$e{\bf f}={\bf f}e^{*}.$$
If for some reason $f$ turns out to be invertible (this will happen
in an interesting special case we consider later on) then we can use
the non-linear relations above to eliminate $g$ and $h$ from these
equations in favor of $e$ and $f$.  In particular one finds,
$$\eqalign{h&=f^{-1}ef\cr
g&=f^{-1}(e^2-M^2).\cr}
$$
\vskip.1in \par\noindent
{\bf Theorem} 5.1. {\sl\ If $f$ is an invertible matrix and we define, }
$$\eqalign{\hat {{\bf F}}&=\hat {F}s(1-|B|^2)\,,\cr
\hat {{\bf G}}&=s^{-1}(1-|B|^2)^{-1}\hat {G},\cr}
\eqno (5.42)$$
{\sl then the deformation equations for $e$ and ${\bf f}$ are }
$$\eqalign{de&={\bf f}\hat {{\bf G}}+\hat {{\bf F}}{\bf g}+[E,e]\,
,\cr
d{\bf f}&=e\hat {{\bf F}}+\hat {{\bf F}}e^{*}+E{\bf f}+{\bf f}E^{
*}.\cr}
\eqno (5.43)$$
{\sl The matrices $e$ and {\bf f} satisfy the conditions, }
$$\eqalign{e{\bf f}&={\bf f}e^{*},\cr
{\bf f}^{*}&={\bf f},\cr}
\eqno (5.44)$$
{\sl with {\bf g} determined by $e$ and ${\bf f}$, }
$${\bf g}={\bf f}^{-1}(e^2-M^2)=e^{*}{\bf f}^{-1}e-M^2{\bf f}^{-1}
,\eqno (5.45)$$
{\sl and $\hat {{\bf G}}$ determined by {\bf g}. }
\vskip.2in \par\noindent\
We leave it to the reader to confirm that (5.43)--(5.45) incorporate
all the information about $e$, $f$, $g$, and $h$ that we have found so far
in the event $f$ is invertible.

\vskip.2in \par\noindent
{\bf The two point deformation theory in a special case}.  In this
subsection we will integrate the deformation equations in a
special case of the two point problem.  Our principal result is
that it is possible to integrate the deformation equations in terms
of a Painlev\'e transcendent of type VI.

It is useful to start by determining what the deformation equations
have to say about the action of the rotational symmetry
$b_j\rightarrow e^{i\theta}b_j$.  This corresponds to the infinitesimal
symmetry,
$$\hbox{rot}={{\partial}\over {\partial\theta_1}}\oplus\cdots\oplus{{
\partial}\over {\partial\theta_n}}=i\left(b_1{{\partial}\over {\partial
b_1}}-\bar b_1{{\partial}\over {\partial\bar {b}_1}}\right)\oplus
\cdots\oplus i\left(b_n{{\partial}\over {\partial b_n}}-\bar b_n{{
\partial}\over {\partial\bar {b}_n}}\right).$$
One sees immediately that
$$\eqalign{dB(\hbox{rot})&=iB\,,\cr
d\bar {B}(\hbox{rot})&=-i\bar {B}.\cr}
$$
It is then a simple matter to calculate,
$$\eqalign{E(\hbox{rot})&=-ie+i(k+\lambda )\,,\cr
F(\hbox{rot})&=if\bar {B}\,,\cr
G(\hbox{rot})&=-igB\,,\cr
H(\hbox{rot})&=-ih+i(k+\lambda ).\cr}
$$
The deformation equations become
$$\eqalign{de(\hbox{rot})&=i[k+\lambda ,e]=i[\lambda ,e]\,,\cr
df(\hbox{rot})&=i[\lambda ,f].\cr}
$$
 From these equations it follows that,
$$\eqalign{e(e^{i\theta}b_1,\ldots ,e^{i\theta}b_n)&=e^{i\lambda\theta}
e(b_1,\ldots ,b_n)e^{-i\lambda\theta}\,,\cr
f(e^{i\theta}b_1,\ldots ,e^{i\theta}b_n)&=e^{i\lambda\theta}f(b_1
,\ldots ,b_n)e^{-i\lambda\theta}.\cr}
\eqno (5.46)$$
Next we want to consider what happens for the infinitesimal
version of the SU(1,1) one parameter group,
$$\left[\matrix{\hbox{ch}\,t&\hbox{sh}\,t\cr
\hbox{sh}\,t&\hbox{ch}\,t\cr}
\right].$$
The vector field associated with the action of this one parameter
group on the disk of radius $R$ is
$$v=v_1\oplus\cdots\oplus v_n$$
with
$$v_j={1\over R}\left(1-b_j^2\right){{\partial}\over {\partial b_
j}}+{1\over R}\left(1-\bar b_j^2\right){{\partial}\over {\partial
\bar {b}_j}}.$$
It follows that
$$\eqalign{dB(v)&={1\over R}\left(1-B^2\right)\,,\cr
d\bar {B}(v)&={1\over R}\left(1-\bar B^2\right)\,.\cr}
$$
One computes,
$$\eqalign{E(v)&={1\over R}\left((k+\lambda )(\bar B-B)-{1\over 2}
(\bar B+B)+eB+Be\right)\,,\cr
F(v)&={1\over R}\left(f+Bf\bar B\right)\,,\cr
G(v)&=-{1\over R}\left(g+\bar BgB\right)\,,\cr
H(v)&={1\over R}\left((k+\lambda )(\bar B-B)-{1\over 2}(\bar B+B)
-h\bar B-\bar Bh\right),\cr}
$$
and
$$\eqalign{\hat {F}(v)&=-{1\over R}\left(f\bar B+Bf\right)\,,\cr
\hat {G}(v)&={1\over R}\left(gB+\bar Bg\right)\,.\cr}
$$
A little computation using the relations among $e$, $f$, $g$, and $
h$ shows
that
$$\eqalign{de(v)&={1\over R}[\lambda -{1\over 2}(\bar {B}+B),e]\,\,
,\cr
df(v)&={1\over R}[\lambda -{1\over 2}(\bar {B}+B),f]\,.\cr}
\eqno (5.47)$$
Since the matrices
$$\lambda -{1\over 2}\left(\bar B+B\right)$$
are diagonal they commute for different values of the matrices
$B$ and it follows that the differential equations for $e$ and $f$ may
be explicitly integrated along the orbits of the one parameter
group action,
$$t\rightarrow\left[\matrix{\hbox{ch}\,t&\hbox{sh}\,t\cr
\hbox{sh}\,t&\hbox{ch}\,t\cr}
\right]\cdot p,$$
for $p\in {\bf D}_R.$  We wish to
parametrize the two points $b_1$ and $b_2$ in the following fashion,
$$\eqalign{b_1&=e^{i\theta_1}\hbox{th}\,{{r_1}\over 2}\,,\cr
b_2&=e^{i\theta_1}{{\hbox{ch}\,\left({{r_1}\over 2}\right)\delta
+\hbox{sh}\,\left({{r_1}\over 2}\right)}\over {\hbox{sh}\,\left({{
r_1}\over 2}\right)\delta +\hbox{sh}\,\left({{r_1}\over 2}\right)}}
,\cr}
$$
where
$$\delta =e^{i\theta}\hbox{th}\,\left({r\over 2}\right).$$
The parameters for $b_1$ and $b_2$ are $\theta$, $\theta_1$, $r$, and $
r_1$.  Note that
$(r,\theta )$ are essentially the geodesic polar coordinates for $
b_2$ relative
to $b_1.$  The reason for this parametrization of $b_1$ and $b_2$ is as
follows.  First rotate both $b_1$ and $b_2$ by $-\theta_1$, then take the
result of
this and flow along the appropriate $v-$orbit by $-{{r_1}\over 2}$.  The result
is
$$\eqalign{&b_1\rightarrow 0,\,\,\,\,b_2\rightarrow\delta .\cr}
$$
Finally rotate 0 and $\delta$ simultaneously by $-\theta .$  One finds that
$$\eqalign{&b_1\rightarrow 0,\,\,\,b_2\rightarrow\hbox{th}\,\left
({r\over 2}\right).\cr}
$$
It follows that the symmetry flows that we have determined above
suffice to determine $e(b_1,b_2)$ and $f(b_1,b_2)$ for all values of $
(b_1,b_2)$
in terms of
$e(0,t)$ and $f(0,t)$ for $t$ real.
Next we will use the deformation equations to write down
ordinary differential equations for $e\left(0,t\right)$ and $f\left
(0,t\right).$
Let
$$u=0\oplus{{\partial}\over {\partial t}}.$$
Now we substitute $(0,t)$ in the deformation equations and
evaluate them along the vector field $u$.  One finds,
$$\eqalign{B=\left[\matrix{0&0\cr
0&t\cr}
\right],\cr
\bar {B}=\left[\matrix{0&0\cr
0&t\cr}
\right],\cr}
\eqno (5.48)$$
and
$$\eqalign{dB(u)&=\left[\matrix{0&0\cr
0&1\cr}
\right],\cr
d\bar {B}(u)&=\left[\matrix{0&0\cr
0&1\cr}
\right].\cr}
\eqno (5.49)$$
To make use of the symmetry conditions to reduce the number of
variables it is desirable to work with $e$ and ${\bf f}$ rather than $
e$ and
$f$.  Indeed at this point we will confine our attention to the case
in which $\lambda_j>0$ for $j=1,2$.  In this case the response functions
$W_{\nu}$ are in $L^2$ and the matrix $-s^{-1}b_{{1\over 2}}(k,\lambda
)$ is the matrix of the inner
products of these reponse functions and is consequently positive
definite.  Using the fact that $|b_j|<1$ one easily sees that this implies
{\bf f} is positive definite (and hence invertible).   Hence the deformation
equations become
$$\eqalign{de&={\bf f}\hat {{\bf G}}+\hat {{\bf F}}{\bf g}+[E,e]\,
,\cr
d{\bf f}&=e\hat {{\bf F}}+\hat {{\bf F}}e^{*}+E{\bf f}+{\bf f}E^{
*},\cr}
\eqno (5.50)$$
with
$${\bf g}=e^{*}{\bf f}^{-1}e-M^2{\bf f}^{-1},\eqno (5.51)$$
and $\hat {{\bf G}}$ determined by {\bf g}.  At this point it is useful to
introduce
$$e=\left[\matrix{k+\lambda_1&e_{+}\cr
\bar {e}_{-}&k+\lambda_2\cr}
\right]\eqno (5.52)$$
and
$${\bf f}=c\left[\matrix{\kappa\hbox{ch}\,\psi&\epsilon\hbox{sh}\,
\psi\cr
\bar{\epsilon}\hbox{sh}\,\psi&\kappa^{-1}(1-t^2)\hbox{ch}\,\psi\cr}
\right]\eqno (5.53)$$
where $c>0$, $\kappa$, and $\psi$ are real and $\epsilon\bar{\epsilon }
=1.$  This parametrization
incorporates the known diagonal for $e$ and the fact that {\bf f} is a
symmmetric positive definite matrix which is related in a particular
fashion to the matrix of the $L^2$ inner product of wave functions.
The last algebraic relation $e{\bf f}={\bf f}e^{*}$ implies,
$$\eqalign{&(\bar{\epsilon }e_{-}-\epsilon\bar {e}_{-})\hbox{sh}\,
\psi =0,\cr
&(\bar{\epsilon }e_{+}-\epsilon\bar {e}_{+})\hbox{sh}\,\psi =0,\cr
&(\kappa^{-1}(1-t^2)e_{+}-\kappa e_{-})\hbox{ch}\,\psi
-\lambda\epsilon\hbox{sh}\,
\psi =0,\cr}
$$
where $\lambda =\lambda_2-\lambda_1.$  Now define
$$\xi =\bar{\epsilon}\kappa^{-1}(1-t^2)e_{+}+\bar{\epsilon}\kappa
e_{-}.$$
If we suppose that $\hbox{sh}\,\psi\ne 0$ then we find that $\bar{
\epsilon }e_{+}$ and $\bar{\epsilon }e_{-}$ are
real and
$$\eqalign{e_{+}&={{\epsilon\kappa}\over {2(1-t^2)}}(\xi +\lambda\hbox{th}\,
\psi ),\cr
e_{-}&={{\epsilon}\over {2\kappa}}(\xi -\lambda\hbox{th}\,\psi ),\cr}
\eqno (5.54)$$
where $\bar{\xi }=\xi$.  Now write
$$\eqalign{k_j&=k+\lambda_j\,,\cr
\mu&=k_1+k_2=2k+\lambda_1+\lambda_2\,,\cr
\lambda&=k_2-k_1=\lambda_2-\lambda_1\,,\cr}
$$
for $j=1,2$ and
$${\bf f}=\left[\matrix{{\bf f}_{11}&{\bf f}_{12}\cr
{\bf f}_{21}&{\bf f}_{22}\cr}
\right]$$
and
$${\bf g}=\left[\matrix{{\bf g}_{11}&{\bf g}_{12}\cr
{\bf g}_{21}&{\bf g}_{22}\cr}
\right].$$
Then one easily calculates,
$$\eqalign{E(u)&={{2t(k_2-\half)}\over {1-t^2}}\left[\matrix{0&0\cr
0&1\cr}
\right]-t^{-1}\left[\matrix{0&e_{+}\cr
\bar {e}_{-}&0\cr}
\right]\,,\cr
{\bf F}(u)&=\left[\matrix{0&{\bf f}_{12}\cr
0&{{1+t^2}\over {1-t^2}}{\bf f}_{22}\cr}
\right]\,,\cr
{\bf G}(u)&=\left[\matrix{0&-{\bf g}_{12}\cr
0&-{{1+t^2}\over {1-t^2}}{\bf g}_{22}\cr}
\right]\,,\cr}
$$
$$\eqalign{\hat {{\bf F}}(u)&=-{{2t}\over {1-t^2}}\left[\matrix{0&
0\cr
0&{\bf f}_{22}\cr}
\right],\cr
\hat {{\bf G}}(u)&={{2t}\over {1-t^2}}\left[\matrix{0&0\cr
0&{\bf g}_{22}\cr}
\right].\cr}
$$
If we now substitute these into the equation for $d{\bf f}(u)$ we find
$$d{\bf f}(u)=N$$
where
$$\eqalign{N_{11}&=-t^{-1}(e_{+}{\bf f}_{21}+\bar {e}_{+}{\bf f}_{
12})\,,\cr
N_{12}&=-{{t+t^{-1}}\over {1-t^2}}e_{+}{\bf f}_{22}-t^{-1}e_{-}{\bf f}_{
11}+{{2t(k_2-\half)}\over {1-t^2}}{\bf f}_{12}\,,\cr
N_{21}&=-{{t+t^{-1}}\over {1-t^2}}\bar {e}_{+}{\bf f}_{22}-t^{-1}
\bar {e}_{-}{\bf f}_{11}+{{2t(k_2-\half)}\over {1-t^2}}{\bf f}_{2
1}\,,\cr
N_{22}&=-t^{-1}(\bar {e}_{-}{\bf f}_{12}+e_{-}{\bf f}_{21})-{{2t}\over {
1-t^2}}{\bf f}_{22}.\cr}
$$
Next we convert this using (5.53) and (5.54) to find,
$$\eqalign{N_{11}&=-c\kappa{{t^{-1}}\over {1-t^2}}(\xi +\lambda\hbox{th}\,
\psi )\hbox{sh}\,\psi\,,\cr
N_{12}&=-c\epsilon\left({{t^{-1}}\over {1-t^2}}\xi\hbox{ch}\,\psi
-{{t(\mu -1)}\over {1-t^2}}\hbox{sh}\,\psi\right)\,,\cr
N_{21}&=-c\bar{\epsilon}\left({{t^{-1}}\over {1-t^2}}\xi\hbox{ch}\,
\psi -{{t(\mu -1)}\over {1-t^2}}\hbox{sh}\,\psi\right)\,,\cr
N_{22}&=-c\kappa^{-1}t^{-1}(\xi -\lambda\hbox{th}\,\psi )\hbox{sh}\,
\psi -c\kappa^{-1}2t\hbox{ch}\,\psi .\cr}
$$
Write $f'={{df}\over {dt}}$ and the equation $d{\bf f}(u)=N$ for $
c'$, $\epsilon'$, $\kappa'$ and
$\psi'$ becomes,
$$\eqalign{c^{-1}c'\hbox{ch}\,\psi +\kappa^{-1}\kappa'\hbox{ch}\,
\psi +\hbox{sh}\,\psi\,\,\psi'&=c^{-1}\kappa^{-1}N_{11}\,,\cr
c^{-1}c'\hbox{sh}\,\psi +\epsilon^{-1}\epsilon'\hbox{sh}\,\psi +\hbox{ch}\,
\psi\,\,\psi'&=c^{-1}\epsilon^{-1}N_{12}\,,\cr
c^{-1}c'\hbox{sh}\,\psi -\epsilon^{-1}\epsilon'\hbox{sh}\,\psi +\hbox{ch}\,
\psi\,\,\psi'&=c^{-1}\epsilon N_{21}\,,\cr
-2t\hbox{ch}\,\psi +(1-t^2)\left(c^{-1}c'\hbox{ch}\,\psi -\kappa^{
-1}\kappa'\hbox{ch}\,\psi +\hbox{sh}\,\psi\,\,\psi'\right)&=c^{-1}
\kappa N_{22}\,.\cr}
$$
The very last equation simplifies somewhat when combined with
the expression for $N_{22}$.  One finds,
$$c^{-1}c'\hbox{ch}\,\psi -\kappa^{-1}\kappa'\hbox{ch}\,\psi +\hbox{sh}\,
\psi\,\,\psi'=-{{t^{-1}}\over {1-t^2}}(\xi -\lambda\hbox{th}\,\psi
)\hbox{sh}\,\psi .$$
Solving these equations for $c^{-1}c'$, $\kappa^{-1}\kappa'$, $\epsilon^{
-1}\epsilon'$ and $\psi'$ one finds,
$$\eqalign{c^{-1}c'&=-{{(\mu -1)t}\over {1-t^2}}\hbox{sh}^2\psi\,
,\cr
\epsilon^{-1}\epsilon'&=0\,,\cr
\kappa^{-1}\kappa'&=-{{\lambda t^{-1}}\over {1-t^2}}\hbox{th}^2\psi\,
,\cr
\psi'&=-{{t^{-1}}\over {1-t^2}}\xi +{{(\mu -1)t}\over {1-t^2}}\hbox{ch}\,
\psi\,\hbox{sh}\,\psi .\cr}
$$
Next we work out the differential equation associated with
$e$.  Substituting the results for $\hat {F}$ and $\hat {G}$ into the
differential equation
for $e$ above, one finds,
$$\eqalign{e'&={\bf f}\left[\matrix{0&0\cr
0&{{2t}\over {1-t^2}}{\bf g}_{22}\cr}
\right]-\left[\matrix{0&0\cr
0&{{2t}\over {1-t^2}}{\bf f}_{22}\cr}
\right]{\bf g}+[E,e]\cr
&=\left[\matrix{0&{{2t}\over {1-t^2}}{\bf f}_{12}{\bf g}_{22}\cr
-{{2t}\over {1-t^2}}{\bf f}_{22}{\bf g}_{21}&0\cr}
\right]+[E,e].\cr}
\eqno (5.55)$$
We also calculate
$$[E,e]=\left({{\lambda +(\mu -1)t^2}\over {t(1-t^2)}}\right)\left
[\matrix{0&-e_{+}\cr
\bar {e}_{-}&0\cr}
\right].\eqno (5.56)$$
Combining (5.55) and (5.56) we find,
$$\eqalign{e_{+}'&={{2t}\over {1-t^2}}{\bf f}_{12}{\bf g}_{22}-{{
\lambda +(\mu -1)t^2}\over {t(1-t^2)}}e_{+}\,,\cr
\bar {e}_{-}'&=-{{2t}\over {1-t^2}}{\bf f}_{22}{\bf g}_{21}+{{\lambda
+(\mu -1)t^2}\over {t(1-t^2)}}\bar {e}_{-}\,.\cr}
\eqno (5.57)$$
Use
$${\bf g}={\bf f}^{-1}(e^2-M^2)$$
to obtain,
$$\eqalign{(1-t^2\hbox{ch}^2\psi )c{\bf g}_{21}&=-\bar{\epsilon}\left
(k_1^2-M^2+\bar e_{-}e_{+}\right)\hbox{sh}\,\psi +\kappa\mu e_{-}\hbox{ch}\,
\psi\cr
(1-t^2\hbox{ch}^2\psi )c{\bf g}_{22}&=-\bar{\epsilon}\mu e_{+}\hbox{sh}\,
\psi +\kappa (k_2^2-M^2+\bar {e}_{-}e_{+})\hbox{ch}\,\psi\,.\cr}
\eqno (5.58)$$
Now substitute this last expression for ${\bf g}_{21}$ in the second equation
in (5.57) along with the parametrization for ${\bf f}_{22}$ and the expression
for $\bar {e}_{-}$.  Then multiply both sides of the resulting equation by
$\kappa$, make use of the equation for $\kappa^{-1}\kappa'$ given above and the
observation that
$$\bar {e}_{-}e_{+}={{\xi^2-\lambda^2\hbox{th}\,^2\psi}\over {4(1
-t^2)}},$$
to obtain,

$$\eqalign{{1\over 2}{d\over {dt}}\left(\xi -\lambda\hbox{th}\,\psi\right
)&={{\lambda (1-\hbox{th}^2\psi )+(\mu -1)t^2}\over {2t(1-t^2)}}\left
(\xi -\lambda\hbox{th}\,\psi\right)\cr
&+{{t\hbox{ch}\,\psi\,\hbox{sh}\,\psi}\over {1-t^2\hbox{ch}^2\psi}}\left
(k_1^2+k_2^2-2M^2+{{\xi^2-\lambda^2\hbox{th}^2\psi}\over {2(1-t^2
)}}\right)\cr
&-{{t\mu\hbox{ch}^2\psi}\over {1-t^2\hbox{ch}^2\psi}}\xi .\cr}
\eqno (5.59)$$
As a check on this equation we remark that the variable $\psi$ in
the Poincar\'e disk has the same significance as the variable $\psi$
in [8,12,15].  Introducing $t={r\over R}$ and letting $R\rightarrow
\infty$ one finds that the
limiting form of (5.59) is
$$r{{\partial}\over {\partial r}}\left(r{{\partial\psi}\over {\partial
r}}\right)=\lambda^2(1-{\rm t}{\rm h}^2\psi ){\rm t}{\rm h}\,\psi
+{{m^2r^2}\over 2}{\rm s}{\rm h}\,2\psi$$
which is equation (3.3.46) in [12].  In the special case $\lambda
=0$
this equation was first found in [8].

Next consider the substitutions
$$\eqalign{s&=t^2,\,\,\,w={1\over {\hbox{ch}^2\psi}}\cr}
.$$
One finds,
$$\eqalign{{{d\psi}\over {ds}}&={1\over {2t}}{{d\psi}\over {dt}}=
-{1\over {2w\sqrt {1-w}}}{{dw}\over {ds}}\cr
\xi&={1\over {w\sqrt {1-w}}}\left(s(1-s){{dw}\over {ds}}+s(\mu -1
)(1-w)\right).\cr}
$$
Now convert (5.59) to a differential equation in $s$ by multiplying
both sides by $t^{-1}$ and substituting $s$ for $t^2$.  In the resulting
equation make the substitution $w=\hbox{1/ch}^2\psi$ and multiply both
sides by $w\sqrt {1-w}$.  One finds,
$$\eqalign{&\left(w\sqrt {1-w}{d\over {ds}}{1\over {w\sqrt {1-w}}}\right
)\left(s(1-s){{dw}\over {ds}}+((\mu -1)s-\lambda w)(1-w)\right)\cr
&+{d\over {ds}}\left(s(1-s){{dw}\over {ds}}+((\mu -1)s-\lambda w)
(1-w)\right)\cr
&={{\lambda w+(\mu -1)s}\over {2s(1-s)}}\left(s(1-s){{dw}\over {d
s}}+((\mu -1)s-\lambda w)(1-w)\right)\cr
&+{{w(1-w)}\over {w-s}}\left(k_1^2+k^2_2-2M^2+{{\xi^2-\lambda^2(1
-w)}\over {2(1-s)}}\right)\cr
&-{{\mu}\over {w-s}}\left(s(1-s){{dw}\over {ds}}+s(\mu -1)(1-w)\right
).\cr}
\eqno (5.60)$$
It is simple to compute,
$$w\sqrt {1-w}{d\over {ds}}{1\over {w\sqrt {1-w}}}={{2-3w}\over {
2w(w-1)}}{{dw}\over {ds}}.$$
Substituting this last expression in (5.60) and simplifying the resulting
equation one finds,
$$\eqalign{&w^{\prime\prime}-{1\over 2}\left({1\over w}+{1\over {
w-1}}+{1\over {w-s}}\right)(w')^2+\left({1\over s}+{1\over {s-1}}
+{1\over {w-s}}\right)w'\cr
&={{w(w-1)(w-s)}\over {s^2(1-s)^2}}\left({{(1-4M^2)s(s-1)}\over {
2(w-s)^2}}-{{(\mu -1)^2s}\over {2w^2}}+{{\lambda^2}\over 2}\right
).\cr}
\eqno (5.61)$$
This is Painlev\'e VI (see, e.g., [4])
$$\eqalign{&w^{\prime\prime}-{1\over 2}\left({1\over w}+{1\over {
w-1}}+{1\over {w-s}}\right)(w')^2+\left({1\over s}+{1\over {s-1}}
+{1\over {w-s}}\right)w'\cr
&={{w(w-1)(w-s)}\over {s^2(1-s)^2}}\left({{\delta s(s-1)}\over {(
w-s)^2}}+{{\gamma (s-1)}\over {(w-1)^2}}+{{\beta s}\over {w^2}}+\alpha\right
),\cr}
$$
with
$$\eqalign{\alpha&={{\lambda^2}\over 2}\,,\cr
\beta&=-{{(\mu -1)^2}\over 2}\,,\cr
\gamma&=0\,,\cr
\delta&={{1-4M^2}\over 2}.\cr}
$$
\vskip.1in \par\noindent
We summarize these developments in the following theorem:
\vskip.1in \par\noindent
{\bf Theorem} 5.2 {\sl Suppose that $\lambda_j>0$ for $j=1,2.$ Let $
b_j={{a_j}\over R}$ and
consider $e(b_1,b_2)$ and $f(b_1,b_2)$ as functions of the scaled variables $
b_j$.
Define $\lambda =\lambda_2-\lambda_1$ and $\mu =2k+\lambda_1+\lambda_
2$.  Write }
$$e(0,t)=\left[\matrix{k+\lambda_1&e_{+}(t)\cr
e_{-}(t)&k+\lambda_2\cr}
\right],$$
{\sl and }
$${\bf f}(0,t)=c(t)\left[\matrix{\kappa (t)\hbox{ch}\,\psi (t)&\epsilon
(t)\hbox{sh}\,\psi (t)\cr
\bar{\epsilon }(t)\hbox{sh}\,\psi (t)&\kappa (t)^{-1}(1-t^2)\hbox{ch}\,
\psi (t)\cr}
\right].$$
{\sl Then the algebraic relation $e{\bf f}={\bf f}e^{*}$ implies, }
$$\eqalign{e_{+}=&{{\epsilon\kappa}\over {2(1-t^2)}}(\xi +\lambda\hbox{th}\,
\psi )\,,\cr
e_{-}=&{{\epsilon}\over {2\kappa}}(\xi -\lambda\hbox{th}\,\psi ),\cr}
$$
{\sl for a real valued function $\xi$ and $\epsilon$ of absolute value 1.  The
deformation equation for
{\bf f} implies, }
$$\eqalign{c^{-1}c'&=-{{(\mu -1)t}\over {1-t^2}}\hbox{sh}^2\psi\,
,\cr
\epsilon^{-1}\epsilon'&=0\,,\cr
\kappa^{-1}\kappa'&=-{{\lambda t^{-1}}\over {1-t^2}}\hbox{th}^2\psi\,
,\cr
\psi'&=-{{t^{-1}}\over {1-t^2}}\xi +{{(\mu -1)t}\over {1-t^2}}\hbox{ch}\,
\psi\,\hbox{sh}\,\psi .\cr}
$$
{\sl The last equation allows us to eliminate $\xi$ in favor of $
\psi$ and
$\psi'$.  Once this is done and the further substitutions, }
$$\eqalign{s&=t^2\,,\cr
w&={1\over {\hbox{ch}^2\psi}}\,,\cr}
$$
{\sl are made, the deformation equation for $e$ becomes the type VI Painlev\'e
equation (5.61) above for $w$. }

\vskip.2in \par\noindent
\centerline{\S 6 {\bf The Tau Function}}
\vskip.2in \par\noindent\
{\bf The Grassmannian formalism}.
In this section we will introduce the $\tau -$function for the Dirac
operator with branched singularities whose Green function was
described in \S 4.  Suppose that $S$ is an open subset of ${\bf D}_
R$ with
a boundary $\partial S$ that consists of the union of smooth simple curves
in ${\bf D}_R$ (we assume that the closure of $S$ is contained in the open
set ${\bf D}_R)$.  Suppose that $f$ is a smooth function in the Sobolev space
$H^1({\bf D}_R)$ which satisfies the Dirac equation
$$(m-D_k)f(x)=0\hbox{  for  }x\in S.$$
Then since the integral operator associated with the Green function
inverts the Dirac operator we have
$$f_j(x)=\int_{{\bf D}_R\backslash S}(m-D_k)f\cdot\overline {G_j(
x,y)}\,d\mu (y)\hbox{  for  }x\in S.$$
The integral can be confined to ${\bf D}_R\backslash S$ since $(m
-D_k)f(y)=0$ for
$y\in S.$  Since the Green function $G_j(x,y)$ is a solution to
$$(m+D_k)_yG_j(x,y)=0$$
in the second variable when $x\ne y$, it follows that the expression
for $f_j(x)$ can be written
$$f_j(x)=\int_{{\bf D}_R\backslash S}\left\{(m-D_k)f(y)\cdot\overline {
G_j(x,y)}-f(y)\cdot\overline {(m+D_k)G_j(x,y)}\right\}\,d\mu (y)$$
which becomes, using (3.5) and Stokes' theorem, the following boundary
integral representation for $f_j(x),$
$$f_j(x)=2\int_{\partial S}\left\{{{\overline {G_{j1}(x,y)}f_2(y)}\over {
1-{{|y|^2}\over {R^2}}}}i\,d\bar y-{{\overline {G_{j2}(x,y)}f_1(y
)}\over {1-{{|y|^2}\over {R^2}}}}i\,dy\right\}\hbox{  for  }x\in
S,\eqno (6.1)$$
in which each component of the boundary, $\partial S,$ is given the standard
counterclockwise orientation.  If $f\in H^{{1\over 2}}(\partial S
)$ is prescribed arbitrarily
then we can also regard (7.1) as a formula for the projection of $
f$
onto the space of boundary values on $\partial S$ of $H^1(S)$ solutions to the
Dirac
equation in the interior of $S.$ The complementary subspace for this
projection is the space of boundary values on $\partial S$ of solutions to the
Dirac equation in $H^1({\bf D}_R\backslash S).$

Our principal tool in the discussion of the tau function
for the Dirac operator is a formula analogous to (6.1) with the
Green function $G$ replaced by $G^{a,\lambda}$.  To explain the significance of
this formula it will be useful to describe the transfer
formalism [12] and its relation to ``localization'' away from
singularities.  It is natural when considering the Dirac operator with
branch type singularities at the points $a_j$ for $j=1,2,\ldots ,
n$ to localize
the operator away from the branch cuts.  One natural way to do
this in the hyperbolic disk is to transform the disk into the upper
half plane by a fractional linear transformation and then to draw
horizontal strips around horizontal branch cuts emanating
from each of the points $a_j$ for $j=1,2,\ldots ,n$.  There are many ways
to transform the disk into the upper half plane and it is clear
that, provided the branch points $a_j$ are all distinct, one can choose
a suitable such transform so that
the {\it second coordinates} of the transformed branch points are distinct.
Supposing this to be done we can then choose the strips $S_j$
containing $a_j$ to be pairwise disjoint by making them
sufficiently narrow.  It is certainly possible to
transform the Dirac operator into upper half plane coordinates and work
exclusively in the upper half plane, so that the description of the
localization is geometrically simple.  However, because we wish
to use the
many formulas which we've written down in the disk but have
not written down in
the upper half plane this is not economical.  Instead
we adopt the expedient of using the geometrically natural terminology
in describing the localization via horizontal strips in the upper half
plane but when doing calculations we transform the strips $S_j$ back
into the disk where they become crescent shaped objects with the
sharp ends meeting at a common point on the boundary of the
disk.  The reader should not find it difficult to keep in mind that
any reference to horizontal branch cuts or strips implicitly assumes
that one has chosen an appropriate upper half plane coordinate
system.

Until now it has been convenient to work
on the simply connected covering $\tilde {{\bf D}}_R(a)$ of ${\bf D}_
R(a).$  However, at
this point it is simpler to draw horizontal branch cuts (rays), $
\ell_j,$
emanating to the right
from each of the branch points $a_j$ and to work with functions on
${\bf D}_R(a)\backslash \{\ell_1,\ell_2,\ldots ,\ell_n\}$ that have appropriate
branching behavior.  We
will now describe the localization of the singular Dirac operator, $
D^{a,\lambda},$
which we wish to consider.  Let $S=\cup S_j$ denote the union of the
strips $S_j.$  In the exterior of the set $S$ the operator, $D^{a
,\lambda},$
will act precisely as the ordinary Dirac operator acts.  However,
in order for this operator to adequately mirror the singular Dirac
operator whose domain contains functions that are branched at
the points $a_j$ with specified monodromy we restrict its domain.
\def\wint{W_{\rm int}}
\def\wext{W_{\rm ext}}
We will now define the subspaces of $H^{{1\over 2}}(\partial S)$ that are
important
in our description of localization of the singular Dirac operator
$D^{a,\lambda}.$  We will say that a function $F$ is locally in the null
space of $D^{a,\lambda}$ at $a_j$ provided that for some $\epsilon
>0$ the function
$F$ has a local expansion,
$$F(x)=\sum_{n\ge{1\over 2}}\left\{a_n(F)w_{n+\lambda_j}(x,a_j)+b_
n(F)w^{*}_{n-\lambda_j}(x,a_j)\right\},$$
valid for $x\in B_{\epsilon}(a_j)$, the ball of radius $\epsilon$ about $
a_j.$  In this description
it is understood that a branch cut $\ell_j$ has been chosen for $
a_j$ and
that branches for $F$ , $w_{n+\lambda_j}$, and $w_{n-\lambda_j}^{
*}$ are fixed.  Suppose now
that $S_j$ is a horizontal strip containing $a_j$ in its interior.  We define
a subspace of $H^{{1\over 2}}(\partial S_j)$ in the following fashion,
\vskip.1in \par\noindent
{\bf Definition}. {\sl A function $f$ on $\partial S_j$ will be in the subspace
$\wint
(a_j)$
provided that it is the boundary value of a function $F$ defined in
$S_j$ with the following properties: }
\vskip.1in \par\noindent
{\sl(1) $F$ is a branched solution to the Dirac equation in $S_j$ with
branch cut $\ell_j$ and monodromy $e^{2\pi i\lambda_j}$. }
\vskip.1in \par\noindent
{\sl(2) $F$ is locally in the null space of $D^{a,\lambda}$ at $a_
j$. }
\vskip.1in \par\noindent
{\sl(3)  $F\in H^1(S_j\backslash \{B_{\epsilon}(a_j)\cup\ell_j\})$ for some $
\epsilon >0$. }
\vskip.2in \par\noindent\
If each branch point $a_j$ is contained in a strip $S_j$ and the
strips $S_j$ are pairwise disjoint then we also write,
$$\wint(a)=\wint(a_1)\oplus\wint(a_2)\oplus\cdots\oplus\wint(a_n)
,$$
to define a subspace of $H^{{1\over 2}}(\partial S).$ Next we define a subspace
of $H^{{1\over 2}}(\partial S_j)$ complementary to $\wint(a_j)$.
\vskip.1in \par\noindent
{\bf Definition}. {\sl We define the subspace $\wext(a_j)$ to consist of
functions $f$ defined on $\partial S_j$ which are the boundary values of
functions $F$ defined in ${\bf D}_R\backslash S_j$ which satisfy the Dirac
equation
$(m-D_k)F=0$ outside $S_j$ and such that $F\in H^1({\bf D}_R\backslash
S_j).$ }
\vskip.1in \par\noindent
Note: For both of the preceeding definitions the functions of
interest are in the first Sobolev space $H^1$ near the boundary and
so the Sobolev embedding theorem implies that the boundary values
are assumed in the $H^{{1\over 2}}$ norm on the boundary.
\vskip.1in \par\noindent
Finally we let $\wext$ denote the subspace of boundary values
on $\partial S$ of functions $F$ defined in ${\bf D}_R\backslash
S$ which satisfy the Dirac
equation $(m-D_k)F=0$ in the exterior of $S$ and such that
$F\in H^1({\bf D}_R\backslash S).$  Note that this subspace is certainly not
the direct
sum of the subspaces $\wext(a_j)$.  The following formula defines a
projection that is fundamental for us,
$$P^{a,\lambda}f_j(x)=2\int_{\partial S}\left\{{{\overline {G_{j1}^{
a,\lambda}(x,y)}f_2(y)}\over {1-{{|y|^2}\over {R^2}}}}i\,d\bar y-{{\overline {
G_{j2}^{a,\lambda}(x,y)}f_1(y)}\over {1-{{|y|^2}\over {R^2}}}}i\,
dy\right\}.\eqno (6.2)$$
\vskip.2in \par\noindent\
{\bf Theorem} 6.0 {\sl The subspaces $\wint(a)$ and $\wext$ are transverse to
one
another in $H^{{1\over 2}}(\partial S).$  The operator $P^{a,\lambda}$ defined
in (6.2) above is the
projection on $\wint(a)$ along $\wext$.}
\vskip.1in \par\noindent
The proof of this theorem is parallel to the calculation used to
establish (4.7) in reference [12] and so we will not repeat it here.

There is another projection that will be important for us.  Let
$P_j$ denote the projection of $H^{{1\over 2}}(\partial S_j)$ on $\wint
(a_j)$ along $\wext(a_j)$.
We define
$$F(a)=P_1\oplus P_2\oplus\cdots\oplus P_n.\eqno (6.3)$$
We are now prepared to discuss the determinant bundle formalism
that will allow us to define a $\tau$-function for the singular Dirac
operator $D^{a,\lambda}$.   Fix a point $a^0=(a_1^0,a_2^0,\ldots
,a_n^0)$ with $a_j^0\ne a_k^0$ if
$j\ne k$.  Fix a collection of pairwise disjoint open strips $S_j$ with
$a_j^0\in S_j.$  Choose $\epsilon >0$ so that for each $j=1,2,\ldots
,n$ the ball
of radius $\epsilon$ about $a_j^0$ is contained in $S_j$.  That is,
$$B_{\epsilon}(a_j^0)\subset S_j.$$
Let $H=H^{{1\over 2}}(\partial S)$ and write
$$H=\wint(a^0)\oplus\wext\eqno (6.4)$$
for a distinguished splitting of the Hilbert space $H$.  Write
$P_0$ for the projection of $H$ on $\wint(a^0)$ along $\wext.$  We are
interested in the Grassmannian, $\hbox{Gr}_0$, of subspaces of $H$ which are
close to $\wint(a^0)$ in the following sense.  A closed subspace $
W$ is
in $\hbox{Gr}_0$ provided that the map,
$$P_0:W\rightarrow\wint(a^0),\eqno (6.5)$$
is Fredholm with index 0, and the map,
$$(I-P_0):W\rightarrow\wext,\eqno (6.6)$$
is compact.  Except that we have additionally specified the index of
the first projection this is essentially the definition of the
Grassmannian that Segal and Wilson consider in [16] and the reader
can find a detailed theory of such Grassmannians in the book
[14] by
Segal and Pressley.  The restriction to index 0 in (6.5) means that we
confine our attention to the {\it connected component} of the
Grassmannian, $\hbox{Gr}_0$, containing $\wint(a^0)$.   It is simpler to
discuss the
$\hbox{det}^{*}$ bundle over the connected component of the Grassmannian and
it will suffice for our purposes.  The first result we require is
\vskip.1in \par\noindent
{\bf Theorem} 6.1 {\sl If $a_j\in B_{\epsilon}(a_j^0)$ then $\wint
(a)\in\hbox{Gr}_0$. }
\vskip.1in \par\noindent
The proof of this theorem is precisely parallel to the proof of
Theorem 4.1 in [12] and so we refer the reader to this paper for
details.
\vskip.1in \par\noindent
The localization of the singular Dirac operator we are interested in
can now be described as follows.  The operator, $D^{a,\lambda},$ acts as the
ordinary
Dirac operator in the exterior of the union of strips $S$ and its
domain is the subspace of $H^1({\bf D}_R\backslash S)$ whose
boundary values belong to the subspace $\wint(a)$.  We define a
determinant for this family of Dirac operators by trivializing the
$\hbox{det}^{*}$ bundle over the family of subspaces $\wint(a)\in\hbox{Gr}_
0$.  This
trivialization is then compared with the canonical section for the
$\hbox{det}^{*}$ bundle and the result defines the $\tau -$function for the
Dirac
operator.
\vskip.1in \par\noindent
Recall that an invertible linear map $F:\wint(a^0)\rightarrow W$ is an {\it
admissable
frame} for the subspace $W\in\hbox{Gr}_0$ provided that
$P_0F:\wint(a^0)\rightarrow\wint(a^0)$ is a trace class perturbation of the
identity.
We will now introduce two different frames for the subspace
$\wint(a).$ The first such frame will define the canonical section of
the $\hbox{det}^{*}$ bundle and the second will be used to provide the
trivialization of $\hbox{det}^{*}$ over the family or subspaces
$$a\rightarrow\wint(a)\in\hbox{Gr}_0.$$
First we show that the restriction $P^{a,\lambda}:\wint(a^0)\rightarrow\wint
(a)$ inverts
the projection $P_0:\wint(a)\rightarrow\wint(a^0)$.  From this it follows that
the restriction of $P^{a,\lambda}$ to $\wint(a^0)$ is an admissable frame for
$\wint(a)$ which defines the {\it canonical section} of the $\hbox{det}^{
*}$ bundle.
The argument is simple.  Suppose that $w\in\wint(a^0)$ and
$w=w_a+w_e$ with $w_a\in\wint(a)$ and $w_e\in\wext.$ Then $P^{a,\lambda}
w=w_a$ and
writing $w_a=w-w_e$ we see that $P_0w_a=w.$ Evidently, this is just
an expression of the fact that the complementary subspace $\wext$ is
the same for both projections $P_0$ and $P^{a,\lambda}.$

The second frame arises from considering the direct sum of ``one
point'' projections:
\vskip.1in \par\noindent
{\bf Proposition} 6.2 {\sl The restriction of $F(a)$ to the subspace $\wint
(a^0)$
is an admissable frame for $\wint(a)$. }
\vskip.1in \par\noindent
Once again this is identical to Proposition 4.2 in [12] and we refer
the reader to that paper for more details.
\vskip.1in \par\noindent
We now recall that the fiber in the $\hbox{det}^{*}$ bundle over a subspace
$W\in\hbox{Gr}_0$ can be identified with equivalence classes of pairs $
(w,\alpha )$
where $w:\wint(a^0)\rightarrow W$ is an admissable frame and $\alpha$ is a
complex
number.  The equivalence relation which defines the fiber is
$(w_1,\alpha_1)=(w_2,\alpha_2)$ if and only if
$$\alpha_1=\alpha_2\hbox{det}(w_2^{-1}w_1).$$
In this representation the canonical section of the $\hbox{det}^{
*}$ bundle is
given by
$$\hbox{Gr}_0\ni W\rightarrow (w,\hbox{det}(P_0w))\in\hbox{det}^{
*}$$
where $w$ is any admissable frame for $W.$  Since $P_0P^{a,\lambda}$ is
the identity on $\wint(a^0)$ it follows that we may regard
$$\hbox{Gr}_0\ni\wint(a)\rightarrow (P^{a,\lambda}|_0,1)$$
as a representation of the canonical section $\sigma (\wint(a)).$  We use
the notation $P^{a,\lambda}|_0$ to signify the restriction of the projection
$P^{a,\lambda}$ to $\wint(a^0)$.  We now use $F(a)$ to define a trivialization
of
the $\hbox{det}^{*}$ bundle over the family of subspaces $\wint(a
).$  Define
$$\delta (\wint(a))=(F(a)|_0,1).$$
We may then define a determinant $\tau (a,a^0)$ for the Dirac operator
as follows,
$$\tau (a,a^0)={{\sigma (\wint(a))}\over {\delta (\wint(a))}}=\hbox{det}_
0(P^{a,\lambda}|_0^{-1}F(a))=\hbox{det}_0(F(a)|_0^{-1}P^{a,\lambda}
)^{-1}.\eqno (6.7)$$
In this formula $\hbox{det}_0$ refers to the determinant of an operator
restricted to the subspace $\wint(a^0)$.  The second form of the
determinant as a reciprocal is included since it will be slightly
simpler for us to calculate the logarithmic derivative of the
$\tau$-function in this form.  We can now state one of the principal
results of this paper:
\vskip.1in \par\noindent
{\bf Theorem} 6.3 {\sl\ The $\tau -$function defined by (6.7) has logarithmic
derivative given by, }
$$d\log\tau =\sum_{j=1}^n{1\over {1-|b_j|^2}}\left\{a_{{1\over 2}
j}^jdb_j+d_{{1\over 2}j}^jd\bar b_j\right\},\eqno (6.8)$$
{\sl where the coefficients $a_{{1\over 2}j}^j$ and $d_{{1\over 2}
j}^j$ are the local expansion
coefficients for $W_j$ and $W_j^{*}$ found in (3.22) and (3.25) above. }
\vskip.1in \par\noindent
Proof.  For simplicity we write $P(a)=P^{a,\lambda}$ in the
following calculation.  We also write $d$ for the exterior derivative
with respect to $\{a_1,a_2,\ldots ,a_n\}.$  The rule for differentiating
determinants when applied to the final term in (6.7) above
yields,
$$d\log\tau (a,a^0)=-\hbox{Tr}_0(d(F(a)^{-1}P(a))P(a)^{-1}F(a)),$$
where $\hbox{Tr}_0$ is the trace on $\wint(a^0)$ and it is understood that both
$F(a)$ and $P(a)$ are regarded as maps from $\wint(a^0)$ to $\wint
(a)$ in
this last formula.  However, we know that $P(a^0)$ inverts $P(a)$
restricted to $\wint(a^0)$.  Thus
$$P(a)^{-1}F(a)=P(a^0)F(a)$$
when both sides are restricted to $\wint(a^0).$  In a precisely similar
fashion $F(a^0)$ inverts $F(a)$ restricted to $\wint(a^0).$  Thus we
find,
$$d\log\tau (a,a^0)=-\hbox{Tr}_0(F(a^0)d(P(a))P(a^0)F(a)).$$
But $P(a)(I-P(a^0)=0$ so that $d(P(a))P(a^0)=dP(a)$ where $dP(a)$ is
now regarded as a map on all of $H.$ Combining this observation with
the fact that the range of $F(a^0)$ is all of $\wint(a^0)$ we can remove
the subspace restriction on the trace to find,
$$d\log\tau (a,a^0)=-\hbox{Tr}(F(a^0)(dP(a))F(a))=-\hbox{Tr}((dP(
a))F(a)F(a^0)).$$
Since $F(a)(I-F(a^0))=0$ we find that $F(a)F(a^0)=F(a).$  Thus
$$d\log\tau (a,a^0)=-\hbox{Tr}((dP(a))F(a)).\eqno (6.9)$$
The reader might note that although $\tau (a,a^0)$ depends on $a^
0$ its
logarithmic derivative does not!  Since it is clear that $\tau (a^
0,a^0)=1$
from the definition, the fact that the logarithmic derivative is
independent of $a^0$ implies that there exists a function $\tau (
a)$ such
that
$$\tau (a,a^0)={{\tau (a)}\over {\tau (a^0)}}.$$
The function $\tau (a)$ is the function which we would like to identify
as a correlation function in a quantum field theory.  This will have
to await further developments.  However, one should observe that
$$d\log\tau (a,a^0)=d\log\tau (a).$$
We will now use the derivative formula (4.50) and the formula
(6.2) for the projection to evaluate the trace (6.9) in terms of the low
order expansion coefficients for the wave functions $W_j$ and $W_
j^{*}.$
The derivative formula (4.50) and the projection formula (6.2)
together imply,
$$\eqalign{\left(\partial_{b_{\nu}}P(a)f\right)(x)&={{m^2RW_{\nu}
(x,m)}\over {4s_{\nu}(1-|b_{\nu}|^2)}}\int_{\partial S}{{\overline {
W_{\nu}^{*}(y,-m)}_1}\over {1-{{|y|^2}\over {R^2}}}}f_2(y)i\,d\bar {
y}\cr
&-{{m^2RW_{\nu}(x,m)}\over {4s_{\nu}(1-|b_{\nu}|^2)}}\int_{\partial
S}{{\overline {W_{\nu}^{*}(y,-m)}_2}\over {1-{{|y|^2}\over {R^2}}}}
f_1(y)i\,dy,\cr}
\eqno (6.10)$$
and
$$\eqalign{\left(\bar\partial_{b_{\nu}}P(a)f\right)(x)&={{m^2RW_{
\nu}^{*}(x,m)}\over {4s_{\nu}(1-|b_{\nu}|^2)}}\int_{\partial S}{{\overline {
W_{\nu}(y,-m)}_1}\over {1-{{|y|^2}\over {R^2}}}}f_2(y)i\,d\bar {y}\cr
&-{{m^2RW_{\nu}^{*}(x,m)}\over {4s_{\nu}(1-|b_{\nu}|^2)}}\int_{\partial
S}{{\overline {W_{\nu}(y,-m)}_2}\over {1-{{|y|^2}\over {R^2}}}}f_
1(y)i\,dy\cr}
\eqno (6.11)$$
The appropriate trace is,
$$\eqalign{-\hbox{Tr}(F(a)\partial_{b_{\nu}}P(a))=&-{{m^2R}\over {
4s_{\nu}(1-|b_{\nu}|^2)}}\int_{\partial S}{{\overline {W^{*}_{\nu}
(y,-m)}_1F(a)W_{\nu}(y,m)_2}\over {1-{{|y|^2}\over {R^2}}}}i\,d\bar {
y}\cr
&{{m^2R}\over {4s_{\nu}(1-|b_{\nu}|^2)}}\int_{\partial S}{{\overline {
W_{\nu}^{*}(y,-m)}_2F(a)W_{\nu}(y,m)_1}\over {1-{{|y|^2}\over {R^
2}}}}i\,dy,\cr}
\eqno (6.12)$$
and
$$\eqalign{-\hbox{Tr}(F(a)\bar{\partial}_{b_{\nu}}P(a))=&-{{m^2R}\over {
4s_{\nu}(1-|b_{\nu}|^2)}}\int_{\partial S}{{\overline {W_{\nu}(y,
-m)}_1F(a)W_{\nu}^{*}(y,m)_2}\over {1-{{|y|^2}\over {R^2}}}}i\,d\bar {
y}\cr
&{{m^2R}\over {4s_{\nu}(1-|b_{\nu}|^2)}}\int_{\partial S}{{\overline {
W_{\nu}(y,-m)}_2F(a)W_{\nu}^{*}(y,m)_1}\over {1-{{|y|^2}\over {R^
2}}}}i\,dy.\cr}
\eqno (6.13)$$
According to (3.5) both of the integrands in (6.12) and (6.13) are
exact away from the singularities at $a_j$ for $j=1,2,\ldots ,n$.  The contours
$\partial S_j$ can be closed about the singularities $a_j$ and the same
asymptotic calculation that leads to (4.4) above shows that,
$$-\hbox{Tr}(F\partial_{b_{\nu}}P)={1\over {1-|b_{\nu}|^2}}a(FW_{
\nu})_{{1\over 2}}^{\nu}\eqno (6.14)$$
and
$$-\hbox{Tr}(F\bar{\partial}_{b_{\nu}}P)={1\over {1-|b_{\nu}|^2}}
b(FW_{\nu}^{*})_{{1\over 2}}^{\nu},\eqno (6.15)$$
where we used the notation,
$$g(x)=\sum_n\left\{a(g)^{\nu}_nw_{n+\lambda_{\nu}}+b(g)^{\nu}_nw^{
*}_{n-\lambda_{\nu}}\right\},$$
for the local expansion coefficients of $g$ about the point $a_{\nu}$.  Note
that the contours $\partial S_j$ with $j\ne\nu$ do not make a contribution in
this
calculation.  To compute $FW_{\nu}$ and $FW_{\nu}^{*}$  on $\partial
S_{\nu}$ we need only
subtract from $W_{\nu}$ and $W_{\nu}^{*}$ the ``one point versions'' of these
two
functions (just the translate of the appropriate multiple of the
function $\hat {w}_{-{1\over 2}+\lambda_{\nu}}$ to $a_{\nu}$).  Doing this we
find,
$$a(FW_{\nu})_{{1\over 2}}^{\nu}=a_{{1\over 2}\nu}^{\nu}$$
and
$$b(FW_{\nu}^{*})_{{1\over 2}}^{\nu}=d_{{1\over 2}\nu}^{\nu}.$$
This finishes the proof of Theorem 6.3. QED
\vskip.1in\par\noindent
We will now provide the connection between the coefficients in the
formula for the logarithmic derivative of $\tau$ given in (6.8) and
the deformation theory of \S 5.  If one collects all the terms in
(5.16) that involve $a_{{1\over 2}j}^j$ on the left hand side using $
e_{jj}=k+\lambda_j$
then one finds,
$$a_{{1\over 2}j}^j=\bar {b}_jm_2(\lambda_j-\half)+\sum_{\mu\ne j}
e_{j\mu}a_{{1\over 2}\mu}^j+\sum_{\mu}f_{j\mu}\bar {b}_{\mu}c_{{1\over
2}\mu}^j.\eqno (6.16)$$
The analogous calculation for $d_{{1\over 2}j}^j$ yield,
$$d_{{1\over 2}j}^j=b_jm_1(-\lambda_j-\half)-\sum_{\mu\ne j}h_{j\mu}
d_{{1\over 2}\mu}^j-\sum_{\mu}g_{j\mu}b_{\mu}b_{{1\over 2}\mu}^j.\eqno
(6.17)$$
The right hand sides of both (6.16) and (6.17) are completely
determined by the deformation variables $e,f,g$ and $h$.
\vskip.1in \par\noindent
{\bf The log derivative of $\tau$ for the two point function}.  In
this subsection we will write out the formula for the logarithmic
derivative of the tau
function in terms of the Painlev\'e transcendent of the sixth
kind that was found in the integration of a special case of the
deformation equations described at the end of \S 5.  We suppose
therefore that $\lambda_j>0$ for $j=1,2$ and that $b_1=0$ and $b_
2=t$.
In this special case (6.8) becomes,
$${{\partial}\over {\partial t}}\log\tau (0,t)={1\over {1-t^2}}\left
\{a^2_{{1\over 2},2}+d_{{1\over 2},2}^2\right\}.$$
 From (6.16) and (6.17) one finds,
$$a_{{1\over 2},2}^2=t\,m_2(\lambda_2-\half)+e_{21}a^2_{{1\over 2}
,1}+tf_{22}c^2_{{1\over 2},2}$$
and
$$d_{{1\over 2},2}^2=t\,m_1(-\lambda_2-\half)-h_{21}d_{{1\over 2}
,1}^2-tg_{22}b_{{1\over 2},2}^2.$$
Expressing the off diagonal elements of $a$, $b$, $c$ and $d$ in terms of the
deformation parameters one finds,
$$\eqalign{a_{{1\over 2},1}^2=&t^{-1}(1-t^2)e_{12}\,,\cr
b_{{1\over 2},2}^2=&-f_{22}\,,\cr
c_{{1\over 2},2}^2=&g_{22}\,,\cr
d_{{1\over 2},1}^2=&-t^{-1}(1-t^2)h_{12}.\cr}
$$
Thus,
$$\eqalign{a_{{1\over 2},2}^2&=t\,m_2(\lambda_2-\half)+t^{-1}(1-t^
2)e_{21}e_{12}+tf_{22}g_{22}\cr
d_{{1\over 2},2}^2&=t\,m_1(-\lambda_2-\half)+t^{-1}(1-t^2)h_{21}h_{
12}+tf_{22}g_{22}.\cr}
$$
Now we observe that
$$m_2(\lambda_2-\half)=m_1(-\lambda_2-\half)={{m^2R^2}\over 4}+k^
2-(k+\lambda_2)^2$$
and since
$$\eqalign{{\bf g}&=s^{-1}(1-|B|^2)^{-1}g\,,\cr
{\bf f}&=fs(1-|B|^2)\,,\cr}
$$
it follows that
$$f_{22}g_{22}={\bf f}_{22}{\bf g}_{22}.$$
Eliminating $h$ in favor of $e^{*}$ one has,
$$\eqalign{h_{12}&={{s_1}\over {s_2}}\left(1-t^2\right)^{-1}e_{12}^{
*}\,,\cr
h_{21}&={{s_2}\over {s_1}}\left(1-t^2\right)e_{21}^{*}\,,\cr}
$$
so that,
$$h_{21}h_{12}=e_{21}^{*}e_{12}^{*}=e_{21}e_{12},$$
where we have used the fact that $e_{21}e_{12}=\bar {e}_{-}e_{+}$ is real.
Thus
$${{\partial}\over {\partial t}}\log\tau (0,t)={1\over {1-t^2}}\left
\{2tm_2(\lambda_2-\half)+2t^{-1}(1-t^2)\bar e_{-}e_{+}+2t{\bf f}_{
22}{\bf g}_{22}\right\}.\eqno (6.18)$$
Using (5.53) (5.57) and (5.58) one finds,
$$\eqalign{{\bf f}_{22}&=\kappa^{-1}(1-t^2)\hbox{ch}\,\psi\,,\cr
{\bf g}_{22}&={{-\bar{\epsilon}\mu e_{+}\hbox{sh}\psi +\kappa (k_
2^2-M^2+\bar {e}_{-}e_{+})\hbox{ch}\psi}\over {1-t^2\hbox{ch}^2\psi}}\,
,\cr
&=\kappa{{-\mu{{\xi +\lambda {\rm t}{\rm h}\,\psi}\over {2(1-t^2)}}\hbox{sh}
\psi +\left(k_2^2-M^2+{{\xi^2-\lambda^2{\rm t}{\rm h}^2\psi}\over {
4(1-t^2)}}\right)\hbox{ch}\psi}\over {1-t^2\hbox{ch}^2\psi}},\cr}
$$
so that,
$${\bf f}_{22}{\bf g}_{22}={{-2\mu (\xi +\lambda\hbox{th}\psi )\hbox{ch}
\psi\,\hbox{sh}\psi +(\xi^2-\lambda^2\hbox{th}^2\psi )\hbox{ch}^2
\psi +4(1-t^2)(k_2^2-M^2)\hbox{ch}^2\psi}\over {4(1-t^2\hbox{ch}^
2\psi )}}.$$
As a check on (6.18) we can, as we did for (5.59), introduce
$t={r\over R}$ and then take the limit $R\rightarrow\infty$.  One finds that
the limit
of $d\log\tau (0,{r\over R})$ as $R\rightarrow\infty$ is
$${1\over {2r}}\left(r^2\left({{\partial\psi}\over {\partial r}}\right
)^2-\lambda^2{\rm t}{\rm h}^2\psi -m^2r^2{\rm s}{\rm h}^2\psi\right
)dr$$
This should be compared with (4.5.42) in [15] and (5.39) in
[12].  The overall sign difference with (4.5.42) is a consequence
of the fact that (4.5.42) is a formula for the log derivative of
the {\it Bosonic} tau function in SMJ.

Next we make the substitutions, $w=1/\hbox{ch}^2\psi$ and  $s=t^2$ in the
expression for ${\bf f}_{22}{\bf g}_{22}$.  Substitute the result in (6.18)
along with (5.54) for $e_{+}$ and $\bar {e}_{-}$.  Eliminate $\xi$ in the
result
with the substitutions
$$\eqalign{\xi\hbox{th}\,\psi =&{1\over w}\left(s(1-s)w'+s(1-\mu
)(w-1)\right),\cr
\xi^2=&{{\left(s(1-s)w'+s(1-\mu )(w-1)\right)^2}\over {w^2(1-w)}}
.\cr}
$$
Noting that
$${1\over {2t}}{{\partial}\over {\partial t}}\log\tau (0,t)={{\partial}\over {
\partial s}}\log\tau (0,\sqrt s),$$
we find after some calculation
$$\eqalign{{{\partial}\over {\partial s}}\log\tau (0,\sqrt s)={{s
(s-1)}\over {4w(w-1)(w-s)}}\left({{dw}\over {ds}}-{{w-1}\over {s-
1}}\right)^2-{{M^2}\over {w-s}}\cr
-{{\mu^2}\over {4(s-1)w}}-{{\lambda^2w}\over {4s(s-1)}}-{{\lambda^
2}\over {4s}}+{{4M^2-\mu^2-\lambda^2}\over {4(1-s)}}.\cr}
\eqno (6.19)$$
The formula on the right hand side of this last equation bears
some striking resemblance to the formula in Okamoto [7] for
a Hamiltonian associated to the Painlev\'e VI equation.  It does
not, however, appear to be a Hamiltonian for $P_{VI}$ for any
simple choice of canonical coordinates.

It would be interesting to study the behavior of this tau
function as $t\rightarrow 0,1$.  The asymptotics for $t$ near 0 should match
onto the asymptotics of the analogous tau function in the Euclidean
domain.  The behavior as $t\rightarrow 1$ would follow if one knew the
appropriate connection formula for $P_{VI}.$  We hope to return to
this question elsewhere.
\vskip.2in\hfill\break
\centerline{{\bf References}}
\vskip.2in \par\noindent
[1] E.~Barouch, B.~M.~McCoy,  T.~T.~Wu,
{\sl\ Zero-field susceptibility of the two-dimensional Ising model
near $T_c$\/}, Phys.\  Rev.\  Lett.\  {\bf 31} (1973), 1409--1411.
\vskip.1in \par\noindent
[2] J.~D.~Fay, {\sl Fourier coefficients for the resolvent for a
Fuchsian group\/}, J.\ Reine Angew.\  Math.\ {\bf 294} (1977), 143--203.
\vskip.1in \par\noindent
[3] P.~Hartman, {\sl Ordinary Differential Equations\/} (John Wiley \& Sons,
1964).
\vskip.1in \noindent
[4] E.~L.~Ince, {\sl Ordinary Differential Equations\/} (Dover, New York,
1947).
\vskip.1in
\noindent
[5] M.~Jimbo, T.~Miwa, K.~Ueno, {\sl Monodromy preserving deformation theory
of differential equations with rational coefficients. I. General theory and the
$\tau$ function\/}, Physica {\bf 2D} (1981), 306--352.
\vskip.1in \par\noindent
[6] R.~Narayanan,  C.~A.~Tracy, {\sl Holonomic quantum field theory of
Bosons in the Poincar\'e disk and the zero curvature limit\/}, Nucl.\  Phys.\
{\bf B340} (1990),  568--594.
\vskip.1in \par\noindent
[7] K.~Okamoto, {\sl On the $\tau -$function of the Painlev\'e equations\/},
Physica {\bf 2D} (1981),  525--535;  {\sl Studies on the Painlev\'e equations.
I. Sixth Painlev\'e $P_{VI}$\/}, Annali di Mathematica Pura and
Applicata, vol. {\bf CXLVI} (1987),  337--381.
\vskip.1in \par\noindent
[8] B.~M.~McCoy, C.~A.~Tracy,  T.~T.~Wu, {\sl Painlev\'e functions of
the third kind\/}, J.\  Math.\  Phys.\  {\bf 18} (1977), 1058--1092.
\vskip.1in
\noindent
[9] J.~Palmer, C.~Tracy, {\sl Two dimensional Ising correlations: the SMJ
analysis\/}, Adv.\ in Appl.\  Math.\ {\bf 4} (1983), 46--102.
\vskip.1in \par\noindent
[10] J.~Palmer, {\sl Monodromy fields on ${\bf Z}^2$\/}, Commun.\  Math.\
 Phys.\ {\bf 102}
(1985), 175--206.
\bigbreak
\vskip.1in \par\noindent
[11] J.~Palmer, {\sl Determinants of Cauchy-Riemann operators as $
\tau$
functions\/}, Acta Applicandae Mathematicae {\bf 18} (1990), 199--223.
\vskip.1in
\noindent
[12] J.~Palmer, {\sl Tau functions for the Dirac operator on the Euclidean
plane\/}, Pacific J.\   Math., to appear
\vskip.1in
\noindent
[13] J.~Palmer, C.~A.~Tracy, {\sl Monodromy preserving deformations of the
Dirac operator on the hyperbolic plane\/}, in ``Mathematics of Nonlinear
Science: proceedings of an AMS special session held January 11--14, 1989,"
M.~S.~Berger ed., Contemporary Mathematics {\bf 108},  119--131.
\vskip.1in
\noindent
[14] A.~Pressley, G.~Segal, {\sl Loop Groups\/}, (Clarendon Press, Oxford
     1986)
\vskip.1in
\noindent
 [15] M.~Sato, T.~Miwa, M.~Jimbo, {\sl Holonomic Quantum Fields
I--V\/},
     Publ.\  RIMS, Kyoto Un., {\bf 14} (1978), 223--267;  {\bf 15} (1979),
201--
     278;  {\bf 15} (1979),  577--629;  {\bf 15} (1979),  871--972;  {\bf 16}
     (1980), 531--584.
\vskip.1in
\noindent
[16] G.~Segal, G.~Wilson, {\sl Loop groups and equations of KdV
     type\/},
      Pub.\  Math.\  I.H.E.S.{\bf 61} (1985), 5--65.
\vskip.1in \par\noindent
[17] C.~A.~Tracy,  B.~M.~McCoy, {\sl\ Neutron scattering and the
correlations of the Ising model near $T_c$\/}, Phys.\  Rev.\
 Lett.\ {\bf 31} (1973),
1500--1504.
\vskip.1in \par\noindent
[18] C.~A.~Tracy, {\sl Monodromy preserving deformation theory for the
Klein-Gordon
equation in the hyperbolic plane\/}, Physica {\bf 34D} (1989), 347--365.
\vskip.1in
\noindent
[19] C.~A.~Tracy, {\sl Monodromy preserving deformations of linear ordinary and
partial differential equations\/} in ``Solitons in Physics, Mathematics, and
Nonlinear Optics," P.~J.~Olver and D.~H.~Sattinger (eds.), Springer-Verlag,
NY (1990), 165--174.
\vskip.1in
\noindent
[20] T.~T.~Wu, B.~M.~McCoy, C.~A.~Tracy,  E.~Barouch, {\sl Spin-
     spin correlation functions for the two dimensional Ising model:
     Exact theory in the scaling region\/}, Phys.\  Rev.\   {\bf B13} (1976),
     316--374.
\bye